\documentclass[aps,prb,10pt,twocolumn,amsfonts,superscriptaddress,showpacs,floatfix]{revtex4-1}

\usepackage{color}
\usepackage{bm}
\usepackage{bbold}
\usepackage{amsmath}
\usepackage{amssymb}
\usepackage{epsfig}

\newcommand{\be}{\begin{equation}}
\newcommand{\ee}{\end{equation}}
\newcommand{\bea}{\begin{eqnarray}}
\newcommand{\eea}{\end{eqnarray}}
\newcommand{\ba}{\begin{aligned}}
\newcommand{\ea}{\end{aligned}}

\newcommand{\la}{\langle}
\newcommand{\ra}{\rangle}
\newcommand{\dg}{^\dagger}

\def\nn{\nonumber\\}

\def\fr#1{(\ref{#1})}

\def\vv{\tilde{V}}

\usepackage[colorlinks,bookmarks=false,citecolor=blue,linkcolor=red,hyperfootnotes=true]{hyperref}

\begin{document}

%%%%%%%%%%%%%%%%%%%%%%%%%%%%%%%%%%%%%%%%%%%%%%%%%%%%%%%%%%%%%%%%%%%
\title{Quench Dynamics in a Model with Tuneable Integrability Breaking}
%%%%%%%%%%%%%%%%%%%%%%%%%%%%%%%%%%%%%%%%%%%%%%%%%%%%%%%%%%%%%%%%%%%
\author{F.H.L. Essler}
\affiliation{The Rudolf Peierls center for Theoretical Physics, Oxford
  University, Oxford, OX1 3NP, United Kingdom}
\author{S. Kehrein}
\affiliation{Institut f\"ur Theoretische Physik,
  Georg-August-Universit\"at 
G\"ottingen, D-37077 G\"ottingen, Germany}
\author{S.R. Manmana}
\affiliation{Institut f\"ur Theoretische Physik,
  Georg-August-Universit\"at 
G\"ottingen, D-37077 G\"ottingen, Germany}
\author{N.J. Robinson}
\affiliation{The Rudolf Peierls center for Theoretical Physics, Oxford
  University, Oxford, OX1 3NP, United Kingdom}

\date{\today}

\begin{abstract}
We consider quantum quenches in an integrable quantum chain with 
tuneable-integrability-breaking interactions. In the case where these
interactions are weak, we demonstrate that at intermediate times after
the quench local observables relax to a prethermalized regime, which
can be described by a density matrix that can be viewed as a
deformation of a generalized Gibbs ensemble. We present explicit
expressions for the approximately conserved charges characterizing
this ensemble. We do not find evidence for a crossover from the
prethermalized to a thermalized regime on the time scales accessible
to us. Increasing the integrability-breaking interactions 
leads to a behavior that is compatible with eventual thermalization.
\end{abstract}

%\keywords{quantum quenches, integrable models}

\maketitle
%%%%%%%%%%%%%%%%%%%%%%%%%%%%
\section{Introduction}
%%%%%%%%%%%%%%%%%%%%%%%%%%%%

Important advances in manipulating cold atomic gases have allowed
recent experiments~\cite{BlochNature02,WeissNature06,SchmiedmayerNature07,
BlochNatPhys12,KuhrNature12,SchmiedmayerScience12} to realize
essentially unitary time evolution for extended periods of
time. Stimulated by such experiments, there has been immense
theoretical effort (see, e.g., Ref.~\onlinecite{PSSV_RMP11} for a recent review) to
understand fundamental questions about the nonequilibrium dynamics of
quantum systems: Do observables in a subsystem relax to stationary
values? If so, can expectation values be reproduced  with a thermal
density matrix? What governs how and to which values observables relax?  

It is generally accepted that conservation laws and dimensionality play
important roles in the time evolution of isolated quantum systems. This
is highlighted by the ground-breaking experiments of Kinoshita, Wenger
and Weiss~\cite{WeissNature06}. There, it was found that a
three-dimensional condensate of $^{87}$Rb atoms driven out of
equilibrium rapidly relaxed to a thermal state (``thermalized''),
whilst a condensate constrained to move in a single spatial dimension
relaxed slowly to a nonthermal ensemble. It is thought that the
presence of additional (approximate) conservation laws in the
one-dimensional case lies at the heart of this difference.

Theoretical investigations on translationally invariant models
have established two central paradigms for the late time behavior
after a quantum quench:
(1) subsystems thermalize and are then described by a Gibbs
ensemble (GE) \cite{nonint};
(2) subsystems do not thermalize, but at late times after the quench
are described by Generalized Gibbs ensembles (GGE). There is
substantial evidence
\cite{BS_PRL08,CrEi,CEF_PRL11,FE_PRB13,CC_JStatMech07,IC_PRA09,BKL_PRL10,FM_NJPhys10,EEF:12,Pozsgay_JStatMech11,CCR_PRL11,CIC_PRE11,CK_PRL12,CE_PRL13,MC_NJPhys12,FE:13b,Pozsgay:13a,Mussardo_arXiv13,CSC:13b,KCC:13a,KSCCI}
that the latter case applies quite generally to quenches in quantum
integrable models, as suggested in a seminal paper by Rigol {\textit{et al}}.\cite{GGE}

The dichotomy in the dependence of stationary behavior after a quench
on integrability then poses an intriguing question: what happens if
integrability is weakly broken? Does the system thermalize, and if so,
how fast does it relax? Might there be an intermediate time scale
still governed by the physics of integrability?

Early numerical studies~\cite{Manmana0709} suggested that 
even with an integrability breaking term the system does not
thermalize on the accessible time scales and system sizes.
Studies using analytical methods\cite{Moeckel080910} for $d>1$ 
and numerical methods in the dynamical mean field limit\cite{KollarPRB11} ($d\to\infty$)
showed that on intermediate time scales the system approaches a nonthermal
quasistationary state (a prethermalization plateau). At later times
the system is expected to thermalize.\cite{BKL_PRL10,thermalize}
Prethermalization plateaus have also been observed in a nonintegrable quantum 
Ising chain with long-range interactions.\cite{MMGS_arXiv13}
It has been suggested recently,\cite{BCK_arXiv13} that the time scale
for integrability breaking (leaving the prethermalization plateau) is
not necessarily related to the strength of the integrability breaking
term. Experimental evidence for the prethermalization plateau in
systems of bosonic cold atoms was reported in
Refs.~\onlinecite{SchmiedmayerNJP11,SchmiedmayerScience12,SchmiedmayerEPJ13}.   
In spite of the aforementioned works exhibiting prethermalization
plateaus in specific models, a general understanding of if, when and
how such plateaus emerge when integrability is broken remains
open. Similarly, a precise characterization of such plateaus in terms
of statistical ensembles has not been achieved.

In this work we study the effects of integrability breaking
interactions on the dynamics following a quantum quench. Our setup
allows us to compare integrable quantum quenches to quenches where an
additional integrability-breaking interaction is added to the post-quench
Hamiltonian. By combining analytical calculations with time-dependent
density matrix renormalization group (t-DMRG) results we 
demonstrate the existence of a prethermalization plateau in the sense
that local observables relax to nonthermal values at intermediate times.
We characterize this prethermalization plateau in terms of a
novel statistical description, that we call the ``deformed GGE''. 

This paper is organized as follows. In Sec.~\ref{Sec:Model} we introduce 
the model under study. In section~\ref{Sec:IntegQuench} we consider
integrable quenches and compare the observed stationary behavior to
thermal and generalized Gibbs ensembles. The continuous unitary
transformation technique is introduced and used to study a weakly
nonintegrable quench of the model in Sec.~\ref{Sec:NonIntQuench}. In
Sec.~\ref{Sec:Pretherm} we establish the existence of the
prethermalized regime and describe the approximately stationary
behavior in this regime by constructing a ``deformed GGE''. The
dynamics in the presence of strong integrability-breaking interactions
is studied numerically in Sec.~\ref{Sec:tdmrg}. Sec.~\ref{Sec:Conc}
contains a summary and discussion of our main results. Technical details
underpinning our analysis are consigned to several appendices.

%%%%%%%%%%%%%%%%%%%%%%%%%%%%
\section{The Model}
\label{Sec:Model}
%%%%%%%%%%%%%%%%%%%%%%%%%%%%
We consider the following Hamiltonian of spinless fermions with
dimerization and density-density interactions
\bea
H(\delta,U) &=&
-J\sum_{l=1}^L\left[1+(-1)^l\delta\right]\left(c^\dagger_lc_{l+1}+{\rm
  h.c.}\right)\nn
  && + U\sum_{l=1}^{L}c\dg_l c_l c\dg_{l+1}c_{l+1}\ ,
\label{H0}
\eea
with periodic boundary conditions. Here $\{c^\dagger_l,c_j\}=\delta_{l,j}$ 
and we restrict our attention to the parameter regime $J>0$, 
$U\geq 0$ and $0<\delta<1$.
We work at half-filling throughout, i.e. the total number of fermions
is $L/2$. When showing results for the time evolution of observables
we measure time in units of $J^{-1}$ throughout.
An important characteristic of $H(\delta,U)$ 
is that fermion number is conserved by virtue of the $U(1)$ symmetry
\be
c_j\longrightarrow e^{i\varphi}c_j\ ,\quad \varphi\in[0,2\pi].
\ee
The presence of the $U(1)$ symmetry is a crucial feature
of our model: on the one hand it leads to dramatic simplifications in
our analytical calculations, while at the same time it enables us to access
very late times in our t-DMRG computations (as compared to existing
studies of other nonintegrable one dimensional models).

We note that the 
Hamiltonian \fr{H0} is equivalent to a spin-1/2 Heisenberg XXZ chain 
with dimerized XX term as can be shown by means of a Jordan-Wigner 
transformation. The model with finite $U,\delta$  has previously been 
studied in order to investigate the effect of interactions
on the equilibrium dimerization of the chain.\cite{SpinlessHubPeierls1,SpinlessHubPeierls2}
Density matrix renormalization group calculations suggest that for large values
of the interaction parameter $U\gtrsim4$, the Peierls transition to a dimerized ground
state is suppressed.\cite{SpinlessHubPeierls2}

There are several limits, in which exact results on the equilibrium
phase diagram of $H(\delta,U)$ are available. Firstly, in the absence of
interactions ($U=0$) and for any value of the dimerization parameter
$\delta$ we obtain a model of a noninteracting Peierls insulator. 
Secondly, for vanishing dimerization $\delta=0$ and $U\ge0$  
a Jordan-Wigner transformation maps the model to the spin-1/2
Heisenberg XXZ chain. Finally, in the regime of small $|\delta|$ and 
$U<J$, the low-energy limit of the model is given by the integrable
sine-Gordon model.\cite{SGM}

%%%%%%%%%%%%%%%%%%%%%%%%%%%%
\subsection{Peierls insulator}
%%%%%%%%%%%%%%%%%%%%%%%%%%%%
The special case $H(\delta,0)$ describes a Peierls insulator and can
be solved by means of a Bogoliubov transformation
\be
c_l=\frac{1}{\sqrt{L}}\sum_{k>0}\sum_{\alpha=\pm}\gamma_\alpha(l,k|\delta)a_\alpha(k)\ .
\label{bogo}
\ee
Here $a_\alpha(k)$ are fermion annihilation operators fulfilling
\be
\{a_\alpha(k),a_\beta(q)\}=0\ ,\quad
\{a_\alpha(k),a^\dagger_\beta(q)\}=\delta_{\alpha,\beta}\delta_{k,q}.
\ee
The coefficients are chosen as
\be
\gamma_\alpha(l,k|\delta)= e^{-ikl} \Big[ u_\alpha(k,\delta)+v_\alpha(k,\delta)(-1)^l\Big],
\label{gamma_def}
\ee
where
\bea
v_\alpha(k,\delta) &=& \left[1 + \left|\frac{2J\cos(k) -
    \epsilon_{\alpha}(k)}{2\delta J\sin(k)} \right|^2\right]^{-1/2}\ ,\nn
u_\alpha(k,\delta) &=& i v_\alpha(k) 
\frac{2J\cos(k) - \epsilon_{\alpha}(k)}{2\delta J\sin(k)}\ ,\\
\epsilon_{\alpha}(k,\delta) &=& 2\alpha J\sqrt{\delta^2 + (1-\delta^2)\cos^2(k)}\ .
\label{dispersion}
\eea
The ``$+$'' and ``$-$'' bands are separated by an energy gap of
$4\delta J$. Finally, $\sum_{k>0}$ is a shorthand notation for the
momentum sum 
\be
\sum_{k>0}f(k)=\sum_{n=1}^{L/2} f\big(\frac{2\pi n}{L}\big).
\ee
In terms of the Bogoliubov fermions the Peierls Hamiltonian is diagonal:
\be
H(\delta,0)=\sum_{k>0}\epsilon_{\alpha}(k,\delta)a^\dagger_\alpha(k)a_\alpha(k).
\ee
 
%%%%%%%%%%%%%%%%%%%%%%%%%%%%
\subsection{Integrability-breaking interactions}
%%%%%%%%%%%%%%%%%%%%%%%%%%%%
Adding interactions to the Peierls Hamiltonian leads to a theory that
is not integrable. An exception is the low-energy limit for
$|\delta|\ll 1$, which is described by a quantum sine-Gordon model.\cite{SGM}
In the following we will be interested in the regime
$0.4\leq \delta\leq 0.8$, which is far away from this limit.
It is useful to express the density-density interaction in $H(\delta,
U)$ in terms of the Bogoliubov fermions diagonalizing $H(\delta, 0)$
\begin{widetext}
\bea
H_{\rm  int}&=&U\sum_{l=1}^Lc^\dagger_lc_lc^\dagger_{l+1}c_{l+1}=
U\sum_{k_j>0}V_{\alpha_1\alpha_2\alpha_3\alpha_4}(k_1,k_2,k_3,k_4)
a^\dagger_{\alpha_1}(k_1)
a_{\alpha_2}(k_2)
a^\dagger_{\alpha_3}(k_3)
a_{\alpha_4}(k_4)\ ,\nn
V_{\bm{\alpha}}(\bm{k})&=&\frac{1}{L^2}\sum_l
\gamma^*_{\alpha_1}(l,k_1|\delta)
\gamma_{\alpha_2}(l,k_2|\delta)
\gamma^*_{\alpha_3}(l+1,k_3|\delta)
\gamma_{\alpha_4}(l+1,k_4|\delta)\ ,\nn
&=& \frac{1}{L}e^{i(k_3-k_4)}\Big\{\delta_{k_1+k_3,k_2+k_4}\left[
w_{\alpha_1\alpha_2}(k_1,k_2)w_{\alpha_3\alpha_4}(k_3,k_4)
-x_{\alpha_1\alpha_2}(k_1,k_2)x_{\alpha_3\alpha_4}(k_3,k_4)\right]\nn
&&\hspace{2cm}+\delta_{k_1+k_3+\pi,k_2+k_4}
\left[
x_{\alpha_1\alpha_2}(k_1,k_2)w_{\alpha_3\alpha_4}(k_3,k_4)
-w_{\alpha_1\alpha_2}(k_1,k_2)x_{\alpha_3\alpha_4}(k_3,k_4)\right]\Big\}.
\label{VertexFns}
\eea
\end{widetext}
Here we have defined
\bea
w_{\alpha\beta}(k,p)&=&u^*_\alpha(k,\delta)u_\beta(p,\delta)+u\rightarrow v,\\
x_{\alpha\beta}(k,p)&=&u^*_\alpha(k,\delta)v_\beta(p,\delta)+u\leftrightarrow v.
\eea

%%%%%%%%%%%%%%%%%%%%%%%%%%%%%%%%%%%%%
\section{Integrable Quantum Quenches}
\label{Sec:IntegQuench}
%%%%%%%%%%%%%%%%%%%%%%%%%%%%%%%%%%%%%
We first consider a quantum quench of the dimerization parameter
$\delta$ in the limit of vanishing interactions $U=0$. The system is
initially prepared in the ground state $|\Psi_0\rangle$ of
$H(\delta_i,0)$, and at time $t=0$ the dimerization is suddenly
quenched from $\delta_i$ to $\delta_f$. At times $t>0$ the system
evolves unitarily with the new Hamiltonian $H(\delta_f,0)$.

\begin{figure}[ht]
\includegraphics[width=0.42\textwidth]{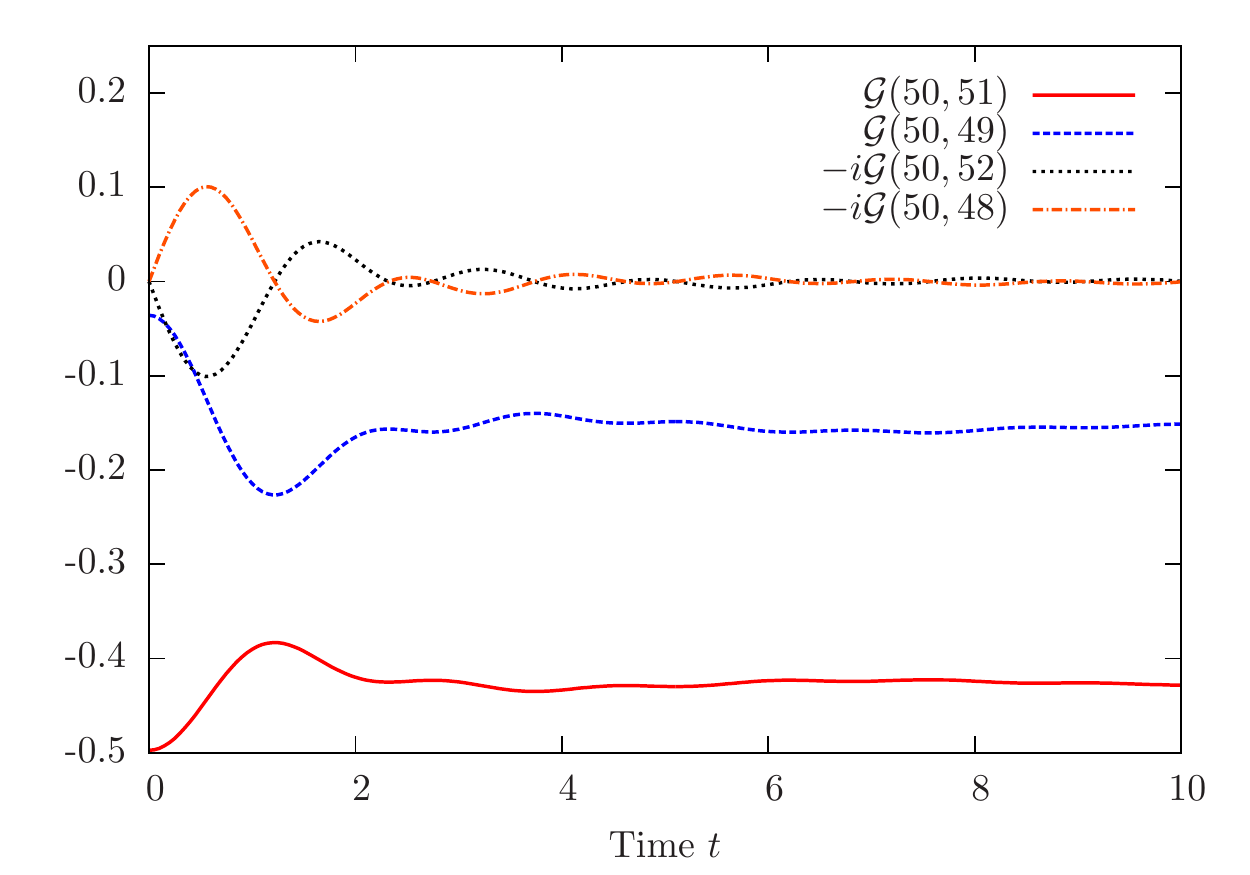}
\caption{(Color online) Green's function $G_0(j,l,t)$ for a
quench with $\delta_i=0.75$, $\delta_f=0.25$ and a lattice with $L=100$
sites. }
\label{fig:GF_U0}
\end{figure}

The diagonal form of our initial Hamiltonian is
\be
H(\delta_i,0) = \sum_{\alpha=\pm}\sum_{k>0} \epsilon_\alpha(k,\delta_i)
b\dg_\alpha(k)b_\alpha(k), 
\ee
and describes two bands of noninteracting fermions. The ground state
is obtained by completely filling the ``$-$'' band; i.e., 
\be
|\Psi_0\ra = \prod_k b\dg_-(k)|0\ra,
\label{GS_PeierlsIns}
\ee
where $|0\ra$ is the fermion vacuum defined by
$b_\alpha(k)|0\rangle=0$, $\alpha=\pm$, $k\in(0,\pi]$. 
At times $t>0$ the system is in the state
\be
|\Psi_0(t)\ra = e^{-iH(\delta_f,0)t}|\Psi_0\rangle.
\ee
The new Hamiltonian is diagonalized by the Bogoliubov transformation
\fr{bogo}
\be
H(\delta_f,0) = \sum_{\alpha=\pm}\sum_{k>0} \epsilon_\alpha(k,\delta_f)
a\dg_\alpha(k)a_\alpha(k), 
\ee
and by virtue of \fr{bogo} the Bogoliubov fermions $a_\alpha(k)$,
$a_\alpha^\dagger(k)$ are {\sl linearly} related to
$b_\alpha(k)$, $b_\alpha^\dagger(k)$. Using this relation it is a
straightforward exercise to obtain an explicit expression for the
time evolution of the fermion Green's function (see Fig.~\ref{fig:GF_U0})
\bea
G_0(j,\ell,t)&=&\la\Psi_0(t)| c\dg_j c_\ell|\Psi_0(t)\ra \nn
&=&
\frac{1}{L}\sum_{k>0}\sum_{\alpha\beta}\gamma^*_\alpha(j,k|\delta_f)
\gamma_\beta(\ell,k|\delta_f)\times\nn  
&&e^{i(\epsilon_\alpha(k)-\epsilon_\beta(k))t}S^-_\alpha(k)S^-_\beta(k)^*
\label{GF0}
\eea
where
\bea
S_\alpha^\beta(k) = u_\alpha(k,\delta_f)u^*_\beta(k,\delta_i) + u \leftrightarrow v.
\label{Sab}
\eea

The late-time behavior can be determined by a stationary phase
approximation, which gives
\be
\lim_{t\to\infty}G_0(j,\ell,t)\sim g_1(j,\ell)+g_2(j,\ell) t^{-3/2}+\ldots
\label{FreeFermionAsymptoticGF}
\ee

%%%%%%%%%%%%%%%%%%%%%%%%%%%%%%%%%%%%%%%%%%%%%%%%%%
\subsection{Generalized Gibbs ensemble (GGE)}
%%%%%%%%%%%%%%%%%%%%%%%%%%%%%%%%%%%%%%%%%%%%%%%%%%

The stationary state of the dimerization quench is described by a
GGE.\cite{GGE} We now briefly review the construction of the GGE
following Refs.~\onlinecite{BS_PRL08,CEF_PRL11,FE_PRB13}. In the thermodynamic limit
the system after the quench possesses an infinite number of
{\sl local} conservation laws $I_a^{(n)}$ ($a=1,2,3,4$, $n\in \mathbb{N}$)
\be
[I^{(n)}_a,I^{(m)}_b]=0\ ,\quad I^{(1)}_1=H(\delta_f,0).
\ee
An explicit construction of these conservation laws is presented in
Appendix \ref{app:HCM}. Given these conserved quantities we defined
a density matrix
\bea
\varrho_{\rm GGE} = \frac{1}{Z_{\rm GGE}}\exp\bigg[-\sum_{a=1}^4\sum_{j\geq 1}
  \lambda_a^{(j)} I^{(j)}_a\bigg],
\eea
where $Z_{\rm GGE}$  ensures normalization.\cite{normalization}
The Lagrange multipliers are fixed by the requirements that the expectation values of the conserved
quantities are the same in the initial state and in the GGE
\be
\lim_{L\to\infty}\frac{1}{L}\langle\Psi_0|I^{(j)}_a|\Psi_0\rangle=
\lim_{L\to\infty}\frac{1}{L}{\rm tr}\left[\varrho_{\rm GGE}I^{(j)}_a\right].
\ee
We then bipartition the system
into a segment B of $\ell$ contiguous sites and its complement $A$ and
form the reduced density matrix
\be
\varrho_{\rm GGE,B}={\rm tr}_A\left[\varrho_{\rm GGE}\right].
\ee
On the other hand the reduced density matrix of segment B after our
quantum quench is simply
\be
\varrho_{B}(t)={\rm tr}_A\Big[|\Psi_0(t)\rangle\langle\Psi_0(t)|\Big].
\ee
At late times after the quench it can be shown by using free fermion
techniques (see, e.g., Ref.~\onlinecite{CEF_PRL11}) that
\be
\lim_{t\to\infty}\lim_{L\to\infty}\varrho_{B}(t)=\varrho_{\rm GGE,B}.
\ee
An alternative \cite{GGE,BS_PRL08,IC_PRA09} but equivalent \cite{FE_PRB13}
construction of the GGE is based on the mode occupation numbers
\be
\hat{n}_\alpha(k)=a^\dagger_\alpha(k)a_\alpha(k).
\ee
By construction these commute with $H(\delta_f,0)$ and among
themselves, and we can express the density matrix in the form
\be
\varrho_{\rm GGE} = \frac{1}{Z_{\rm
    GGE}}\exp\bigg[-\sum_{k>0}\sum_{\alpha=\pm}
\beta_k^{(\alpha)}\hat{n}_\alpha(k)\bigg].
\ee
The Lagrange multipliers are fixed by the conditions
\be
\langle\Psi_0|\hat{n}_\alpha(k)|\Psi_0\rangle=
{\rm tr}\left[\varrho_{\rm GGE}n_\alpha(k)\right],
\ee
which are solved by
\bea
e^{-\beta^{(+)}_k} &=& \frac{|S_+^-(k)|^2}{1-|S_+^-(k)|^2}\ ,\nn
e^{-\beta^{(-)}_k} &=& \frac{|S_-^-(k)|^2}{1-|S_-^-(k)|^2}\ .
\eea
Here the functions $S^\alpha_\beta(k)$ are defined in \fr{Sab}.
%%%%%%%%%%%%%%%%%%%%%%%%%%%%%%%%%%%%%%%%%%%%%%%
\subsection{GGE vs. thermal expectation values}
\label{GGEvsGE}
%%%%%%%%%%%%%%%%%%%%%%%%%%%%%%%%%%%%%%%%%%%%%%%
In the following it will be important to quantify the difference
between the GGE constructed above and a Gibbs ensemble (GE)
\be
\varrho_{G}=\frac{1}{Z_{\rm G}}\exp(-\beta_{\rm eff}H(\delta_f,0))\ ,
\ee
constructed by requiring that the average thermal energy density is
equal to the energy density in the initial state
\be
\lim_{L\to\infty}\frac{\langle\Psi_0|H(\delta_f,0)|\Psi_0\rangle}{L}=
\lim_{L\to\infty}\frac{{\rm tr}\left[\varrho_{\rm G}(\beta_{\rm eff})\ H(\delta_f,0)\right]}{L}.
\label{fixbeta}
\ee
Using the fact that the fermions diagonalizing $H(\delta_f,0)$ and 
$H(\delta_i,0)$ are linearly related by
\be
a\dg_\alpha(k)=S_\alpha^\beta(k)\ b\dg_\beta(k),
\ee
we can rewrite \fr{fixbeta} in the form
\bea
\sum_{k>0}\epsilon_+(k,\delta_f)\left[|S_-^-(k)|^2-|S_+^-(k)|^2\right]\nn
=\sum_{k>0}\epsilon_+(k,\delta_f)\tanh\left[\frac{\beta_{\rm eff}}{2}|\epsilon_+(k,\delta_f)|^2\right].
\eea
%%%%%%%%%%%%%%%%%%%%%%%%%
\subsubsection{Mode occupation numbers}
%%%%%%%%%%%%%%%%%%%%%%%%%
In order to exhibit the difference between Gibbs and generalized Gibbs
ensembles it is useful to consider the mode occupation numbers, which
are given by
\bea
\langle\hat{n}_\alpha(p)\rangle&=&
\begin{cases}
\frac{1}{1+\exp\big(\beta_{\rm eff}\epsilon_\alpha(k,\delta_f)\big)} &
\text{for GE,}\\
\frac{1}{1+\exp\big(\beta_k^{(\alpha)}\big)} &
\text{for GGE}.
\end{cases}
\label{nofk}
\eea
\begin{figure}[ht]
\includegraphics[width=0.45\textwidth]{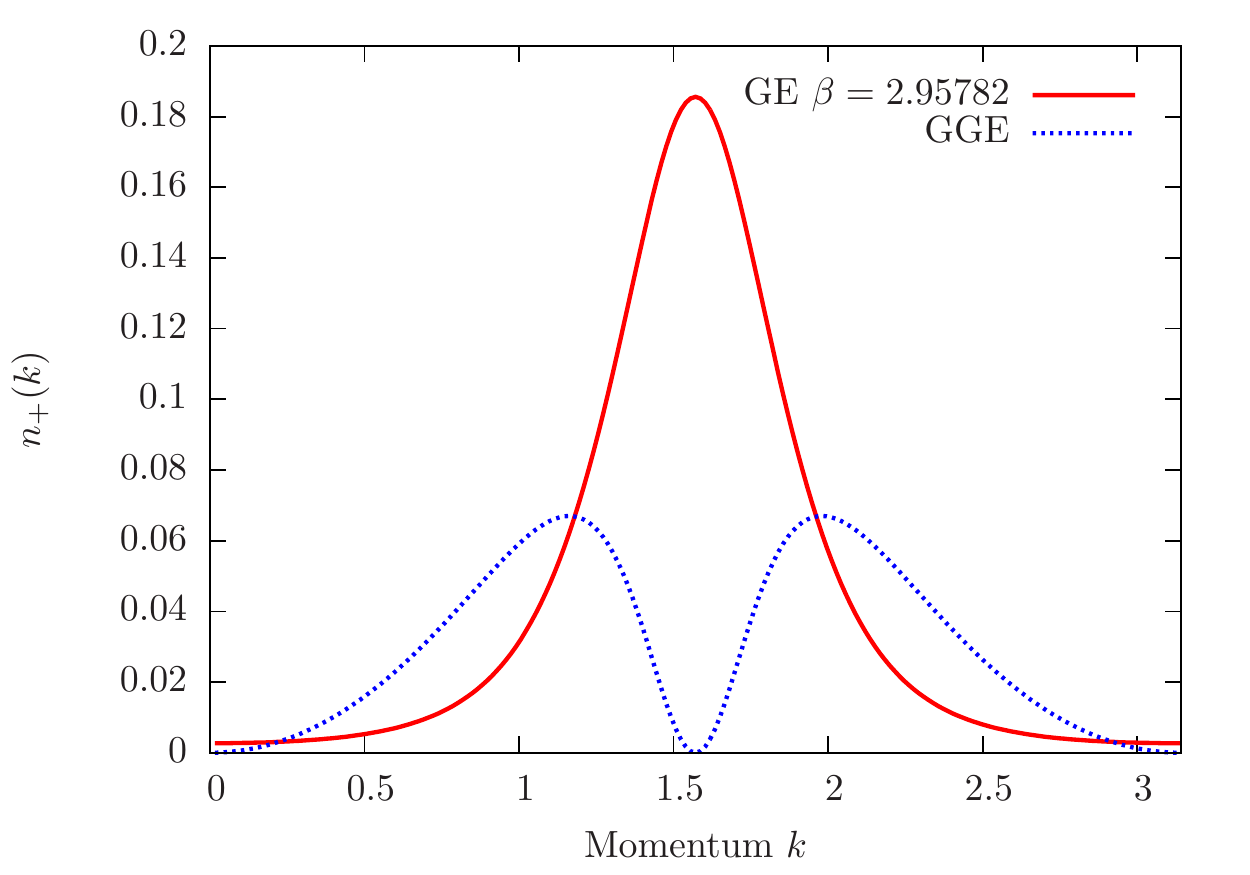}
\caption{(Color online) Comparison between the mode occupation numbers $\langle
  n_+(k)\rangle$ for Gibbs and generalized Gibbs ensembles for a
  quench with $\delta_i=0.75$,
  $\delta_f=0.25$. The effective inverse temperature for this quench is
$\beta_{\rm eff}=2.95782J$.}
\label{fig:npofk}
\end{figure} 
\begin{figure}[ht]
\includegraphics[width=0.45\textwidth]{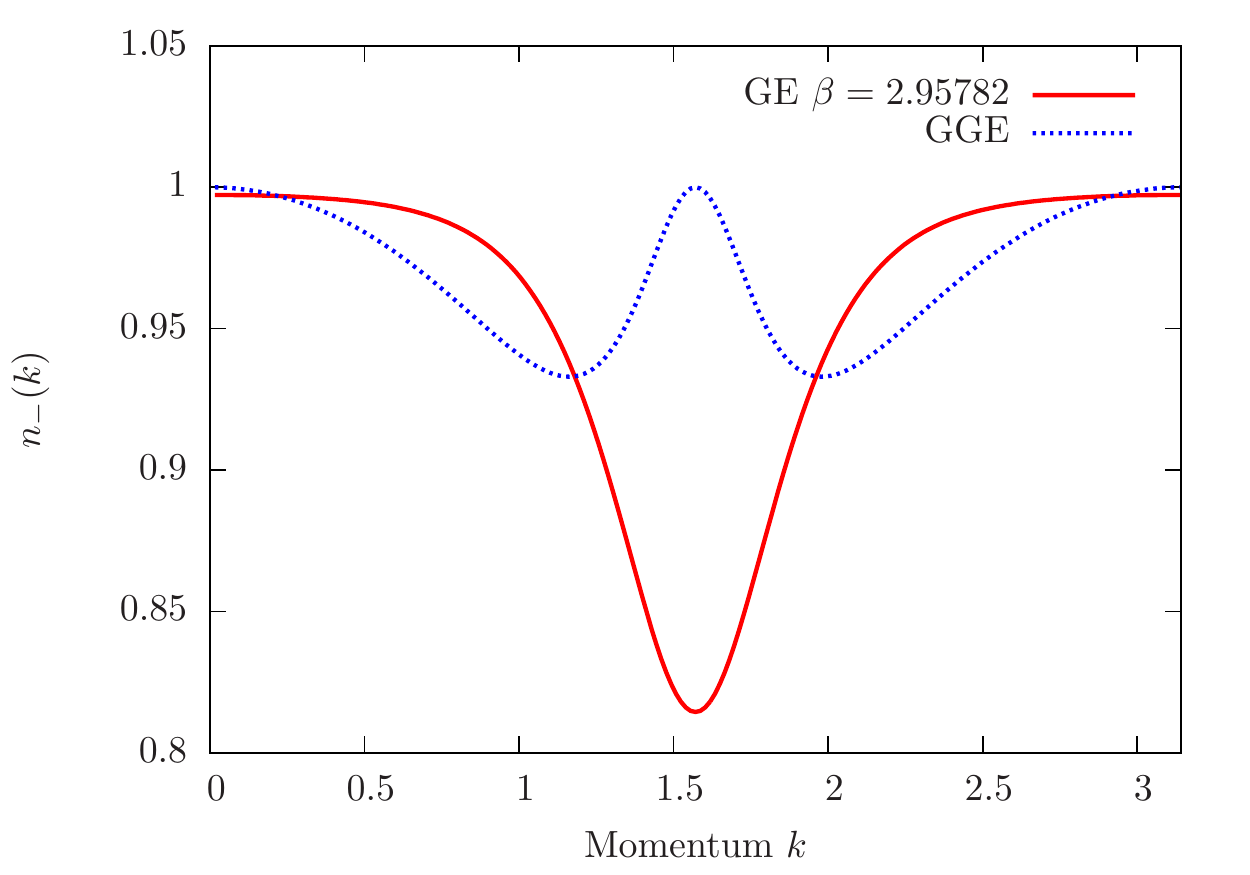}
\caption{(Color online) Comparison between the mode occupation numbers $\langle
  n_-(k)\rangle$ for Gibbs and generalized Gibbs ensembles for a
  quench with $\delta_i=0.75$,
  $\delta_f=0.25$. The effective inverse temperature for this quench is
$\beta_{\rm eff}=2.95782J$.}
\label{fig:nmofk}
\end{figure} 
Clearly the mode occupation numbers shown in Figs.~\ref{fig:npofk}~\&~\ref{fig:nmofk}
are very different in the two ensembles. 
%%%%%%%%%%%%%%%%%%%%%%%%%
\subsubsection{Green's function}
%%%%%%%%%%%%%%%%%%%%%%%%%
As has been emphasized in Ref.~\onlinecite{CEF_PRL11}, as we are dealing with the
nonequilibrium dynamics of an {\it isolated} quantum system, we
should focus on the expectation values of local (in space) operators,
as descriptions in terms of statistical ensembles most naturally apply
to them (see also \cite{FE_PRB13,sirker13}). We therefore consider the
fermionic Green's function in position space, and furthermore focus on
its short-distance properties. The Green's functions in the GGE and
thermal ensembles are 
\bea
\langle c^\dagger_j c_l\rangle&=&\frac{1}{L}\sum_{p>0}\sum_\alpha
\gamma^*_\alpha(j,p|\delta_f)\gamma_\alpha(l,p|\delta_f)\langle\hat{n}_\alpha(p)\rangle,
\eea
where the mode occupation numbers are given by \fr{nofk}.
\begin{figure}[ht]
\includegraphics[width=0.42\textwidth]{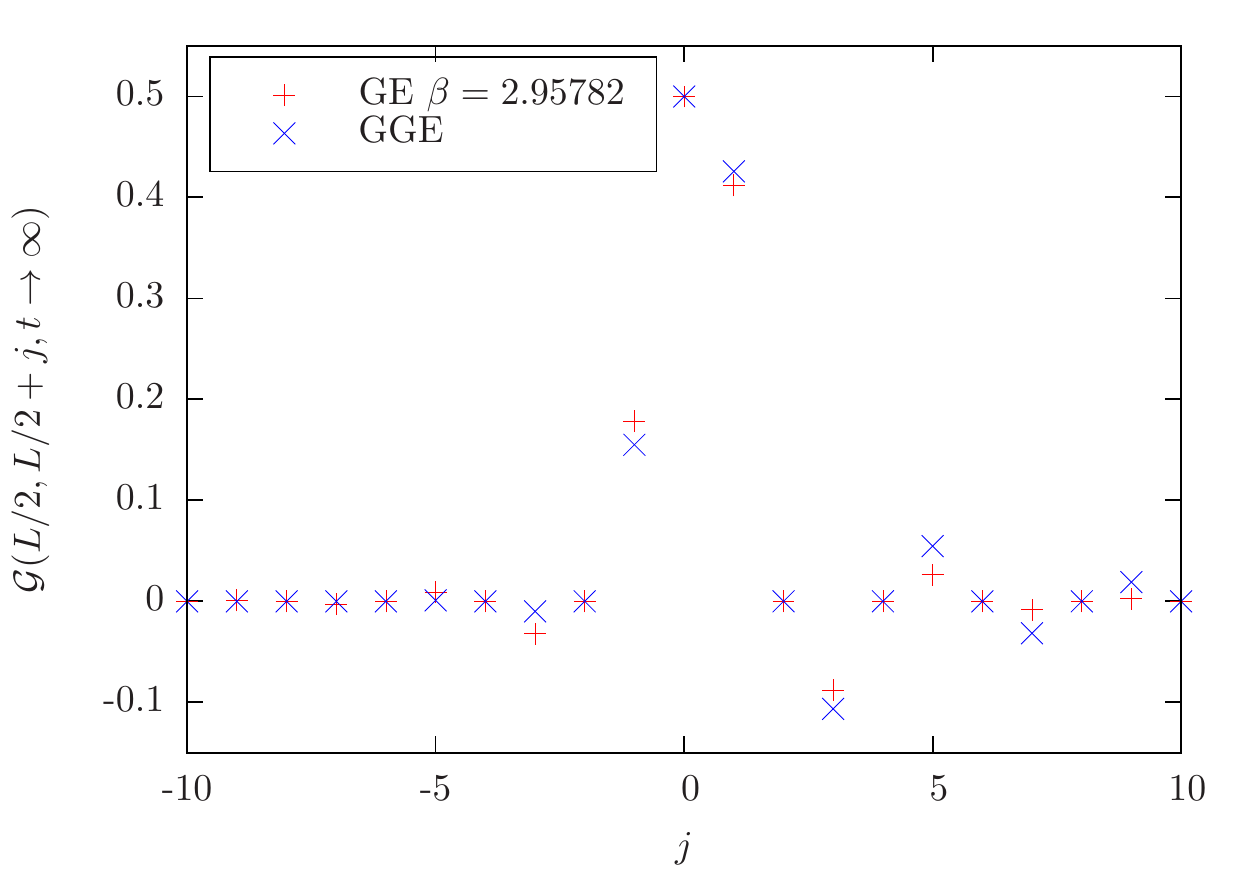}
\caption{(Color online) Green's function $\langle c^\dagger_{L/2}c_{L/2+j}\rangle$
calculated in the Gibbs and generalized Gibbs ensembles for a
quench with $\delta_i=0.75$, $\delta_f=0.25$ and a lattice with $L=100$
sites. The effective inverse temperature for this quench is
$\beta_{\rm eff}=2.95782J$.} \label{fig:g075025}
\end{figure} 
In Fig.~\ref{fig:g075025} we show a comparison between the results for
the fermion Green's function calculated in the appropriate Gibbs and
generalized Gibbs ensembles. We observe that in contrast to the mode
occupation numbers, the difference between the short-distance
behavior of the Green's function in the two ensembles is fairly small. 

%%%%%%%%%%%%%%%%%%%%%%%%%%%%%%%%%%%%%%%%%%%%%%
\section{Quenching to a weakly interacting model}
\label{Sec:NonIntQuench}
%%%%%%%%%%%%%%%%%%%%%%%%%%%%%%%%%%%%%%%%%%%%%%
We now modify our quantum quench as follows. We still start out our
system in the ground state $|\Psi_0\rangle$ of the pure Peierls
Hamiltonian $H(\delta_i,0)$ given by Eq.~\fr{GS_PeierlsIns},
but we now quench to $H(\delta_f,U)$, where we consider $U/J$ to be
small compared to ${\rm min}(\delta_i,\delta_f)$. Our main interest is
to quantify how a non-zero value of $U$ modifies the dynamics after
the quench. 
%{\color{blue} Herein we measure time $t$ in units of the inverse
%  hopping amplitude $J^{-1}$ and set the hopping amplitude $J=1$}.  

To tackle the quench problem in the nonintegrable weakly interacting
model we employ the continuous unitary transformation (CUT) technique~\cite{CUT,KehreinBook} 
which has been applied extensively to nonequilibrium problems 
(see, for example, Refs.~\onlinecite{CUT_NonEq,Moeckel080910}). We provide a brief overview of 
the CUT technique for out-of-equilibrium many-body systems and proceed to calculate the 
time-dependent Green's function and the four-point function. 

%%%%%%%%%%%%%%%%%%%%%%%%%
\subsection{Time evolution of observables by CUT}
%%%%%%%%%%%%%%%%%%%%%%%%%

For a nonintegrable interacting model it is no longer possible to
calculate the time evolution induced by the Hamiltonian~\fr{H0}
exactly. We use the CUT technique to obtain a perturbative expansion
in $U$ of the time-evolved observables.

The central idea of the CUT method is to construct a sequence of
infinitesimal unitary transformations, chosen such that the
Hamiltonian becomes successively more energy-diagonal. A family of
unitarily equivalent Hamiltonians $H(B)$ characterized by the
parameter $B$ can be constructed from the solutions of the
differential equation
\bea
\frac{{\rm d}H(B)}{{\rm d} B} = \Big[ \eta(B),H(B)\Big],
\label{UnitaryTransform}
\eea
where $\eta(B)$ is the anti-Hermitian generator of the unitary transformation. 
Wegner~\cite{CUT} showed that the Hamiltonian in the final basis $H(B=\infty)$
is energy diagonal if $\eta(B)=[H_0(B),H_{\rm int}(B)]$, where $H_0$
is the quadratic part of the Hamiltonian and $H_{\rm int}$ is the
remainder. In practice \fr{UnitaryTransform} is used by expanding all
operators in power series in an appropriate small parameter, which in
our case will be the interaction strength $U$. 

\begin{figure}[ht]
\includegraphics[width=0.45\textwidth]{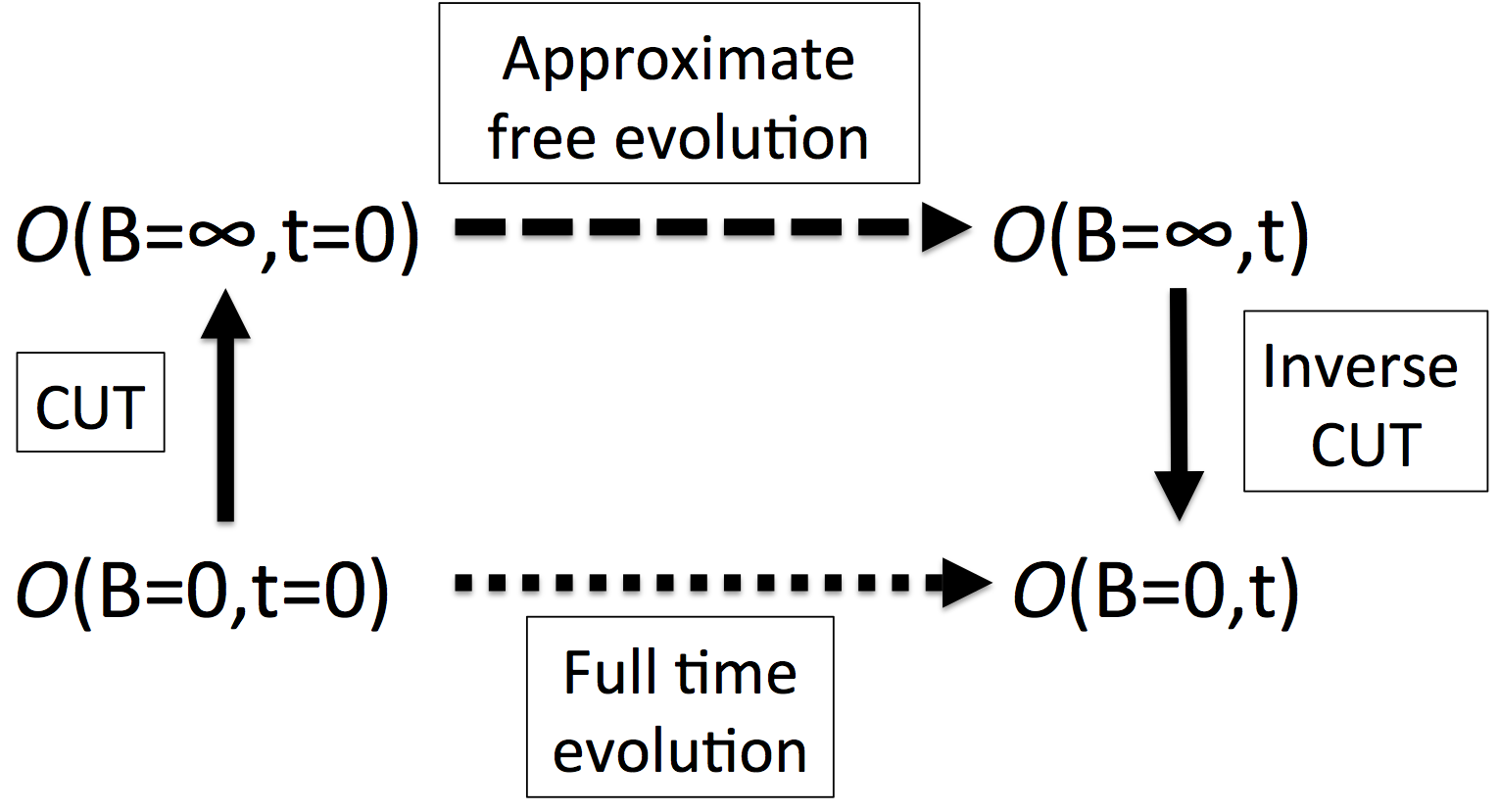} 
\caption{(Color online) A schematic of the CUT method for finding
the approximate time evolution of the operator \textit{O} to order $U$.}
\label{fig:motiv}
\end{figure}

Following the transformation with an appropriate choice of generator,
the Hamiltonian is energy diagonal (but not integrable).
To perform the time evolution we must introduce an additional
approximation: We normal order the interaction term with respect to
the initial state $|\Psi_0\ra$ and neglect the normal-ordered quartic
(and higher order) terms
\bea
H(B=\infty) &=& H_0(B=\infty)+H_{\rm int}(B=\infty)\nn 
&=& H' + :H_{\rm int}(B=\infty):\ ,\nn
{\cal U}(t) &\approx& \exp(-iH't)\ ,\nonumber
\eea
where the time evolution operator ${\cal U}(t)$ depends only on the
quadratic Hamiltonian $H'$ whose single particle energies have 
${\cal O}(U)$ contributions. By construction this approximation
introduces a \emph{maximal time scale}, on which we expect our
calculations to be accurate by virtue of the smallness of
$U$. Estimating this time scale within the CUT formalism is
difficult, as it requires a reliable treatment of the neglected energy
diagonal interaction terms. For this reason we extensively compare our
CUT results to t-DMRG computations (see
Sec.~\ref{CUT_end}). Importantly, we can perform our CUT calculations
for very large systems of hundreds of sites, for which we have
verified that finite-size effects do not play a role on time scales
less then the revival time (the results shown below are for times less
than the revival time). The procedure for calculating the approximate
time evolution of observables is shown schematically in
Fig.~\ref{fig:motiv}.    

%%%%%%%%%%%%%%%%%%%%%%%%%
\subsection{The canonical generator and flow equations for the Hamiltonian}
\label{CUT_beginning}
%%%%%%%%%%%%%%%%%%%%%%%%%

We start by constructing the ``canonical generator'' of the unitary transformation~\cite{KehreinBook}
given by
\be
\eta(B)=[H_0(B),H_{\rm int}(B)].
\ee
The flow-dependent operators are defined by
\bea
H_0(B) &=& \sum_{\alpha=\pm}\sum_{k>0}\epsilon_{\alpha}(k|B)a\dg_{\alpha}(k)a_{\alpha}(k),\\
H_{\rm int}(B) &=& \sum_{k_j>0}V_{\bm\alpha}(\bm k|B)
a\dg_{\alpha_1}(k_1)a_{\alpha_2}(k_2)a\dg_{\alpha_3}(k_3)a_{\alpha_4}(k_4)\nn
&&+\ldots\ .
\eea
where the parameters in the Hamiltonian have been promoted to
functions of the flow parameter $B$ and where the dots indicate terms
sextic and higher in creation and annihilation operators.
The canonical generator is given by
\bea
\eta&=&U\sum_{k_j>0}W_{\bm{\alpha}}({\bm{k}}|B)
a^\dagger_{\alpha_1}(k_1)
a_{\alpha_2}(k_2)
a^\dagger_{\alpha_3}(k_3)
a_{\alpha_4}(k_4)\nn
& +&{\cal O}(U^2),\label{generator}
\eea
where
\bea
W_{\bm{\alpha}}({\bm{k}}|B)&=& V_{\bm{\alpha}}({\bm{k}}|B)E_{\bm{\alpha}}({\bm{k}}|B),\nn
E_{\bm{\alpha}}({\bm{k}}|B)&=& \epsilon_{\alpha_1}(k_1|B)-\epsilon_{\alpha_2}(k_2|B)\nn
&&+ \epsilon_{\alpha_3}(k_3|B)-\epsilon_{\alpha_4}(k_4|B).\ \nonumber
\eea

By inserting the canonical generator~\fr{generator} and the flow Hamiltonian
\be
H(B)=H_0(B)+H_{\rm int}(B)\ ,
\ee 
into the flow equation~\fr{UnitaryTransform} and integrating the resulting differential equations, 
we find the flow-dependent single particle energies and interaction vertices
\bea
\epsilon_{\alpha}(k|B)&=& \epsilon_{\alpha}(k|B=0),\label{SP_Flow}\\
V_{\bm{\alpha}}({\bm{k}}|B) &=&
V_{\bm{\alpha}}({\bm{k}}|B=0)e^{-BE^2_{\bm{\alpha}}({\bm{k}})}\ .\label{VertexB}
\eea
Setting $B=\infty$ we obtain the Hamiltonian in the energy-diagonal basis
\begin{widetext}
\bea
H(B=\infty) &=&
\sum_{\alpha=\pm}\sum_{k>0}\epsilon_{\alpha}(k)a\dg_{\alpha}(k)a_{\alpha}(k) +
\sum_{k_j>0} \breve V_{\bm \alpha}(\bm
k)a\dg_{\alpha_1}(k_1)a_{\alpha_2}(k_2)a\dg_{\alpha_3}(k_3)a_{\alpha_4}(k_4)+{\cal
  O}(U^2)\ ,
\label{Ham_Binf}
\eea
\end{widetext}
where indeed the interaction vertices conserve energy
\be
\breve V_{\bm \alpha}(\bm k) \equiv V_{\bm \alpha}(\bm k|B=\infty) =
V_{\bm \alpha}(\bm k)
\delta_{E_{\bm \alpha}(\bm k),0}\ .
\label{DiagV}
\ee
We note that to leading order in $U$ the single particle energies $\epsilon_{\alpha}(k)$ remain
unchanged by the unitary transformation. Having found the energy-diagonal form of the Hamiltonian 
to leading order we now consider the unitary transformation induced by the canonical generator~\fr{generator} 
on the Green's function.

%%%%%%%%%%%%%%%%%%%%%%%%%
\subsection{Green's function}
%%%%%%%%%%%%%%%%%%%%%%%%%
Our main objective is to determine the fermion Green's function
on the time-evolved initial state
\be
G(j,l;t)=\la\Psi_0(t)| c\dg_j c_l|\Psi_0(t)\ra.
\ee
Using the expression for the original fermions in terms of the Bogoliubov
fermions $a_\alpha(k)$, we see that
\bea
c\dg_j c_l &=& \frac{1}{L}\sum_{k,q>0}\sum_{\alpha,\beta=\pm}
\gamma_{\alpha}^*(j,k|\delta_f)\gamma_\beta(l,q|\delta_f)\nn
&&\qquad\times\ \hat{n}_{\alpha\beta}(k,q|B=0)\ ,
\label{GreensFn}
\eea
where $\gamma_\alpha(j,k|\delta)$ are defined in Eq.~\fr{gamma_def} and
$\hat{n}_{\alpha\beta}(k,q|B=0)=a\dg_{\alpha}(k)a_\beta(q)$. Hence the
basic objects we need to calculate are expectation values of
$\hat n_{\alpha\beta}(p,q|B=0)$. This is done by following the
procedure set out in Fig.~\ref{fig:motiv}.
The flow equations
\be
\frac{{\rm d} \hat{n}_{\alpha\beta}(p,q|B)}{{\rm d} B} =
\Big[\eta(B),\hat{n}_{\alpha\beta}(p,q|B)\Big]\ 
\ee
are easily constructed to order ${\cal O}(U)$ and integrating them
gives 
\begin{widetext}
\bea
\hat n_{\alpha\beta}(k,p|B)&=&
\hat n_{\alpha\beta}(k,p|B=0)
+U\sum_{q_j>0}
N_{\alpha\beta}^{\bm{\alpha}}(\bm{q}|k,p,B)
a^\dagger_{\alpha_1}(q_1)
a_{\alpha_2}(q_2)
a^\dagger_{\alpha_3}(q_3)
a_{\alpha_4}(q_4)+{\cal O}(U^2),\label{nab}
\eea
where we have defined
\bea
N_{\alpha\beta}^{\bm{\alpha}}(\bm{q}|k,p,B)&=&
\delta_{q_4,p}\delta_{\alpha_4,\beta}\vv_{\alpha_1\alpha_2\alpha_3\alpha}(q_1,q_2,q_3,k|B)
+\delta_{q_2,p}\delta_{\alpha_2,\beta}\vv_{\alpha_1\alpha\alpha_3\alpha_4}(q_1,k,q_3,q_4|B)\nn
&&
-\delta_{q_3,k}\delta_{\alpha_3,\alpha}\vv_{\alpha_1\alpha_2\beta\alpha_4}(q_1,q_2,p,q_4|B)
-\delta_{q_1,k}\delta_{\alpha_1,\alpha}\vv_{\beta\alpha_2\alpha_3\alpha_4}(p,q_2,q_3,q_4|B)
\ ,\nn
\vv_{\bm{\alpha}}({\bm{q}}|B)&=&\frac{1-e^{-B[E_{\bm{\alpha}}({\bm{q}})]^2}}
{E_{\bm{\alpha}}({\bm{q}})}V_{\bm{\alpha}}({\bm{q}}).
\eea
\end{widetext}

%%%%%%%%%%%%%%%%%%
\subsubsection{Approximate time evolution}
%%%%%%%%%%%%%%%%%%
In the next step of the procedure sketched in Fig.~\ref{fig:motiv}
we consider the time evolution induced by the $B=\infty$
Hamiltonian~\fr{Ham_Binf}. We approximate the time evolution operator
${\cal U}(t)$ by 
\bea
{\cal U}(t) &=& e^{-iH(B=\infty)t} \approx e^{-iH't}\ , \label{timeevo}
\eea
where the Hamiltonian $H(B=\infty)$ has been replaced by the free fermion Hamiltonian
\bea
H' &=& \sum_{\alpha=\pm}\sum_{k>0} \tilde \epsilon_\alpha(k) a\dg_{\alpha}(k)a_\alpha(k),\nonumber
\eea
with single particle energies
\bea
\tilde \epsilon_\alpha(k) &=& \epsilon_{\alpha}(k) + U P_{\alpha}(k)\ .
\eea 
The additional term $P_{\alpha}(k)$ is given by
\bea
P_{\alpha}(k) &=& \sum_{\gamma,\delta}\sum_{q>0}\bigg[ \breve V_{\alpha\alpha\gamma\delta}(k,k,q,q)+ \breve V_{\gamma\delta\alpha\alpha}(q,q,k,k)\nn
&&-\breve V_{\alpha\delta\gamma\alpha}(k,q,q,k)-\breve V_{\gamma\alpha\alpha\delta}(q,k,k,q)\bigg] n_{\gamma\delta}(q) \ ,\nn
\eea
where $\breve V_{\bm\alpha}(\bm k)$ is defined in Eq.~\fr{DiagV}. The
expectation values $n_{\gamma\delta}(q)=\la \Psi_0| \hat
n_{\gamma\delta}(q,q)|\Psi_0\ra$ taken in the initial state are given by 
\be
\begin{aligned}
 n_{--}(k) &= |S_-^-(k)|^2\\
 n_{++}(k) &= |S_{+}^{-}(k)|^2\\
 n_{+-}(k) &= S_+^-(k)S_-^-(k)^*\\
 n_{-+}(k) &= S_-^-(k)S_+^-(k)^*\ ,
\label{npm}
\end{aligned}
\ee
where functions $S_{\alpha}^{\beta}(k)$ are defined by Eq.~\fr{Sab}. The correction to 
the single-particle energies $P_{\alpha}(k)$ arises from normal ordering the interaction term 
with respect to the initial state $|\Psi_0\ra$. The normal ordering prescription for the quartic term
is given by
\begin{widetext}
\bea
a^\dagger_{\alpha_1}a_{\alpha_2}
a^\dagger_{\alpha_3}a_{\alpha_4}
&=&:a^\dagger_{\alpha_1}a_{\alpha_2}a^\dagger_{\alpha_3}a_{\alpha_4}: 
+n_{\alpha_1\alpha_2}(k_1)\delta_{k_1,k_2}:a^\dagger_{\alpha_3}a_{\alpha_4}:
+n_{\alpha_3\alpha_4}(k_3)\delta_{k_3,k_4}:a^\dagger_{\alpha_1}a_{\alpha_2}:\nn
&-&n_{\alpha_1\alpha_4}(k_1)\delta_{k_1,k_4}:a^\dagger_{\alpha_3}a_{\alpha_2}:
-[n_{\alpha_3\alpha_2}(k_3)-\delta_{\alpha_2,\alpha_3}]
\delta_{k_2,k_3}:a^\dagger_{\alpha_1}a_{\alpha_4}:\nn
&+&n_{\alpha_1\alpha_2}(k_1)n_{\alpha_3\alpha_4}(k_3)\delta_{k_1,k_2}\delta_{k_3,k_4}
-n_{\alpha_1\alpha_4}(k_1)[n_{\alpha_3\alpha_2}(k_2)-\delta_{\alpha_2,\alpha_3}]
\delta_{k_1,k_4}\delta_{k_2,k_3}\ ,
\label{NO}
\eea
\end{widetext}
The normal-ordered quartic interaction term on the right hand side of ~\fr{NO}
has been neglected for the time evolution in Eq.~\fr{timeevo}. Following this approximation, 
the time evolution of fermion operators results only in additional phase factors 
\bea
{\cal U}\dg(t) a\dg_\alpha(k) {\cal U}(t) =
e^{i\tilde\epsilon_{\alpha}(k)t}a\dg_{\alpha}(k). 
\label{time-ev}
\eea
Using \fr{time-ev} in \fr{nab} provides an explicit expression
for the time-evolved operators $\hat n_{\alpha\beta}(k,p|B=\infty,t)$.
In the final step shown in Fig.~\ref{fig:motiv} we
reverse the CUT. Integrating back to the initial basis $B=0$, and then
taking the expectation value with respect to the initial state $|\Psi_0\rangle$ we obtain
\bea
\la \hat n_{\alpha\beta}(p,q|B=0,t) \ra &=& \delta_{p,q} e^{i(\tilde\epsilon_{\alpha}(p)-\tilde\epsilon_\beta(q))t} n_{\alpha\beta}(p)\nn
&+& U c_{\alpha\beta}(p,q|t) + {\cal O}(U^2),
\label{numops}
\eea
Here the order $U$ piece is
\begin{widetext}
\bea
c_{\alpha\beta}(p,q|t) &=& \sum_{q,r>0}
N_{\alpha\beta}^{\bm{\gamma}}(r,r,q,q|p,q|t)n_{\gamma_1\gamma_2}(r)
n_{\gamma_3\gamma_4}(q)
-N_{\alpha\beta}^{\bm{\gamma}}(r,q,q,r|p,q|t)
n_{\gamma_1\gamma_4}(r)\big[n_{\gamma_3\gamma_2}(q)-\delta_{\gamma_2,\gamma_3}\big].
\eea
where we have defined
\bea
N_{\alpha\beta}^{\bm{\gamma}}(\bm{k}|p,q|t) &=&
 N_{\alpha\beta}^{\bm{\gamma}}(\bm{k}|p,q,B=\infty)\Big[
e^{i\tilde
  E_{\bm{\gamma}}(\bm{k})t}-e^{i(\tilde\epsilon_{\alpha}(p)-\tilde\epsilon_\beta(q))t}\Big]\ ,\nn
\tilde E_{\bm{\gamma}}(\bm{k}) &=& \tilde\epsilon_{\gamma_1}(k_1)-\tilde\epsilon_{\gamma_2}(k_2)
+\tilde\epsilon_{\gamma_3}(k_3)-\tilde\epsilon_{\gamma_4}(k_4)\ . 
\label{def_Nt_E}
\eea

Substitution of the observables~\fr{numops} into Eq.~\fr{GreensFn} and
imposing the momentum conserving delta-functions in the
vertices~\fr{VertexFns} gives the time-dependent Green's function 
\bea
{ G}(j,l;t) &=& \la \Psi_0(t) |c\dg_j c_l |\Psi_0(t)\ra\nn
&=& \frac{1}{L}\sum_{k>0}\sum_{\alpha,\beta=\pm}\gamma_{\alpha}^*(j,k|\delta_f) \gamma_\beta(l,k|\delta_f) \Big[  e^{i(\tilde\epsilon_{\alpha}(k)-\tilde\epsilon_\beta(k))t} n_{\alpha\beta}(k)+ U c_{\alpha\beta}(k,k|t)\Big] +{\cal O}(U^2).
\label{GF_CUT}
\eea
The remaining momentum sum $\sum_{k>0}$ has to be evaluated numerically. 
\end{widetext}

%%%%%%%%%%%%%%%%%%%%%%%%%%%%%%%%%%%%%%%%%%%%%%%%%%
\subsection{CUT results for the Green's function}
%%%%%%%%%%%%%%%%%%%%%%%%%%%%%%%%%%%%%%%%%%%%%%%%%%

\begin{figure}[ht]
\includegraphics[width=0.45\textwidth]{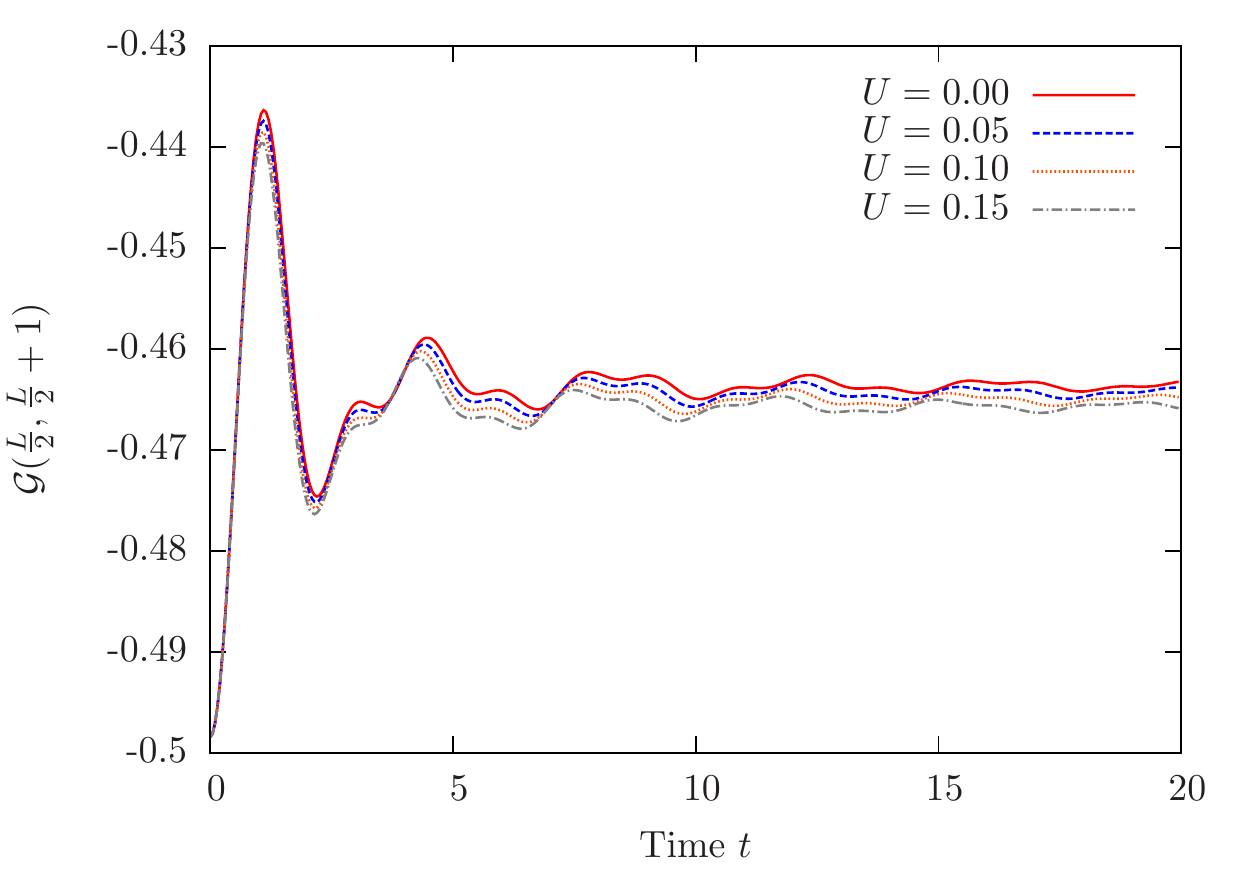}\\
\caption{(Color online) Comparison of exact (solid) $U=0$ nearest-neighbour Green's function 
${\cal G}(L/2,L/2+1) = \langle c_{L/2}c\dg_{L/2+1}\rangle$ with the CUT results for the quench $\delta_i=0.8\to\delta=0.4$ and
$U_i=0\to U$ on the $L=100$ chain.} 
\label{Comp_U0_CUT_GF1}
\end{figure}

\begin{figure}[ht]
\includegraphics[width=0.45\textwidth]{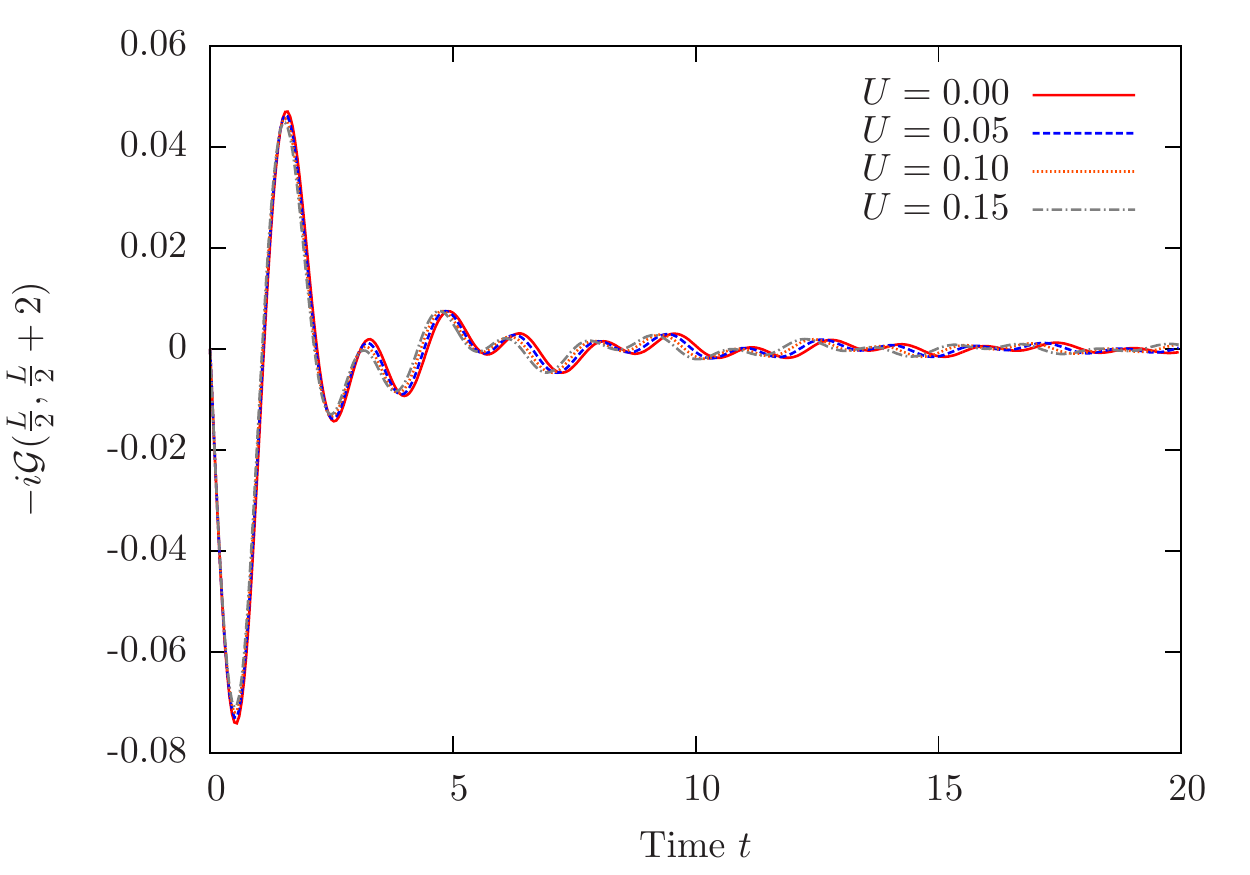}\\
\caption{(Color online) Comparison of exact (solid) $U=0$ next-nearest-neighbour Green's function 
${\cal G}(L/2,L/2+2) = \langle c_{L/2}c\dg_{L/2+2}\rangle$ with
the CUT results for the quench $\delta_i=0.8\to\delta=0.4$ and
$U_i=0\to U$ on the $L=100$ chain.} 
\label{Comp_U0_CUT_GF2}
\end{figure}

We first compare the $U\neq0$ CUT results to the exactly solvable
$U=0$ case. Figures~\ref{Comp_U0_CUT_GF1} and \ref{Comp_U0_CUT_GF2} 
show the nearest-neighbour and next-nearest-neighbour Green's
functions for the quench $\delta_i=0.8\to\delta_f=0.4$ for several
values of $U$. With increasing $U$ the periodicity of the oscillations
and the asymptotic value of the nearest neighbour Green's function are
continuously deformed away from the non-interacting result. The
next-nearest-neighbour Green's function is an imaginary quantity that
decays asymptotically to zero for both the non-interacting and CUT
result.  


\begin{figure}[ht]
\includegraphics[width=0.45\textwidth]{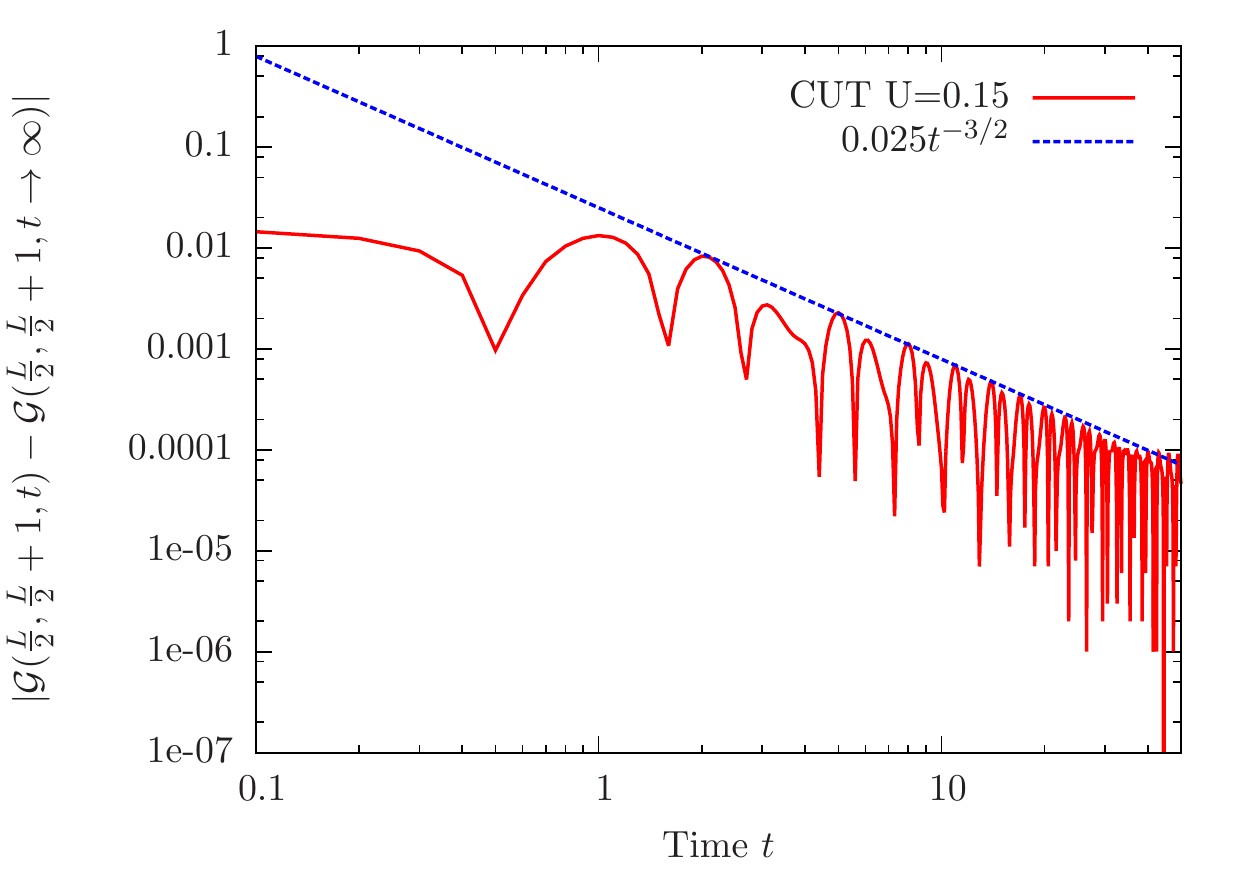}
\caption{(Color online) A comparison of the CUT Green's function 
$|{\cal G}(100,101,t)-{\cal G}(100,101,t\to\infty)|$ and the free
fermion asymptotic form, Eq.~\fr{FreeFermionAsymptoticGF}, on the
$L=200$ chain for the quench $\delta_i=0.75\to\delta=0.5$ and
$U_i=0\to U=0.15$. The prefactor of the power law $t^{-3/2}$ is used
as a fit parameter. The revival time of the $L=200$ chain is $t\sim50$
and the asymptotic value ${\cal G}(100,101,t\to\infty)=-0.482275$.}
\label{CUT_L200_GF_a}
\end{figure}

\begin{figure}[ht]
\includegraphics[width=0.45\textwidth]{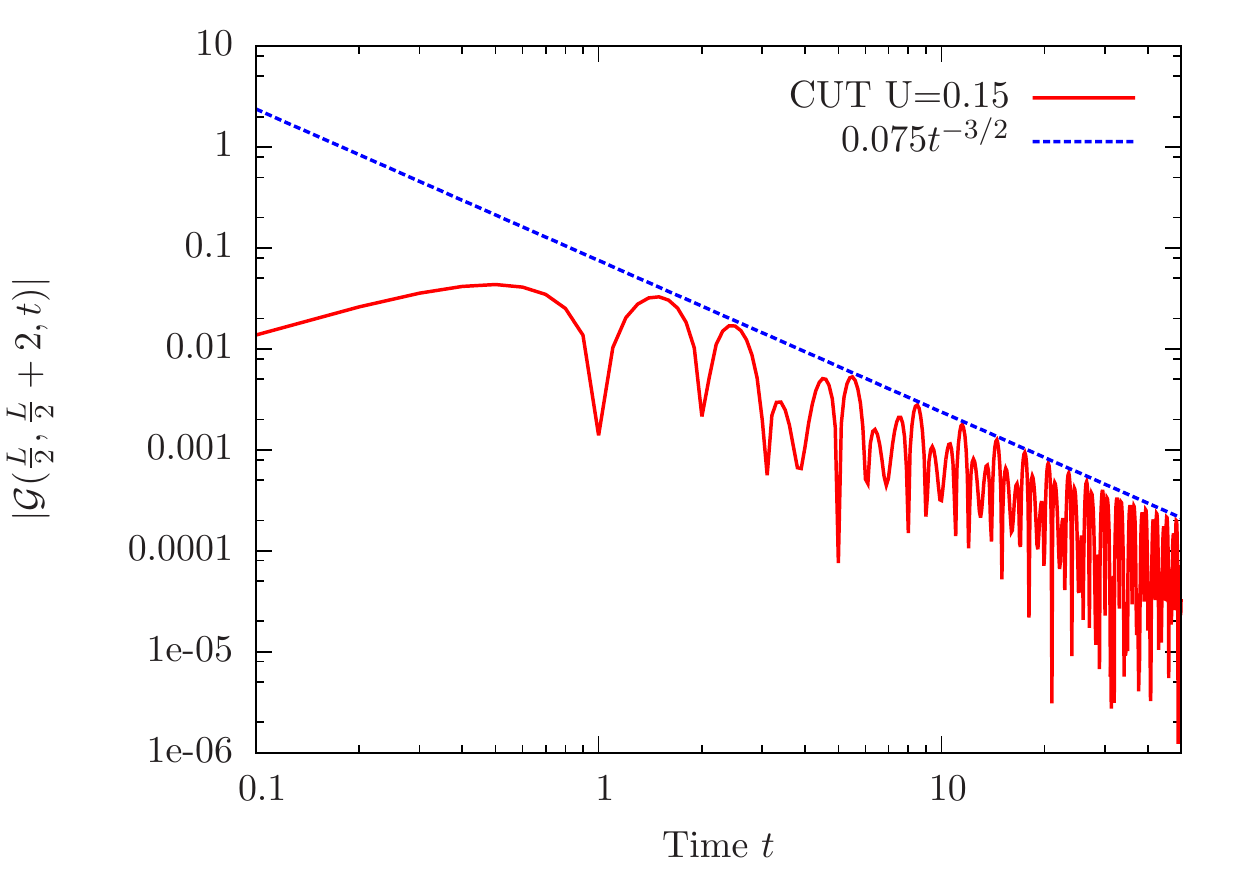}
\caption{(Color online) A comparison between the free fermion asymptotic form of the Green's
function, Eq.~\fr{FreeFermionAsymptoticGF}, and the CUT result for the quench
$\delta_i=0.75\to\delta=0.5$ and $U_i=0\to U=0.15$ on the $L=200$ chain. 
The prefactor of the power law $t^{-3/2}$ is used as a fit parameter.} 
\label{CUT_L200_GF_b}
\end{figure}

In Figs.~\ref{CUT_L200_GF_a} and \ref{CUT_L200_GF_b} we show the
fermion Green's function ${\cal G}(L/2,L/2+j) = \la
c_{L/2}c\dg_{L/2+j}\ra$ for separations $j=1,2$ for the quench
$\delta_i = 0.75 \to \delta = 0.5$ and $U_i = 0 \to U = 0.15$ for the
$L=200$ chain. In both cases the long-time decay of the CUT result is
compatible with the non-interacting $t^{-3/2}$ power-law decay. This
is a consequence of the fact that the CUT result~\fr{GF_CUT} has the
same general $t$-dependence as the non-interacting case~\fr{GF0}.


%%%%%%%%%%%%%%%%%%%%%%%%%
\subsection{Accuracy of the CUT approach: comparison to time-dependent density
matrix renormalization group at small $U/t$}
\label{CUT_end}
%%%%%%%%%%%%%%%%%%%%%%%%%
In order to assess the accuracy of the CUT approach we have carried
out extensive comparisons to numerical results obtained by the
time-dependent density matrix renormalization group (t-DMRG)
algorithm. As is customary in density matrix renormalization group
studies, we impose open boundary conditions. We have carried out
computations for systems up to $L=200$ lattice sites, but for the
puposes of comparing to our CUT results we choose a system size of $L=50$. 
Up to $1500$ density matrix states were kept in the course of the time evolution, 
and a discarded weight of $\varepsilon = 10^{-9}$ was targetted. 
In order to assess the accuracy of the results at later times, we
carried out comparisons to results obtained with a target discarded
weight of $\varepsilon = 10^{-11}$, and in addition compared to
simulations using different time steps of $\delta t = 0.005$ or
$\delta t = 0.01$, respectively. Some details are presented in Appendix
\ref{appendix:tDMRGerrors}. As shown there, the difference between the
results at the end of the time evolution is $\sim10^{-4}$ or smaller
for $L=100$ sites, which means t-DMRG errors are negligible in our
comparison to the CUT results.
\begin{figure}[ht]
\includegraphics[width=0.4\textwidth]{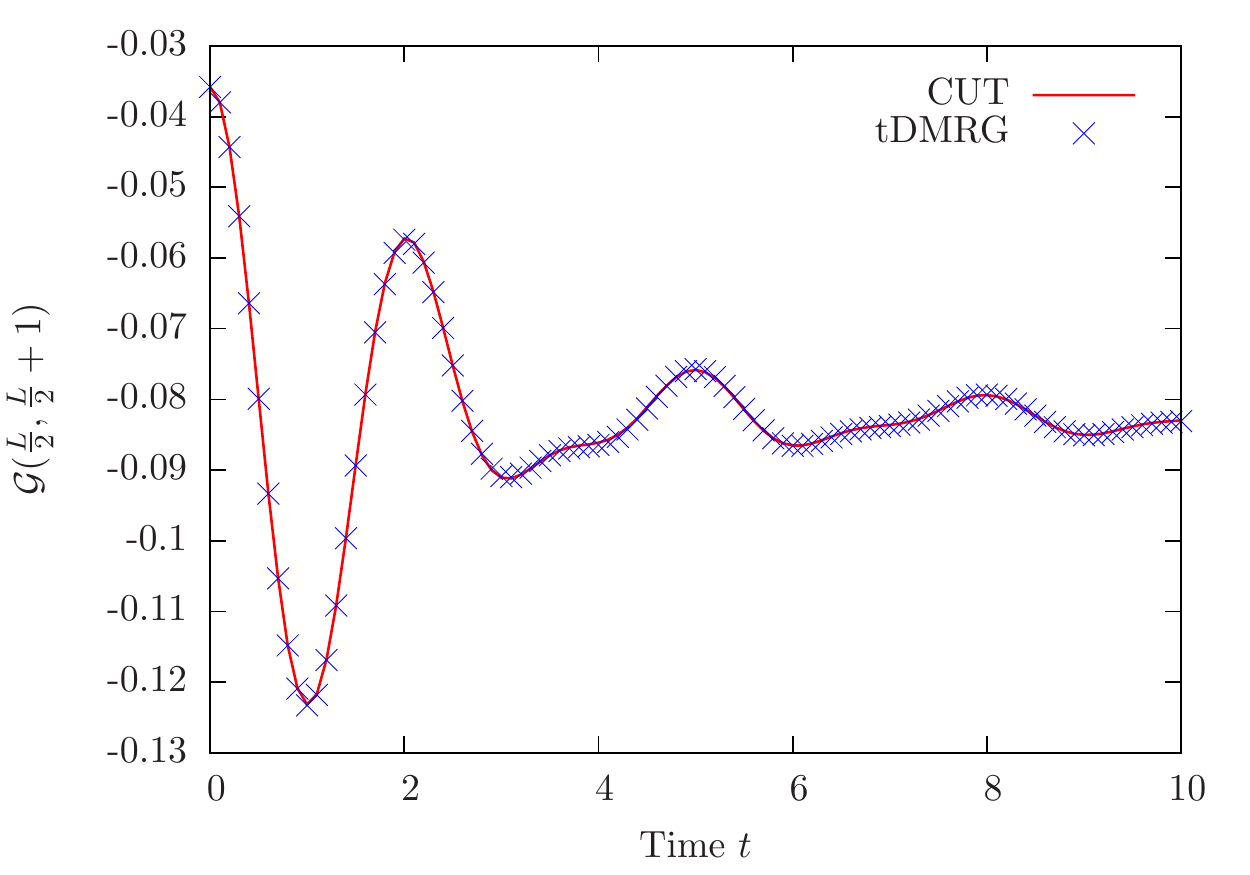}
\caption{(Color online) Comparison of the CUT and t-DMRG results for 
${\cal G}(L/2,L/2+1)= \la c_{L/2} c\dg_{L/2+1} \ra$ for the quench
  $\delta_i = 0.75\to\delta=0.5$ and  $U_i=0 \to$ $U=0.15$ on a $L=50$
  chain. The revival time for the $L=50$ system is $\tau_r\sim13$.}
\label{GF_Ua}
\end{figure} 
\begin{figure}[ht]
\includegraphics[width=0.4\textwidth]{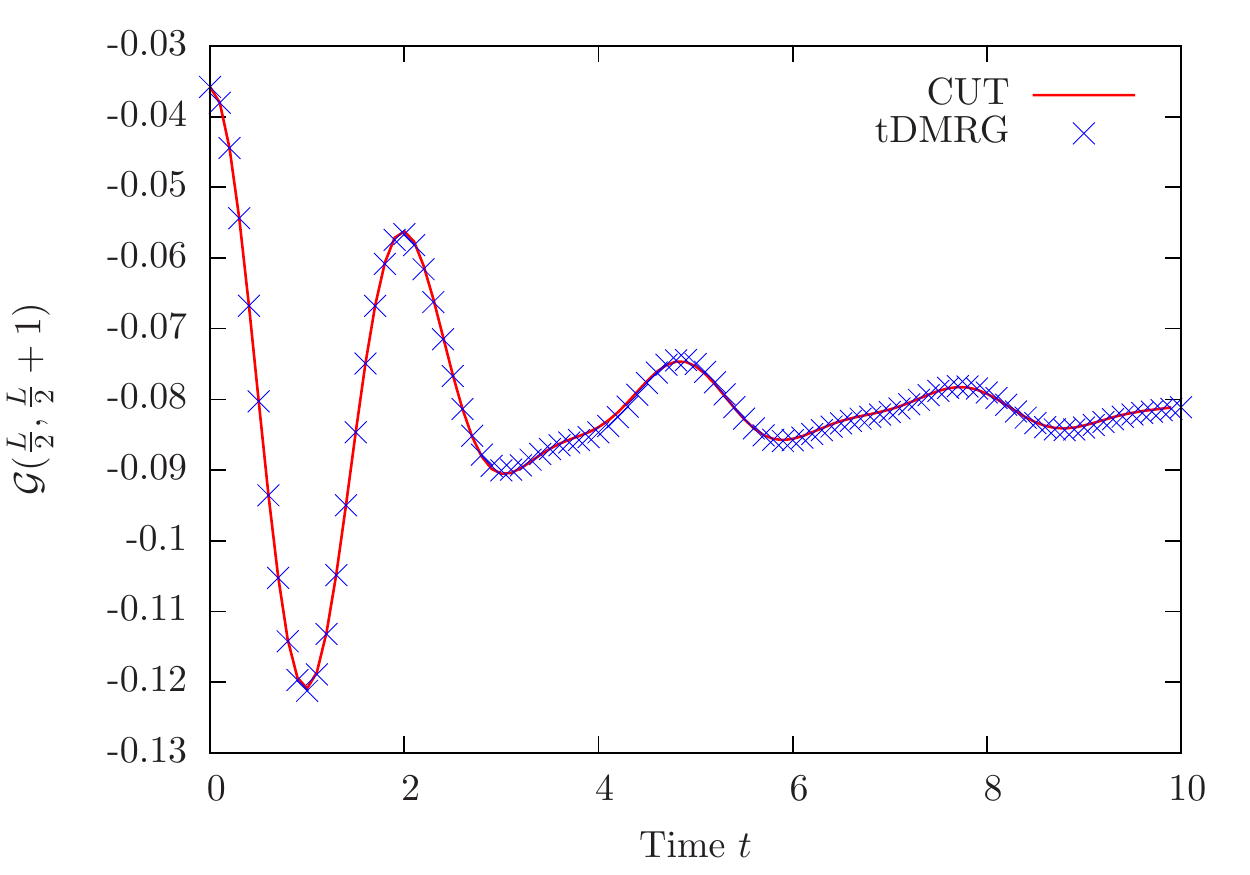}
\caption{(Color online) Comparison of the CUT and t-DMRG results for 
${\cal G}(L/2,L/2+1)= \la c_{L/2} c\dg_{L/2+1} \ra$ for the quench
  $\delta_i = 0.75\to\delta=0.5$ and  $U_i=0 \to$ $U=0.25$ on a $L=50$ chain.}
\label{GF_Ub}
\end{figure} 
The revival time $\tau_r$ for
measurements in the center of a finite chain of noninteracting
particles is $L/2v_{\rm max}$, where $L$ is the system size and
$v_{\rm max}$ is the maximal velocity. In the small-$U$ regime 
of interest here we can obtain a good estimate of $\tau_r$ by
calculating it in the $U=0$ limit. The estimate can be improved by
searching for features associated with revivals at times close to the
free fermion estimate. By comparing data with different
systems sizes $L$, we have verified that finite-size effects 
are negligible in the t-DMRG data for times less than the
revival time $\tau_r$. Finally, we carry out a comparison between CUT
and t-DMRG results only for times $t$ sufficiently smaller than $\tau_r$.
We note that as far as the t-DMRG computations are concerned, we have
been able to reach times $\sim200$ for system size $L=50$. 
Whilst for short enough times the error in the observable can
be estimated as $\sim \sqrt{\varepsilon}$, at longer times, even if
the discarded weight is kept constant, the accumulation of errors in
the course of the sweeps needs to be taken into account. Therefore,
for the situations in which times $>20$ are discussed, a more detailed
error analysis is necessary, which is presented in
Appendix~\ref{appendix:tDMRGerrors}.  
\begin{figure}[ht]
\includegraphics[width=0.42\textwidth]{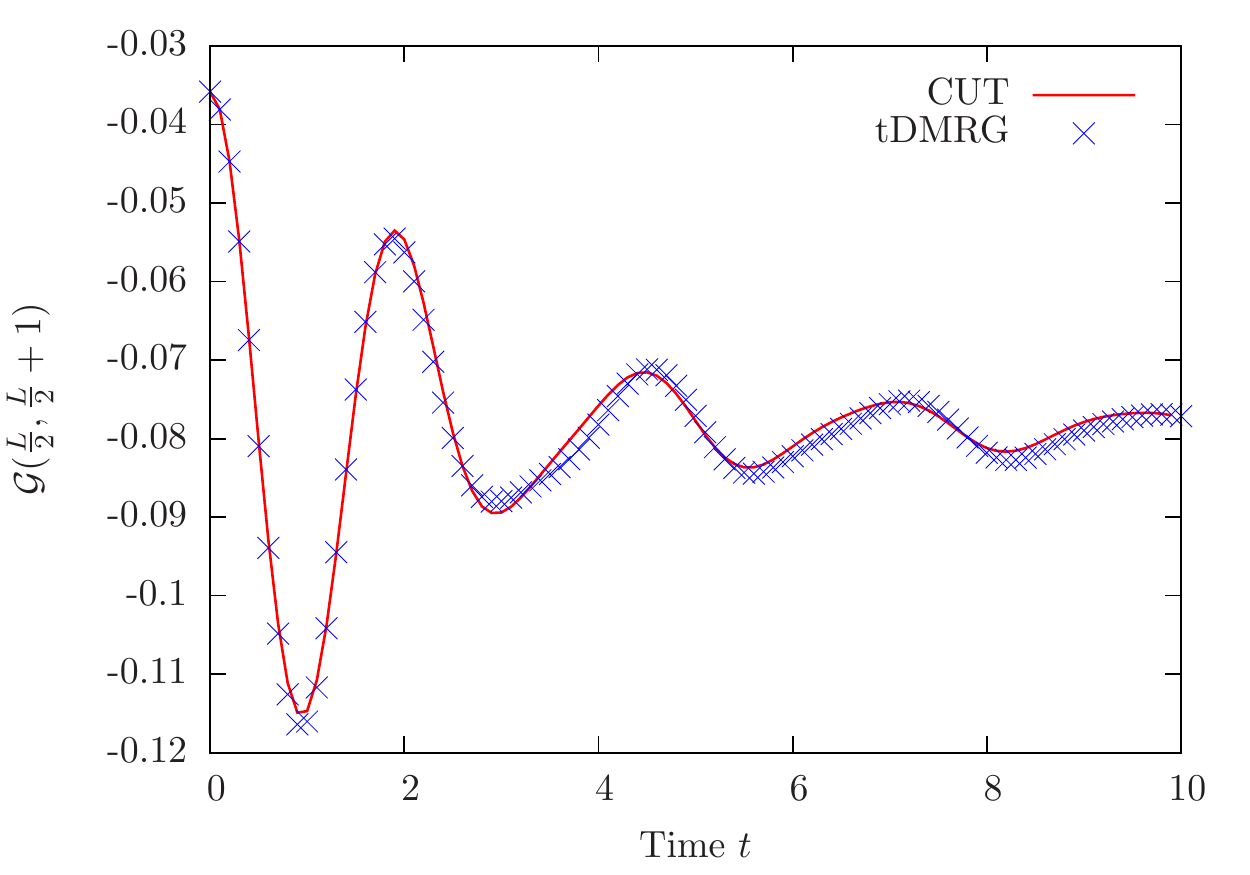}
\caption{(Color online) Comparison of the CUT and t-DMRG results for 
${\cal G}(L/2,L/2+1)= \la c_{L/2} c\dg_{L/2+1} \ra$ for the quench
  $\delta_i = 0.75\to\delta=0.5$ and  $U_i=0 \to$ $U=0.5$ on a $L=50$ chain.}
\label{GF_Uc}
\end{figure} 
\begin{figure}[ht]
\includegraphics[width=0.42\textwidth]{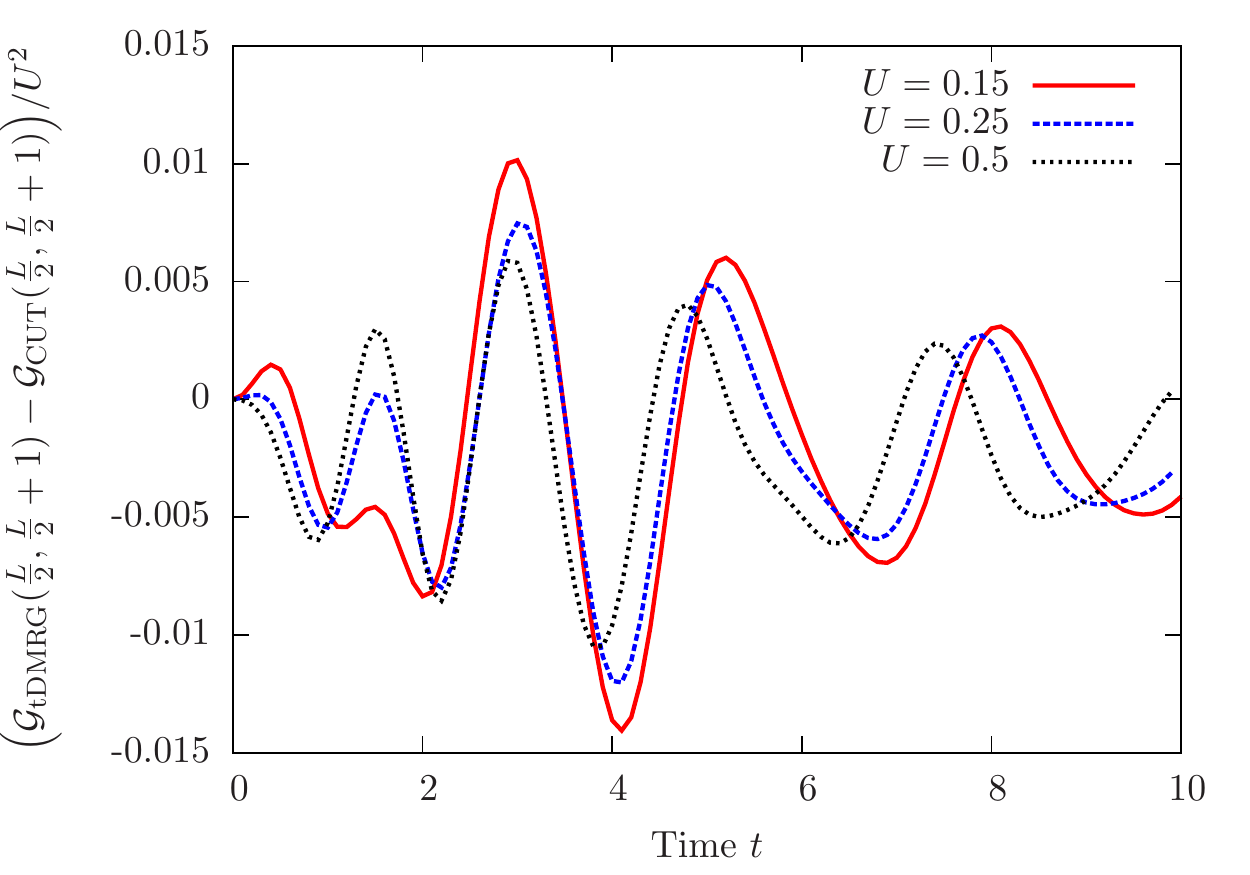}
\caption{(Color online) Rescaled difference between the t-DMRG and CUT data for
${\cal G}(25,26)$ and different values of $U$.}
\label{GF_Ud}
\end{figure} 
In Figs.~\ref{GF_Ua}--\ref{GF_Uc} we show a comparison of the CUT and t-DMRG
results for the time-dependence of the nearest-neighbour Green's function
${\cal G}(25,26)$ for the length $L=50$ chain. We quench the dimerization 
parameter $\delta_i = 0.75 \to \delta = 0.5$ and the interaction strength
$U = 0 \to U = 0.15,0.25,0.5$, respectively. There is good,
quantitative agreement between the CUT and t-DMRG results provided $U$
is small. The remaining discrepancies have their origin in the order
${\cal O}(U^2)$ corrections to the CUT results as is shown in
Fig.~\ref{GF_Ud}, where we plot the rescaled difference between the
t-DMRG data and the CUT result for three values of $U$.
The oscillatory nature of these differences can be explained as a "beat
frequency" arising from subtracting two oscillatory data sets where
the frequencies don't match exactly. 

\begin{figure}[ht]
\includegraphics[width=0.4\textwidth]{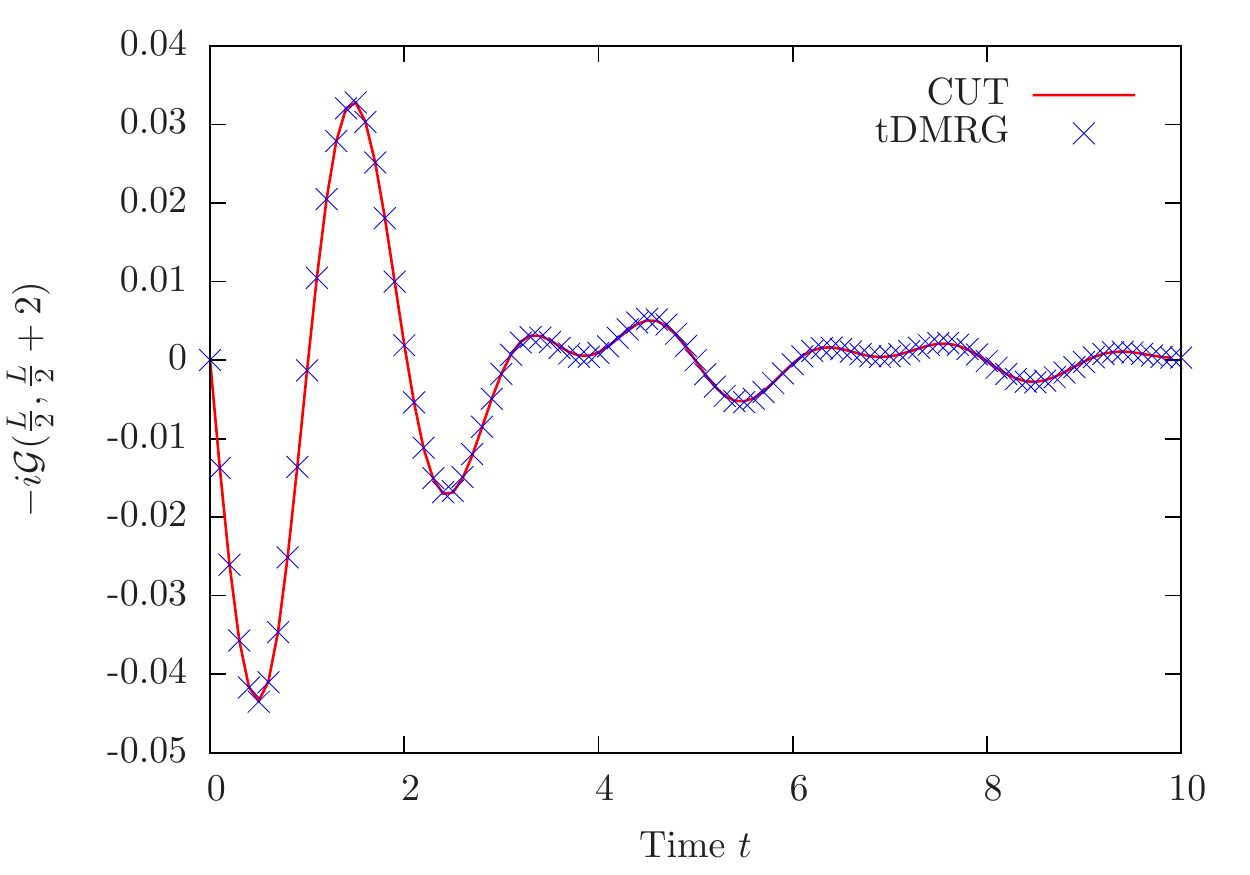}
\caption{(Color online) ${\cal G}(L/2,L/2+2)$ for the quench $\delta_i =
  0.75\to\delta = 0.5$, $U_i=0\to U=0.15$ on a $L=50$ chain.} 
\label{GF_U015_dxa}
\end{figure}
\begin{figure}[h]
\includegraphics[width=0.4\textwidth]{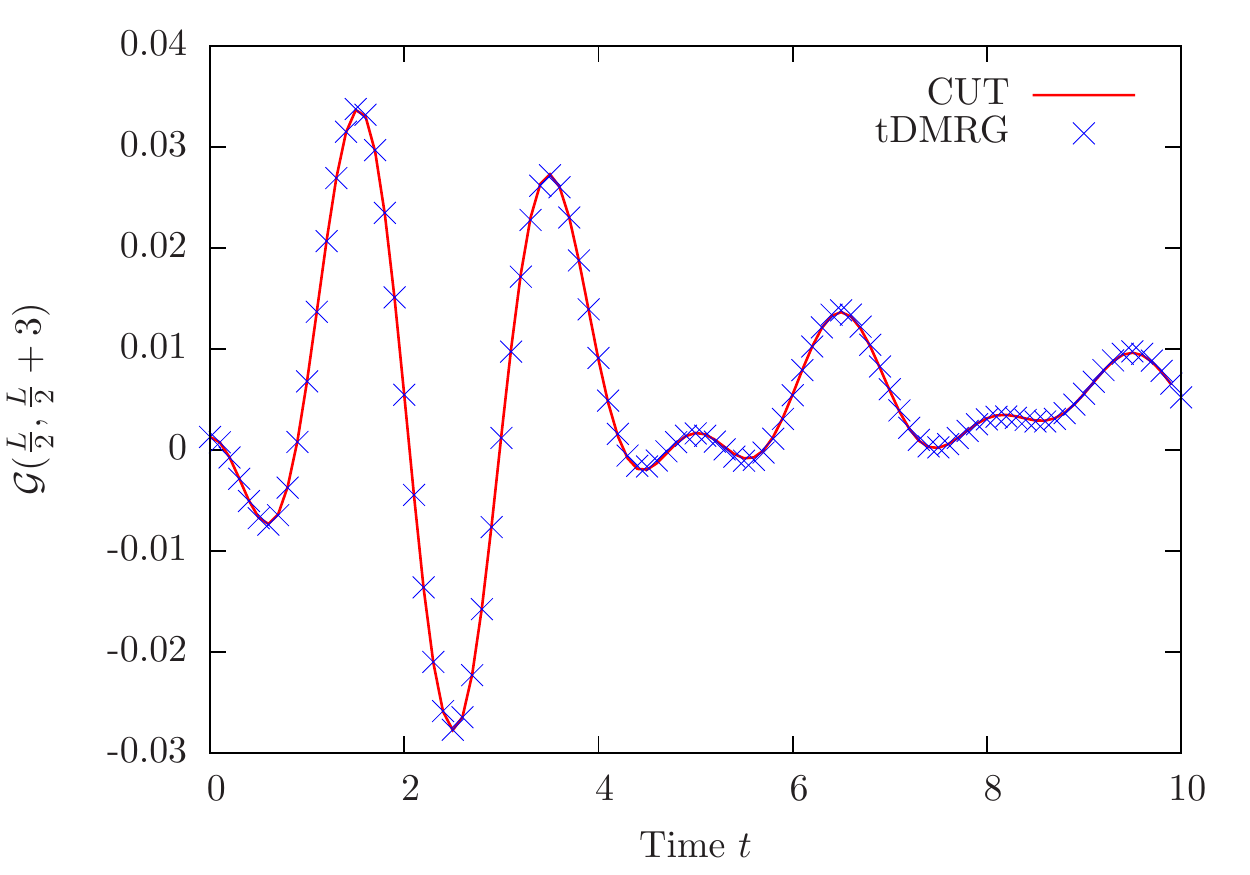}
\caption{(Color online) ${\cal G}(L/2,L/2+3)$ after the quench
  $\delta_i = 0.75\to\delta = 0.5$, $U_i=0\to U=0.15$ on a $L=50$ chain.}
\label{GF_U015_dxb}
\end{figure}

\begin{figure}[h]
\includegraphics[width=0.4\textwidth]{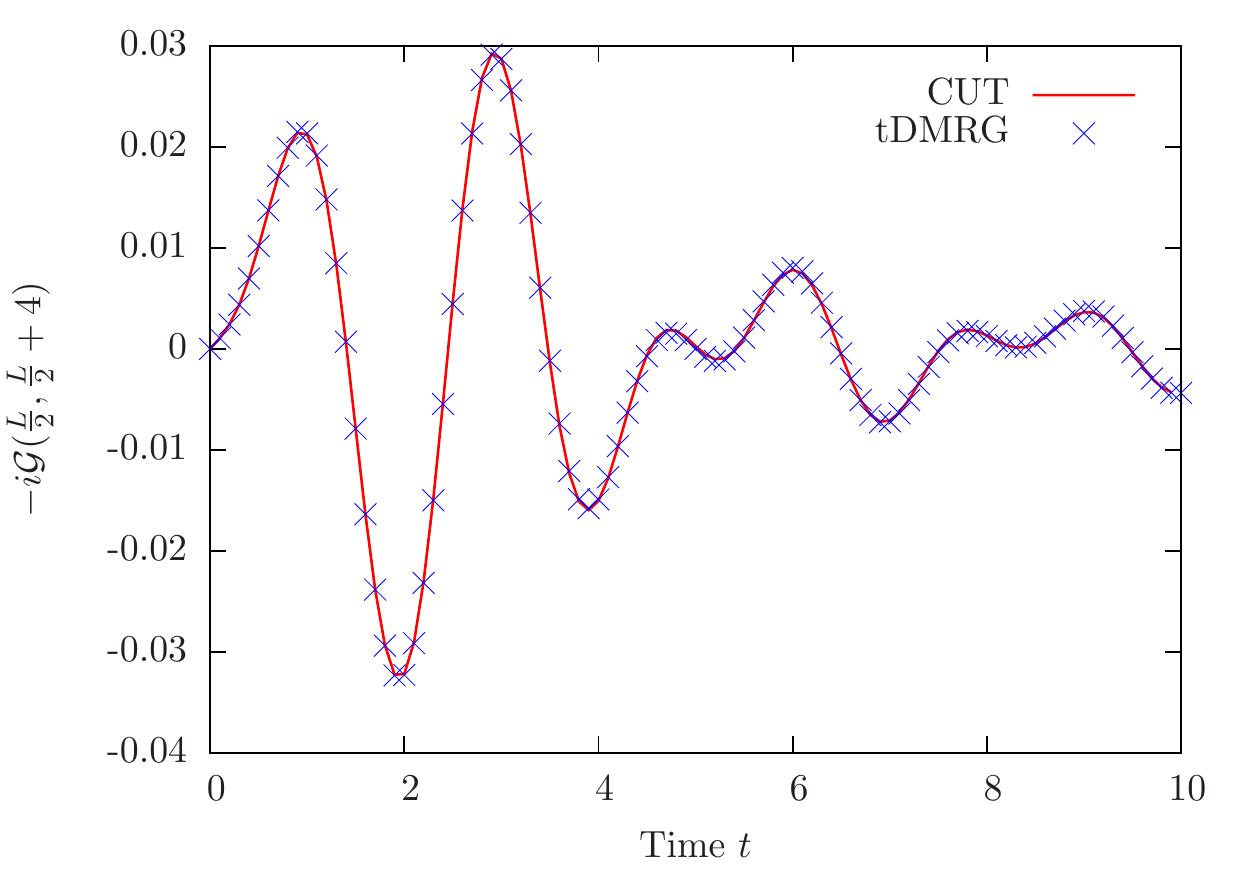}
\caption{(Color online) Comparison of the CUT and t-DMRG results for 
${\cal G}(L/2,L/2+4)$ for the quench
  $\delta_i = 0.75\to\delta = 0.5$, $U_i=0\to U=0.15$ on a $L=50$ chain.}
\label{GF_U015_dxc}
\end{figure}

 
Figures~\ref{GF_U015_dxa}--\ref{GF_U015_dxc} show that the good agreement
between CUTs and t-DMRG is not limited to the nearest-neighbour
Green's function by comparing results for $\la(c_{L/2}
c\dg_{L/2+j})(t)\ra$ with $j=2,3,4$ for the case of $U=0.15$.  

%%%%%%%%%%%%%%%%%%%%
\subsection{CUT results for the four-point function}
%%%%%%%%%%%%%%%%%%%%

The procedure which we have outlined above for the single-particle
Green's function can be generalized to $N$-point functions. The next
non-vanishing correlation function is the four point function
\bea
&&\la \Psi(t)| c\dg_{j}c_{j'}c\dg_{l}c_{l'}|\Psi(t)\ra = 
\frac{1}{L^2}\sum_{q_j > 0}\sum_{\alpha_j=\pm} \langle\Psi_0|\hat
A_{\bm\alpha}(\bm q,t)|\Psi_0\rangle\nn 
&&\qquad\qquad\times\
\gamma_{\alpha_1}^*(j,q_1)\gamma_{\alpha_2}(j',q_2)\gamma^*_{\alpha_3}(l,q_3)\gamma_{\alpha_4}(l',q_4),\nn 
\label{def_fpf}
\eea
where $\gamma_{\alpha}(j,k)$ are defined in Eq.~\fr{gamma_def} and
\be
\hat A_{\bm\alpha}(\bm q,t) = a\dg_{\alpha_1}(q_1,t)a_{\alpha_2}(q_2,t)a\dg_{\alpha_3}(q_3,t)a_{\alpha_4}(q_4,t).
\ee
Going to the $B=\infty$ basis by applying the CUT and then time
evolving with \fr{timeevo}, we obtain
\begin{widetext}
\bea
\hat A_{\bm\alpha}(\bm q,t|B=\infty) &=&e^{i\tilde E_{\bm\alpha}(\bm q)t} a\dg_{\alpha_1}(q_1)a_{\alpha_2}(q_2)a\dg_{\alpha_3}(q_3)a_{\alpha_4}(q_4)\nn
&+& U \sum_{k_j>0}\sum_{\gamma_j=\pm} e^{i(\tilde\epsilon_{\alpha_1}(q_1)-\tilde\epsilon_{\alpha_2}(q_2))t} N_{\alpha_3\alpha_4}^{\bm\gamma}(\bm k|q_3,q_4|t)a\dg_{\alpha_1}(q_1)a_{\alpha_2}(q_2)a\dg_{\gamma_1}(k_1)a_{\gamma_2}(k_2)a\dg_{\gamma_3}(k_3)a_{\gamma_4}(k_4)\nn
&+& U\sum_{k_j>0}\sum_{\gamma_j=\pm}e^{i(\tilde\epsilon_{\alpha_3}(q_3)-\tilde\epsilon_{\alpha_4}(q_4))t} N_{\alpha_1\alpha_2}^{\bm\gamma}(\bm k|q_1,q_2|t)a\dg_{\gamma_1}(k_1)a_{\gamma_2}(k_2)a\dg_{\gamma_3}(k_3)a_{\gamma_4}(k_4)a\dg_{\alpha_3}(q_3)a_{\alpha_4}(q_4)\nn
&+& {\cal O}(U^2),
\label{a_fpf}
\eea
where $\tilde E_{\bm\alpha}(\bm q)$ and $N_{\alpha\beta}^{\gamma}(\bm k|p,q|t)$ are defined in Eq.~\fr{def_Nt_E}. 
Taking the expectation value of Eq.~\fr{a_fpf} on the initial state using Wick's theorem and substituting in to Eq.~\fr{def_fpf}
yields the real-space four-point function. 
\end{widetext}

%%%%%%%%%%%%%%%%%%%%%%%%%%%%%%%%
\section{Prethermalized Regime}
\label{Sec:Pretherm}
%%%%%%%%%%%%%%%%%%%%%%%%%%%%%%%%
The combination of CUT and t-DMRG results establish that at
intermediate times the fermion Green's function $G(j,l,t)$ after a
quench $(\delta_i,0)\rightarrow(\delta_f,U)$ decays in a power-law
fashion with approximate exponent $-3/2$ to a stationary value; i.e., 
\be
G(j,\ell,t)\longrightarrow g(j,\ell)+{\cal O}(t^{-3/2})\ ,\quad Jt\alt \tau_0.
\ee
It is very instructive to compare this to the result
\fr{FreeFermionAsymptoticGF} for the quench
$(\delta_i,0)\rightarrow(\delta_f,0)$. By virtue of the perturbative
nature of the CUT approach, and its excellent agreement with t-DMRG for
small $U$ and the time scales relevant to the present discussion, we
obtain the following relation between the asymptotic values of the
Green's function for the two quenches
\be
g(j,\ell)-g_1(j,\ell)={\cal O}(U).
\label{g1g}
\ee
We will now show that $g(j,\ell)$ {\sl cannot} be described by a
thermal ensemble, which implies that the stationary regime observed by
t-DMRG is in fact a prethermalization plateau.
\begin{enumerate}
\item{}For the quench $(\delta_i,0)\rightarrow(\delta_f,0)$ the
observed plateau corresponds to the true stationary state and
is characterized by a GGE, i.e.
\bea
g_1(j,\ell) = {\rm tr}[\varrho_{\rm GGE} c\dg_j c_\ell].
\label{g1GGE}
\eea
\item{}
As we showed in Sec.~\ref{GGEvsGE}, the GGE expectation values for
the Green's function are generally different from the thermal
expectation values at the appropriate effective inverse temperature
$\beta_0$ characterizing the quench 
\be
{\rm tr}[\varrho_{\rm GGE} c\dg_j c_\ell] - {\rm tr}[\varrho_{\rm
    G}(\beta_0)
  c\dg_j c_\ell ] = {\cal O}(1). 
\label{Diff_GGE_Gibbs}
\ee
\item{} If the stationary state after the quench
  $(\delta_i,0)\rightarrow(\delta_f,U)$ was described by a thermal
  distribution, its effective inverse temperature $\beta_{\rm eff}$
  would be determined by 
\be
\lim_{L\to\infty}\frac{\langle\Psi_0|H(\delta_f,U)|\Psi_0\rangle}{L} = 
\lim_{L\to\infty}\frac{{\rm tr}\left[\varrho_{\rm G}(\beta_{\rm eff})\ H(\delta_f,U)\right]}{L}.
\label{fixbetaU}
\ee
On the other hand, given that Wick's theorem holds in the state
$|\Psi_0\rangle$, we conclude that
\be
\langle\Psi_0|H(\delta_f,U)|\Psi_0\rangle=
\langle\Psi_0|H(\delta_f,0)|\Psi_0\rangle+{\cal O}(U).
\ee
Hence
\be
\beta_{\rm eff}=\beta_0+{\cal O}(U).
\label{betaeff}
\ee
\item{}
Combining \fr{betaeff} with \fr{Diff_GGE_Gibbs} we conclude that
\be
{\rm tr}[\varrho_{\rm GGE} c\dg_j c_\ell] - {\rm tr}[\varrho_{\rm
    G}(\beta_{\rm eff})\  c\dg_j c_\ell ] = {\cal O}(1). 
\label{Diff_GGE_Gibbs2}
\ee
\item{} 
Finally, combining \fr{g1g}, \fr{g1GGE} and \fr{Diff_GGE_Gibbs2}, we
conclude that
\be
g(j,\ell)-{\rm tr}[\varrho_{\rm G}(\beta_{\rm eff})\  c\dg_j c_\ell ] = {\cal O}(1),
\ee
and hence $g(j,\ell)$ is not described by a thermal distribution.
\end{enumerate}

%%%%%%%%%%%%%%%%%%%%%%%%%%%%%%%%%%%%%%%%%%%%%%%%
\subsection{Characterization of the prethermalized regime through
  approximate conservation laws}
%%%%%%%%%%%%%%%%%%%%%%%%%%%%%%%%%%%%%%%%%%%%%%%%

In the previous section we have shown that the CUT result cannot
produce an effective thermal Gibbs ensemble in the long time
limit. Given that the CUT results for the Green's function are in
excellent agreement with t-DMRG data at intermediate times, this
establishes the existence of a ``prethermalized stationary
regime''. An obvious question is then how to characterize the
statistical ensemble describing the corresponding plateau values of
local observables.

%%%%%%%%%%%%%%%%%%%%%%%%%%%%%%%%%%%%%%%%%%%%%
\subsubsection{Approximate conservation laws}
%%%%%%%%%%%%%%%%%%%%%%%%%%%%%%%%%%%%%%%%%%%%%
In our CUT analysis of the nonequilibrium dynamics the generator of time evolution
was taken to be 
\be
H'=\sum_{\alpha=\pm}\sum_{k>0}\tilde{\epsilon}_\alpha(k)a^\dagger_\alpha(k)a_\alpha(k).
\ee
Clearly the mode occupation number operators $n_{\alpha\alpha}(k)$
commute with $H'$, and hence constitute conservation laws
(to first order in $U$) within our CUT approach. Their pre-images
under the CUT, accurate to order ${\cal O}(U)$, are simply
\bea
{\cal Q}_\alpha(k) &=& a^\dagger_{\alpha}(k)a_\alpha(k)
- U\sum_{q_j>0}N_{\alpha\alpha}^{\bm{\gamma}}(\bm{q}|k,k,B=\infty)\nn
&&\qquad\times\
a\dg_{\gamma_1}(q_1)a_{\gamma_2}(q_2)a\dg_{\gamma_3}(q_3)a_{\gamma_4}(q_4).
\label{newQ}
\eea
By construction these operators approximately commute with one another
\be
[{\cal Q}_\alpha(k),{\cal Q}_\beta(p)]={\cal O}(U^2).
\ee
However, the commutator with the Hamiltonian is in fact 
\be
[{\cal Q}_\alpha(k),H(\delta_f,U)]={\cal O}(U),
\ee
i.e. the charges \fr{newQ} are not (approximately) conserved on an
operator level, but only their expectation values with respect to
$|\Psi_0(t)\rangle$ are (approximately) time independent. This is a
fundamental difference to the proposal put forward in
Ref.~\onlinecite{KollarPRB11} for describing prethermalization plateaus. The charges
${\cal Q}_\alpha(k)$ have a very transparent physical meaning: they
are the number operators for approximately conserved
``quasiparticles'', and the quartic terms describe the leading
contribution to the dressing of the non-interacting fermions.

%%%%%%%%%%%%%%%%%%%%%%%%%%%%%%%%%%%%%%%%%%%%%
\subsubsection{Approximate description by a ``deformed GGE''}
%%%%%%%%%%%%%%%%%%%%%%%%%%%%%%%%%%%%%%%%%%%%%
It is natural to attempt a description of the prethermalized regime
in terms of a statistical ensemble of the form
\be
\varrho_{\rm PT}=\frac{1}{Z_{\rm PT}}
\exp\left(\sum_{k,\alpha} \lambda^{(\alpha)}_k{\cal Q}_\alpha(k)\right).
\label{rhoPT}
\ee
Here the Lagrange multipliers $\lambda^{(\alpha)}_k$ are fixed by the
requirements
\bea
{\rm tr}\left[\varrho_{\rm PT}\ {\cal Q}_\alpha(k)\right]=
\langle\Psi_0|{\cal Q}_\alpha(k)|\Psi_0\rangle.
\label{fixlambda}
\eea 
The left-hand side of \fr{fixlambda} 
is most easily evaluated in the $B=\infty$ basis, where it becomes
\bea
\frac{1}{Z_{\rm PT}}{\rm tr}\left[
e^{\sum_{k,\alpha}\lambda^{(\alpha)}_ka^\dagger_{\alpha}(k)a_{\alpha}(k)}
a^\dagger_{\alpha}(k)a_{\alpha}(k)\right]=\frac{1}{1+e^{-\lambda_k^{(\alpha)}}}.\nn
\label{lhs}
\eea
The right-hand side of \fr{fixlambda} is equal to
\begin{widetext}
\bea
n_{\alpha\alpha}(k) -
U\sum_{q_j>0}N_{\alpha\alpha}^{\bm{\gamma}}(\bm{q}|k,B=\infty)
\left[n_{\gamma_1\gamma_2}(q_1)n_{\gamma_3\gamma_4}(q_3)\delta_{q_1,q_2}
\delta_{q_3,q_4}+
n_{\gamma_1\gamma_4}(q_1)\left[\delta_{\gamma_2,\gamma_3}-n_{\gamma_3\gamma_2}(q_2)\right]
\delta_{q_1,q_4}
\delta_{q_2,q_3}\right].
\label{rhs}
\eea
\end{widetext}
Equating \fr{rhs} with \fr{lhs} and using \fr{npm} we obtain an
explicit expression for the Lagrange multipliers $\lambda^{(\alpha)}_k$.
The fermion Green's function evaluated with respect to the density
matrix \fr{rhoPT} is
\bea
G_{\rm PT}(j,\ell)&=&{\rm tr}\left[\varrho_{\rm PT}c^\dagger_jc_\ell\right]\nn
&=&\frac{1}{L}\sum_{q>0}\sum_{\alpha=\pm}
\gamma_{\alpha}^*(j,q|\delta_f)\gamma_\alpha(\ell,q|\delta_f)\nn
&&\qquad\times\ 
{\rm tr}\left[\varrho_{\rm PT}a^\dagger_\alpha(q)a_\alpha(q)\right].
\label{GPT}
\eea
We wish to show that this is equal to the infinite-time limit of the
CUT result up to order ${\cal O}(U^2)$ corrections, i.e.
\be
G_{\rm PT}(j,\ell)=\lim_{t\to\infty}G(j,\ell;t)+{\cal O}(U^2).
\label{PTGGE}
\ee
The trace in \fr{GPT} is most easily evaluated in the $B=\infty$ basis
\begin{widetext}
\bea
{\rm tr}\left[\varrho_{\rm PT}a^\dagger_\alpha(q)a_\alpha(q)\right]&=&
\frac{1}{Z_{\rm PT}}{\rm tr}\left[
e^{\sum_{k,\alpha}\lambda^{(\alpha)}_ka^\dagger_{\alpha}(k)a_{\alpha}(k)}
\hat{n}_{\alpha,\alpha}(q,q|B=\infty)\right]\nn
&=&n_{\alpha\alpha}(q)-U\sum_{k_{1,2}> 0}
N^{\bm{\gamma}}_{\alpha\alpha}(k_1,k_1,k_2,k_2|q,q,B=\infty)
n_{\gamma_1\gamma_2}(k_1)n_{\gamma_3\gamma_4}(k_2)[1-\delta_{\gamma_1,\gamma_2}\delta_{\gamma_3,\gamma_4}]\nn
&-&U\sum_{k_{1,2} > 0}
N^{\bm{\gamma}}_{\alpha\alpha}(k_1,k_2,k_2,k_1|q,q,B=\infty)
n_{\gamma_1\gamma_4}(k_1)\big(\delta_{\gamma_2,\gamma_3}-n_{\gamma_3\gamma_2}(k_2)\big)
[1-\delta_{\gamma_1,\gamma_4}\delta_{\gamma_2,\gamma_3}]\ .
\label{adaPT}
\eea
\end{widetext}
Substituting \fr{adaPT} into \fr{GPT} we obtain an expression that
indeed agrees with the infinite-time limit of \fr{GF_CUT} in the
thermodynamic limit $L\to\infty$. This establishes \fr{PTGGE}.
Hence the Green's function $G(j,\ell)$ (for fixed $j,\ell$ in the
thermodynamic limit) on the prethermalization plateau is described by
the GGE~\fr{rhoPT} with deformed charges~\fr{newQ}. This observation
is consistent with a description of {\it local observables} on the
prethermalization plateau in terms of a deformed GGE. On the other
hand there are non-local operators, $n_{+-}(k)$ being a simple
example, which in fact do not relax at intermediate times and are
therefore not described by the ensemble $\varrho_{\rm PT}$ (without
time-averaging).

%%%%%%%%%%%%%%%%%%%%%%%%%%
\subsubsection{``Deformed GGE'' description of the four-point function}
%%%%%%%%%%%%%%%%%%%%%%%%%%
The preceding section shows that the value of the Green's function on
the prethermalization plateau is given by the deformed GGE
$\varrho_{PT}$. We now show that the deformed GGE also reproduces the
$t\to\infty$ expectation value of the CUT result for the four-point
function~\fr{def_fpf}. We wish to calculate
\begin{widetext}
\bea
{\rm tr}\Big[ \varrho_{PT} c\dg_{j}c_{j'}c\dg_{l}c_{l'}\Big]= 
\frac{1}{L^2} \sum_{q_j > 0}\sum_{\alpha_j=\pm} \gamma_{\alpha_1}^*(j,q_1)\gamma_{\alpha_2}(j',q_2)\gamma^*_{\alpha_3}(l,q_3)\gamma_{\alpha_4}(l',q_4){\rm tr}\Big[\varrho_{PT} a\dg_{\alpha_1}(q_1)a_{\alpha_2}(q_2)a\dg_{\alpha_3}(q_3)a_{\alpha_4}(q_4)\Big],
\label{fpf_trace}
\eea
with $\varrho_{PT}$ given in~\fr{rhoPT}. As in the previous section, this trace is most easily
performed in the $B=\infty$ basis
\bea
{\rm tr}\Big[\varrho_{PT}\hat A_{\bm\alpha}(\bm q) \Big] &=& 
\frac{1}{Z_{PT}}{\rm tr}\Big[e^{\sum_{k,\alpha}\lambda^{(\alpha)}_ka^\dagger_{\alpha}(k)a_{\alpha}(k)} \hat A_{\bm\alpha}(\bm q,B=\infty)\Big]\nn
&=& \frac{1}{Z_{PT}}{\rm tr}\Big[e^{\sum_{k,\alpha}\lambda^{(\alpha)}_ka^\dagger_{\alpha}(k)a_{\alpha}(k)}
\hat A_{\bm\alpha}(\bm q)\Big]\nn
 &&+\frac{U}{Z_{PT}}\sum_{k_j>0}N^{\bm\gamma}_{\alpha_3\alpha_4}(\bm k|q_3,q_4,\infty)
 {\rm tr}\Big[e^{\sum_{k,\alpha}\lambda^{(\alpha)}_ka^\dagger_{\alpha}(k)a_{\alpha}(k)}
 a\dg_{\alpha_1}(q_1)a_{\alpha_2}(q_2)\hat A_{\bm\gamma}(\bm k)\Big]\nn
 &&+\frac{U}{Z_{PT}}\sum_{k_j>0}N^{\bm\gamma}_{\alpha_1\alpha_2}(\bm k|q_1,q_2,\infty)
 {\rm tr}\Big[e^{\sum_{k,\alpha}\lambda^{(\alpha)}_ka^\dagger_{\alpha}(k)a_{\alpha}(k)}
\hat A_{\bm\gamma}(\bm k)a\dg_{\alpha_3}(q_3)a_{\alpha_4}(q_4)\Big]+{\cal O}(U^2),
\eea
\end{widetext}
where $\hat A_{\bm\gamma}(\bm k) =
a\dg_{\gamma_1}(q_1)a_{\gamma_2}(q_2)a\dg_{\gamma_3}(q_3)a_{\gamma_4}(q_4)$. The
GGE expectation values are easily calculated using Wick's theorem and
\fr{fixlambda}. Retaining only terms up to ${\cal O}(U)$ and
substituting the result back into \fr{fpf_trace}, we obtain the deformed
GGE value for the four-point function on the prethermalization plateau. 

\begin{figure}
\includegraphics[width=0.46\textwidth]{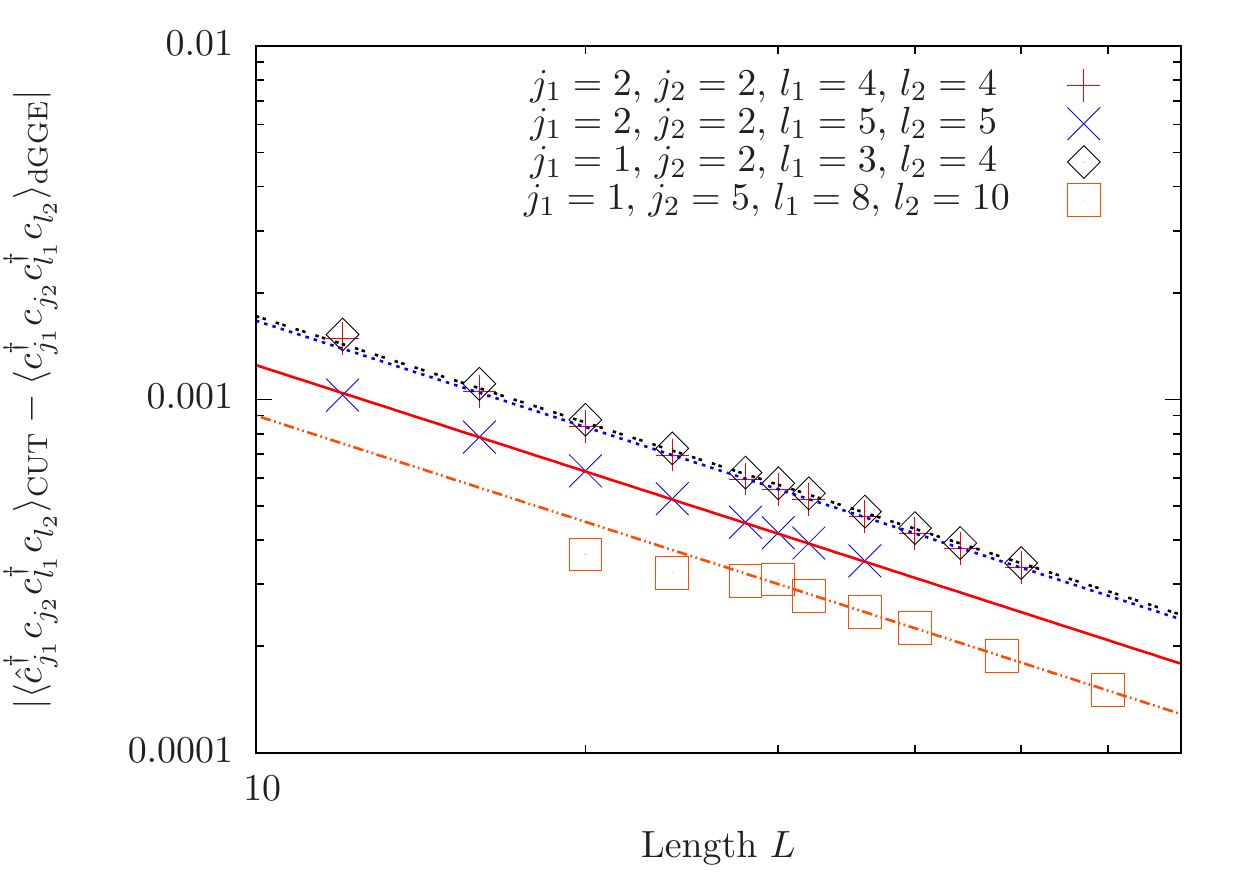}
\caption{(Color online) The $L$ dependence of the difference between the deformed GGE 
and the $t\to\infty$ CUT result for the four point function for a
number of separations. The solid lines are linear fits $c
L^{-1}$ to the data.}
\label{fig:FPF_CUT_dGGE} 
\end{figure}

In Fig.~\ref{fig:FPF_CUT_dGGE} we plot the difference between the
deformed GGE result obtained in this way and the stationary value of
the CUT result (found by projecting on to the stationary terms of
Eq.~\fr{def_fpf}) for a number of system sizes and separations. In all
cases the difference between the CUT and deformed GGE results scales
as $\frac{1}{L}$ and vanishes in the thermodynamic limit
$L\to\infty$. This confirms that the $t\to\infty$ stationary value of
the CUT four-point function is reproduced by the deformed
GGE~\fr{rhoPT}. This is a rather non-trivial check of our proposal
that prethermalization plateaus can be described in terms of a deformed GGE.

\begin{figure}
\includegraphics[width=0.45\textwidth]{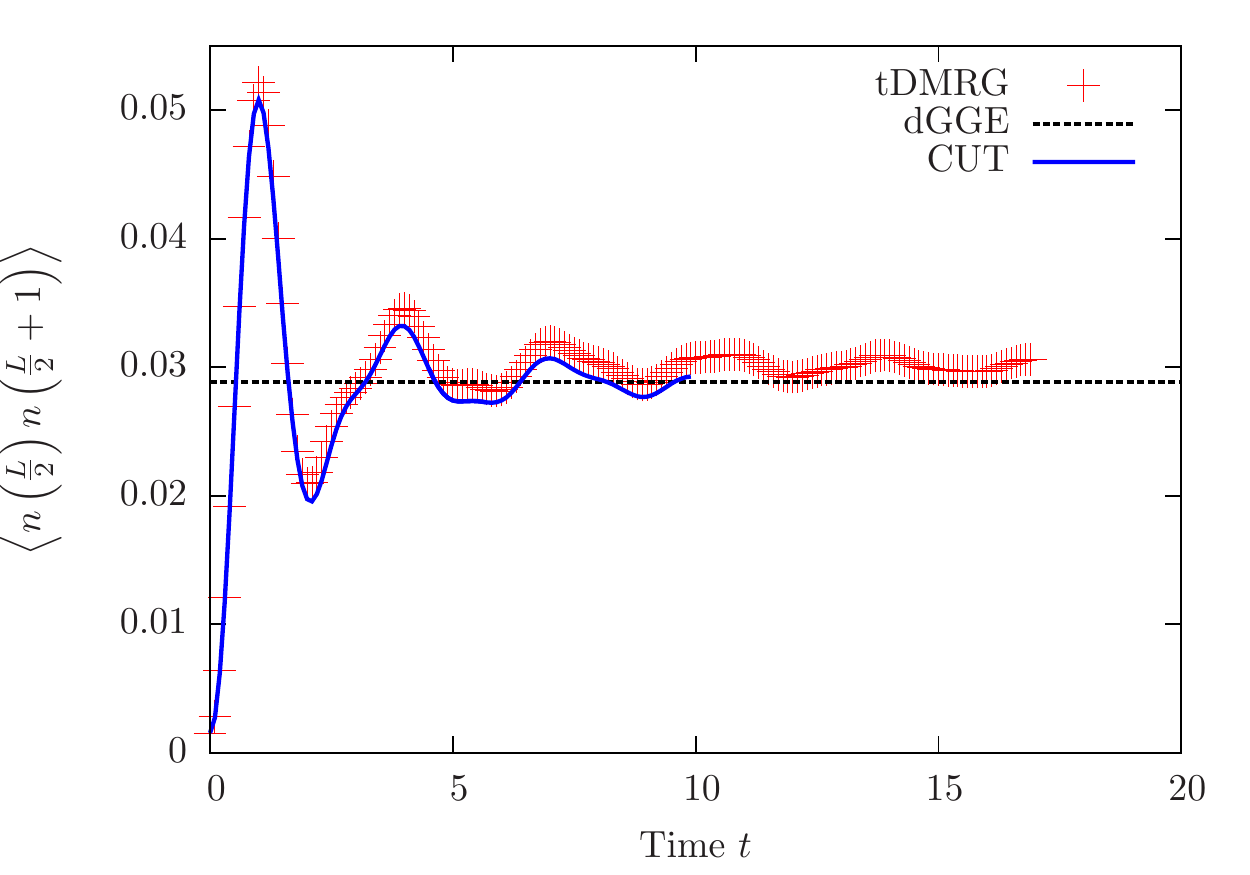}
\caption{(Color online) Nearest neighbour density-density correlation function $\la
  n(\frac{L}{2})n(\frac{L}{2}+1)\ra$ for a quench from $\delta_i = 0.8
\to \delta_f = 0.4$ and $U = 0\to0.4$ computed by t-DMRG for system
size $L=100$. For comparison we show CUT results for $L=40$ and the
asymptotic value predicted by the $L=50$ deformed GGE.}
\label{fig:tDMRG_dGGE_densdens}
\end{figure}
\begin{figure}
\includegraphics[width=0.45\textwidth]{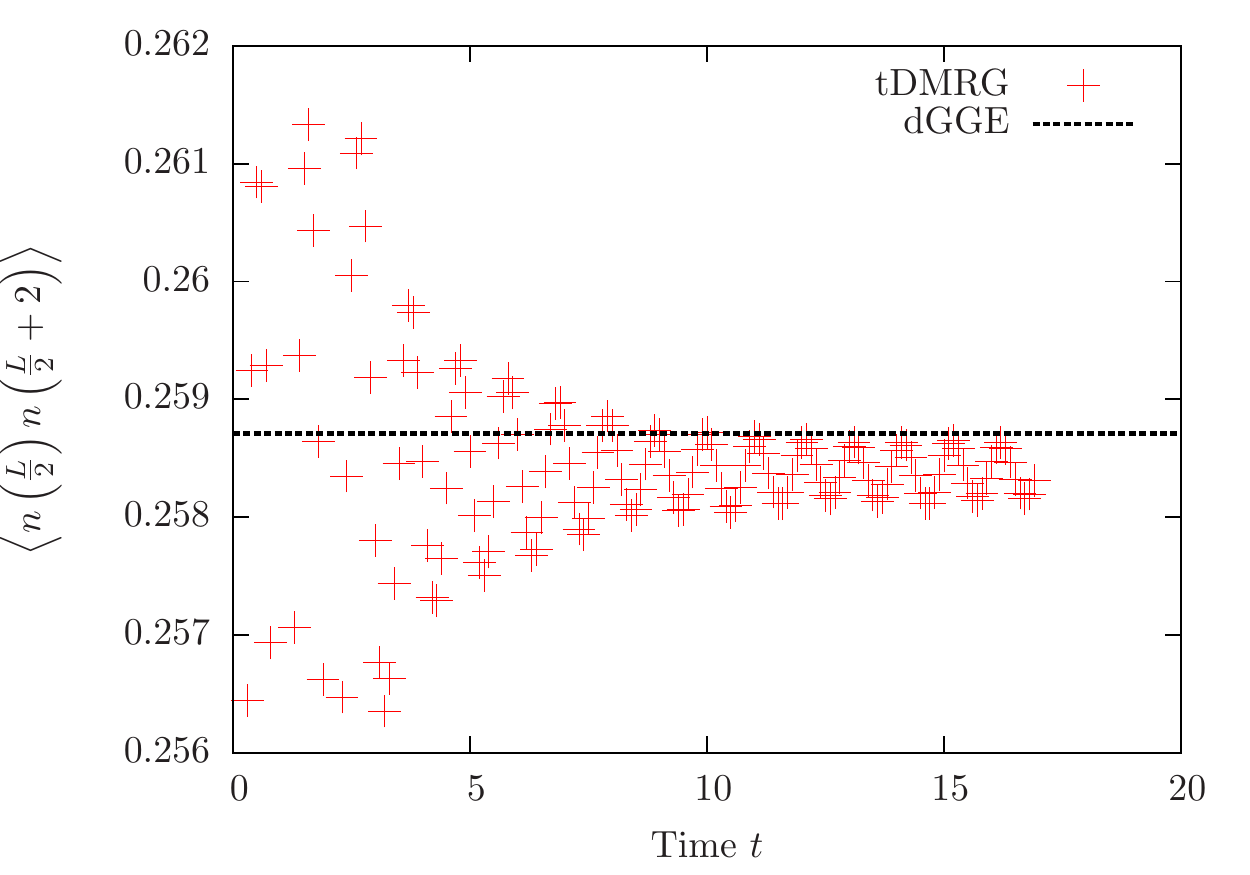}
\caption{(Color online) Next-nearest-neighbor density-density correlation function $\la
  n(\frac{L}{2})n(\frac{L}{2}+2)\ra$ for a quench from $\delta_i = 0.8
\to \delta_f = 0.4$ and $U = 0\to0.4$ computed by t-DMRG for system
size $L=100$. The correlator relaxes to a stationary value
consistent with the deformed GGE prediction (evaluated for $L=50$).}
\label{fig:tDMRG_dGGE_densdens2}
\end{figure}

In Figs.~\ref{fig:tDMRG_dGGE_densdens}~and~\ref{fig:tDMRG_dGGE_densdens2} 
we present comparisons between t-DMRG results and predictions of the
deformed GGE for nearest-neighbor and next-nearest-neighbor density-density
correlation functions~\fr{fpf_trace} for the quench $\delta_i
=0.8\to\delta_f = 0.4$ and $U=0\to0.4$. Taking into account that $U_f$
is not particularly small, the observed agreement between the two results is
quite satisfactory. This supports our assertion that the deformed GGE provides
a good description of higher-order correlation functions on the
prethermalization plateau. We see similarly good agreement for all
separations (up to $4$ sites) that we explicitly checked. The deformed
GGE predictions and the CUT result of
Fig.~\ref{fig:tDMRG_dGGE_densdens}  are calculated for system sizes
$L=40,50$ rather than $L=100$, because the computational cost of carrying
out the momentum sums in the expression for the four-point
function~\fr{def_fpf} increases very rapidly with system size. 

%%%%%%%%%%%%%%%%%%%%%%%%%%%%%%%%%%%%%%%%%%%%%%%%%%%%%%%%%%%%%%
\section{t-DMRG results for larger values of $U$ and absence of thermalization
on accessible times scale}
\label{Sec:tdmrg}
%%%%%%%%%%%%%%%%%%%%%%%%%%%%%%%%%%%%%%%%%%%%%%%%%%%%%%%%%%%%%%

\begin{figure}[ht]
\includegraphics[width=0.46\textwidth]{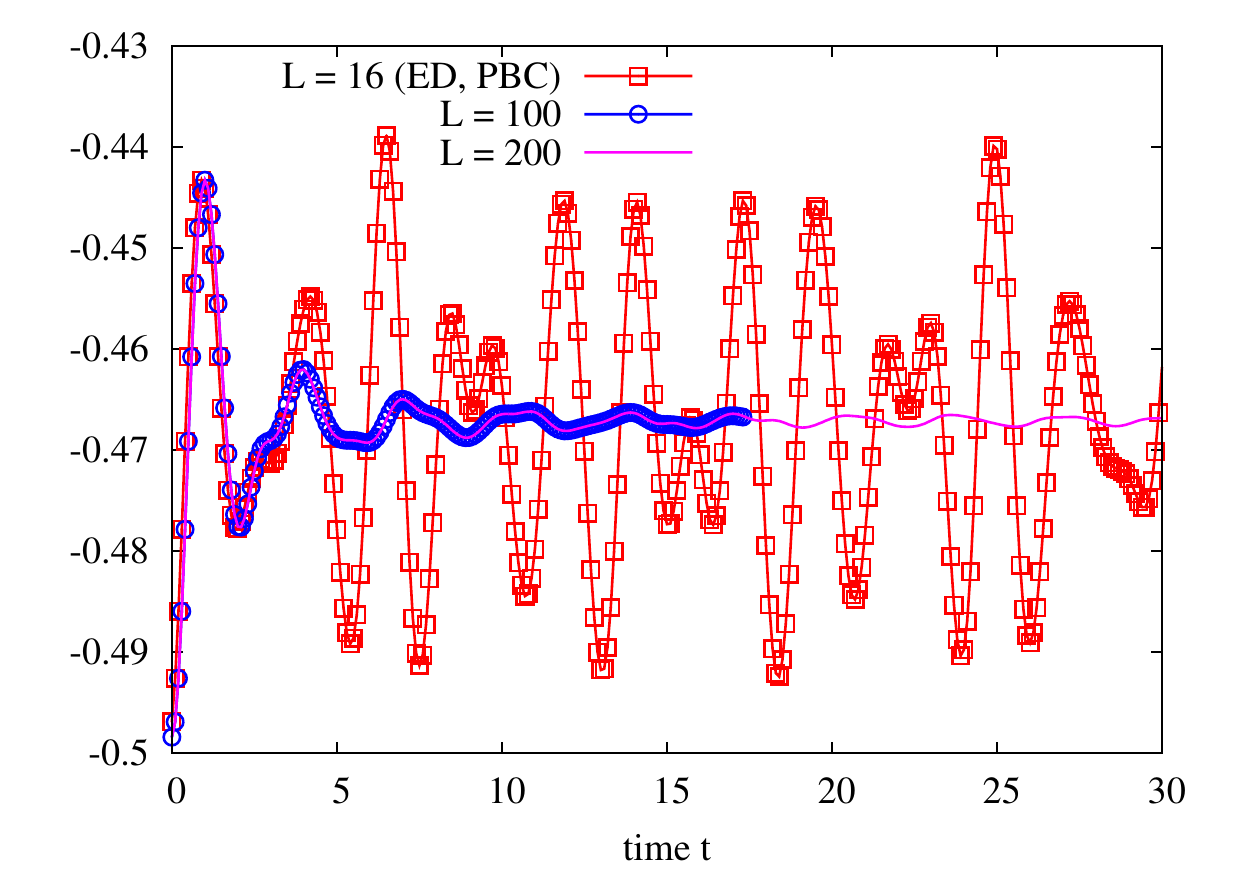}
\caption{(Color online) Time evolution of $\mathcal G(L/2,L/2+1)$ for quenches with
$\delta_i = 0.8 \to \delta_f = 0.4$ and $U_i = 0 \to U = 0.4$ and
system sizes $L=16$, $L=100$ and $L=200$ sites. The data for $L=16$
are ED results for systems with periodic boundary conditions
(PBC) and are seen to exhibit many revivals. }
\label{fig:longtimes}
\end{figure}

In this section we turn to numerical results obtained for quenches to
final Hamiltonians with both weak and strong interactions, i.e., when
$U \gtrsim |\delta_i - \delta_f|$. As can be seen, in all cases the
time evolution seems to reach a plateau and remains - on the
accessible time scales - on this plateau. This is observed for
quenches starting from a non-interacting initial state as well as when
$U_{\rm ini} = 5$. 

%%%%%%%%%%%%%%%%%%%%%%%%%%%%%%%%%%%%%%%%%%%%%%%%%%%%%%%%%%%%
\subsection{Extent of prethermalization plateaus}
%%%%%%%%%%%%%%%%%%%%%%%%%%%%%%%%%%%%%%%%%%%%%%%%%%%%%%%%%%%%
The first issue we want to address is the time scale over which we
observe prethermalization plateaus. In Figs.~\ref{GF_Ua}--\ref{GF_Uc}
and \ref{GF_U015_dxa}-\ref{GF_U015_dxc} results are shown only up to
$t\approx 10$ in order to avoid revivals. The prethermalization
plateau for $U=0.4$ persists to much later times of at least $t\approx
30$, as can be seen in Fig.~\ref{fig:longtimes}, where we present data
for $L=16,\ L =100, \, L=200$.  On the accessible time scales there is
no sign that the $L=200$ system starts to deviate from the plateau at
late times.  

%%%%%%%%%%%%%%%%%%%%%%%%%%%
\subsection{Time averages}
\begin{figure}[th]
\includegraphics[width=0.46\textwidth]{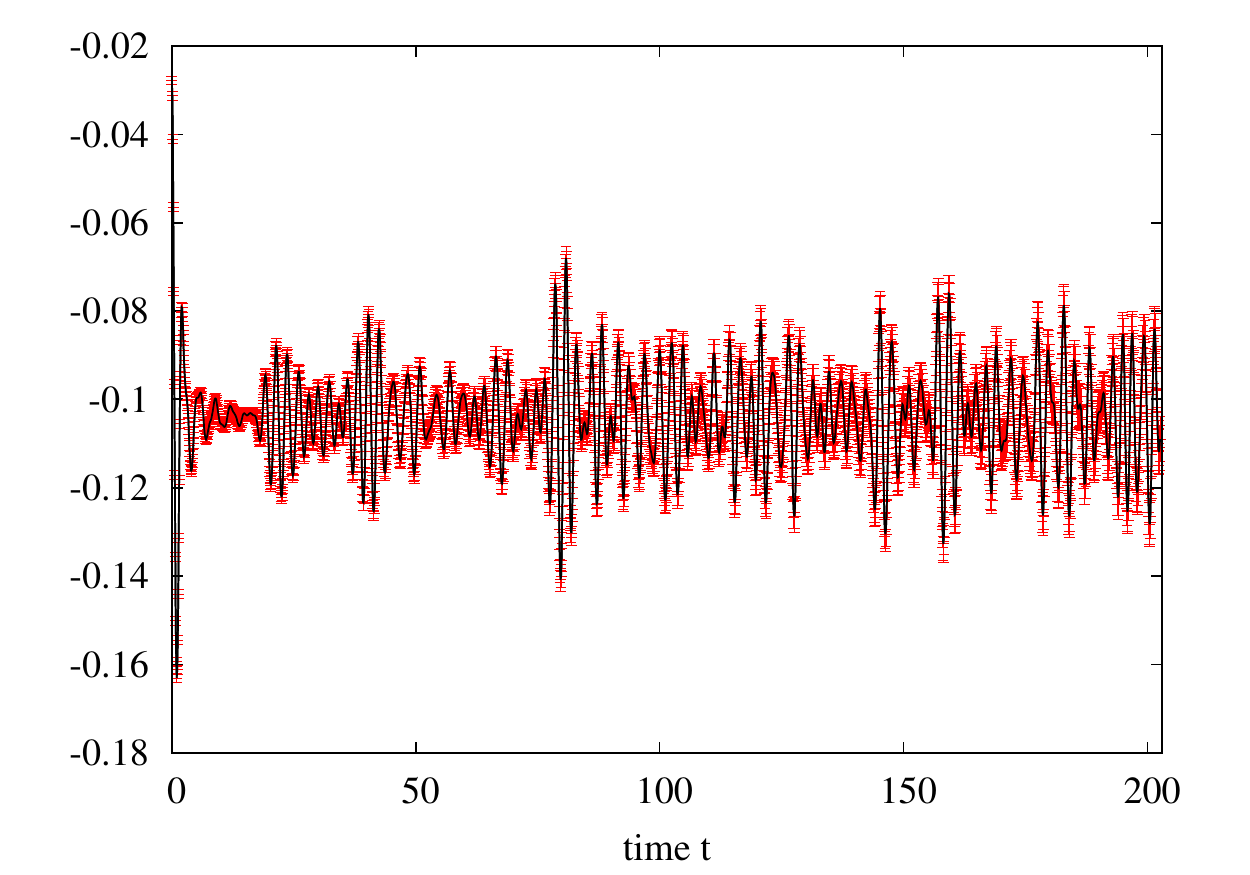} 
\caption{(Color online) Time evolution of $\mathcal G(L/2,L/2+1)$ for quenches with
  $\delta_i = 0.8 \to \delta_f = 0.4$ and $U_i = 0 \to U =
  0.4$. $L=50$ site system up to $t\sim200$, with error bars estimated
  in Appendix~\ref{appendix:tDMRGerrors}.}
\label{fig:longtimes2}
\end{figure}

%%%%%%%%%%%%%%%%%%%%%%%%%%%
A standard method for extracting stationary values of observables from
finite systems is to consider {\it time-averaged} quantities, e.g.
\be
\frac{1}{T}\int_0^Tdt\ G(L/2,L/2+1).
\ee
For the $L=16$ system shown in Fig.~\ref{fig:longtimes} the average
over long times is in good agreement with the plateau value for the
$L=100$ and $L=200$ data. One question that can be asked is whether 
time averages may reveal signs of the system deviating from the
prethermalization plateau. In order to investigate this issue, we have
carried out t-DMRG simulations for a $L=50$ system up to very late
times $t=200$. The results are shown in Fig.~\ref{fig:longtimes2}.
Time averages of the t-DMRG data do not reveal any signs of deviations
from the plateau value at late times.

%%%%%%%%%%%%%%%%%%%%%%%%%%%%%%%%%%%%%%%
\subsection{The role of interactions in the pre-quench and post-quench Hamiltonians}
%%%%%%%%%%%%%%%%%%%%%%%%%%%%%%%%%%%%%%%
In this section we present results for a variety of interaction
strengths $0.4\leq U\leq10$ in the post-quench Hamiltonian, as well as
for quenches from the ground state at a finite value of the interactions.
We provide two benchmarks for comparison:

\subsubsection{Gibbs Ensemble}

\begin{figure}
\includegraphics[width=0.46\textwidth]{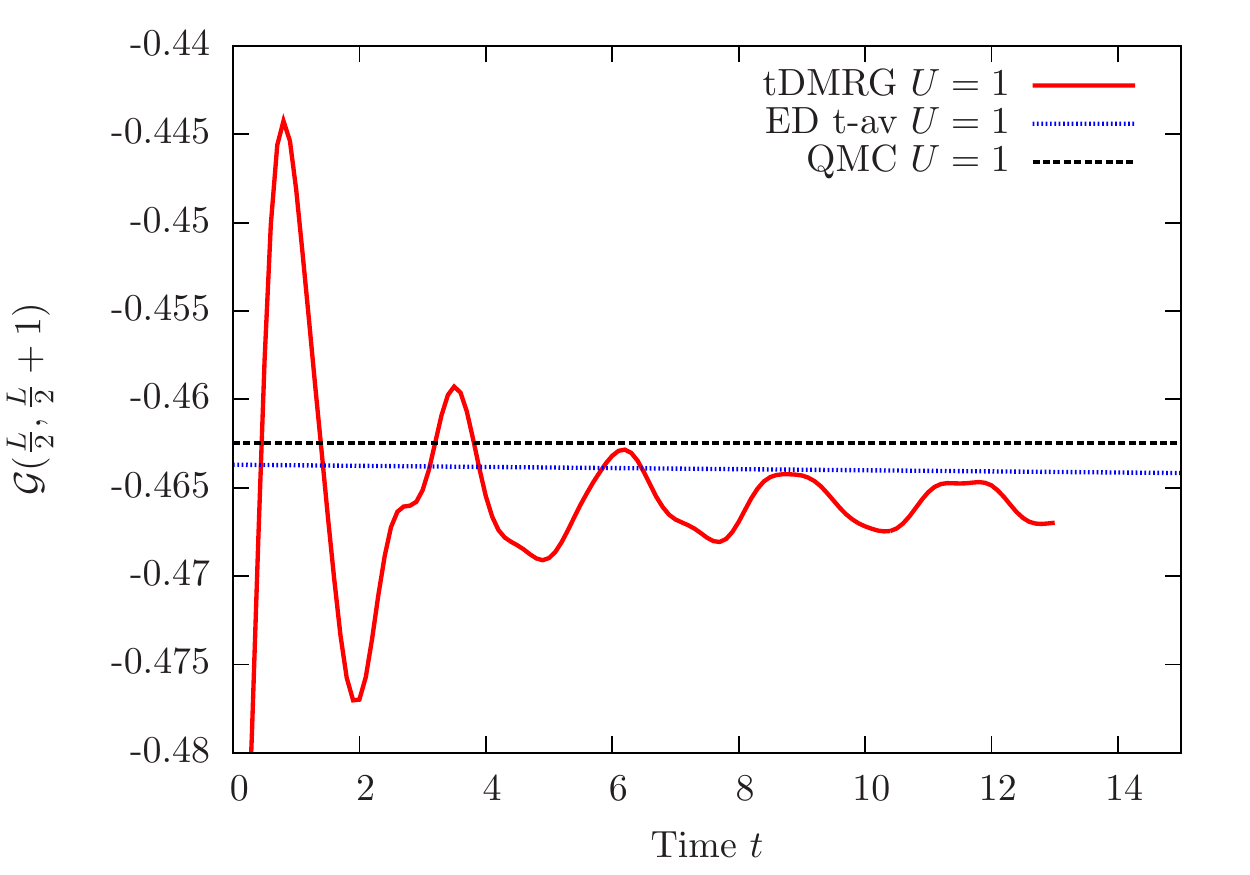}
\caption{(Color online) Comparison of the t-DMRG, time-averaged (t-av) ED and QMC results
for the nearest-neighbour Green's function at time $t$ after the quench
$\delta_i = 0.8\to\delta_f=0.4$ $U=0\to1$. t-DMRG and QMC simulations
are performed on the $L=100$ chain, whilst ED studies the $L=16$ chain.}
\label{fig:U1_thermcomp}
\end{figure}

\begin{figure}
\includegraphics[width=0.46\textwidth]{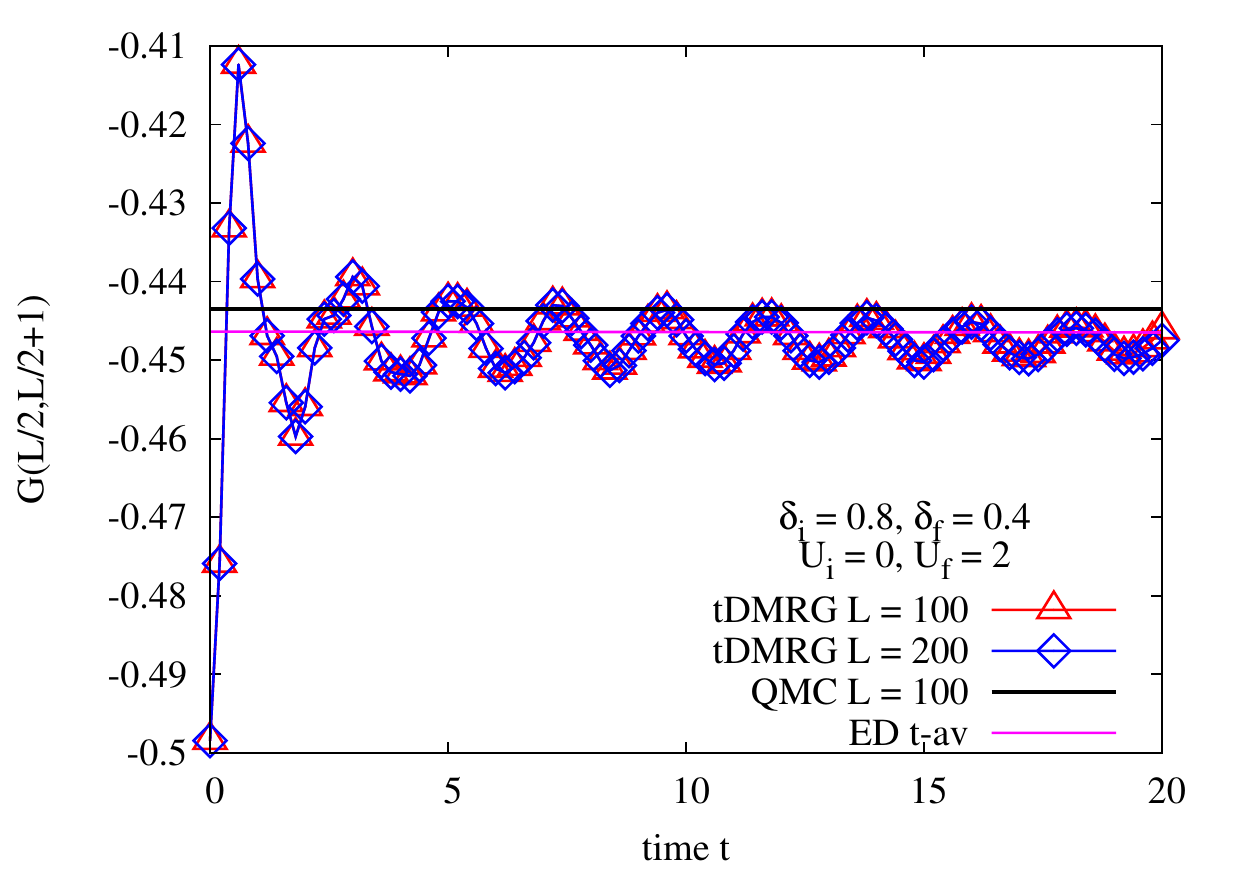}
\caption{(Color online) Comparison of the t-DMRG, time-averaged (t-av) ED and QMC results
for the nearest-neighbour Green's function at time $t$ after the quench
$\delta_i = 0.8\to\delta_f=0.4$ $U=0\to2$. t-DMRG and QMC simulations
are performed on the $L=100$ chain, whilst ED studies the $L=16$ chain.}
\label{fig:U2_thermcomp}
\end{figure}

\begin{figure}
\includegraphics[width=0.46\textwidth]{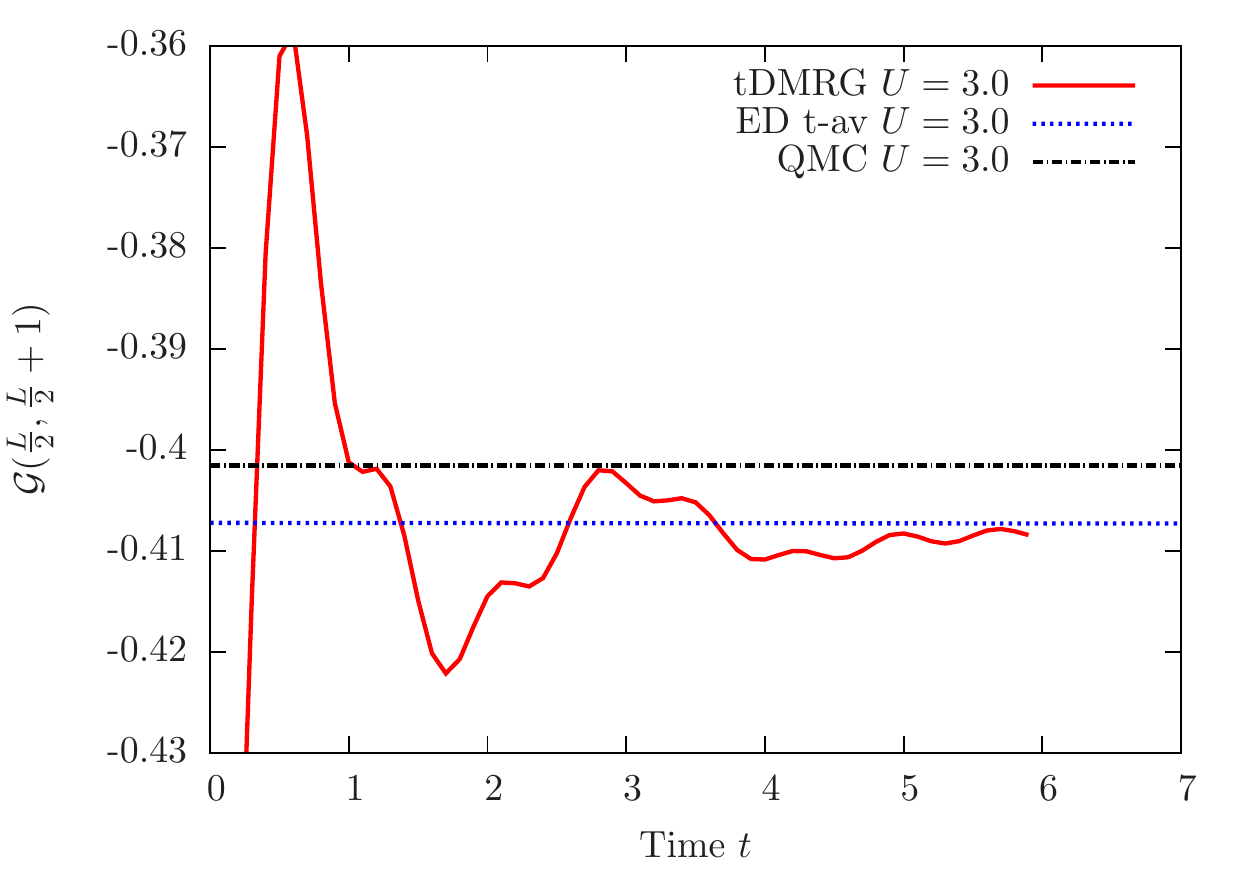}
\caption{(Color online) Comparison of the t-DMRG, time-averaged (t-av) ED and QMC results
for the nearest-neighbour Green's function at time $t$ after the quench
$\delta_i = 0.8\to\delta_f=0.4$ $U=0\to3$. t-DMRG and QMC simulations
are performed on the $L=100$ chain, whilst ED studies the $L=16$ chain.}
\label{fig:U3_thermcomp}
\end{figure}

\begin{figure}
\includegraphics[width=0.46\textwidth]{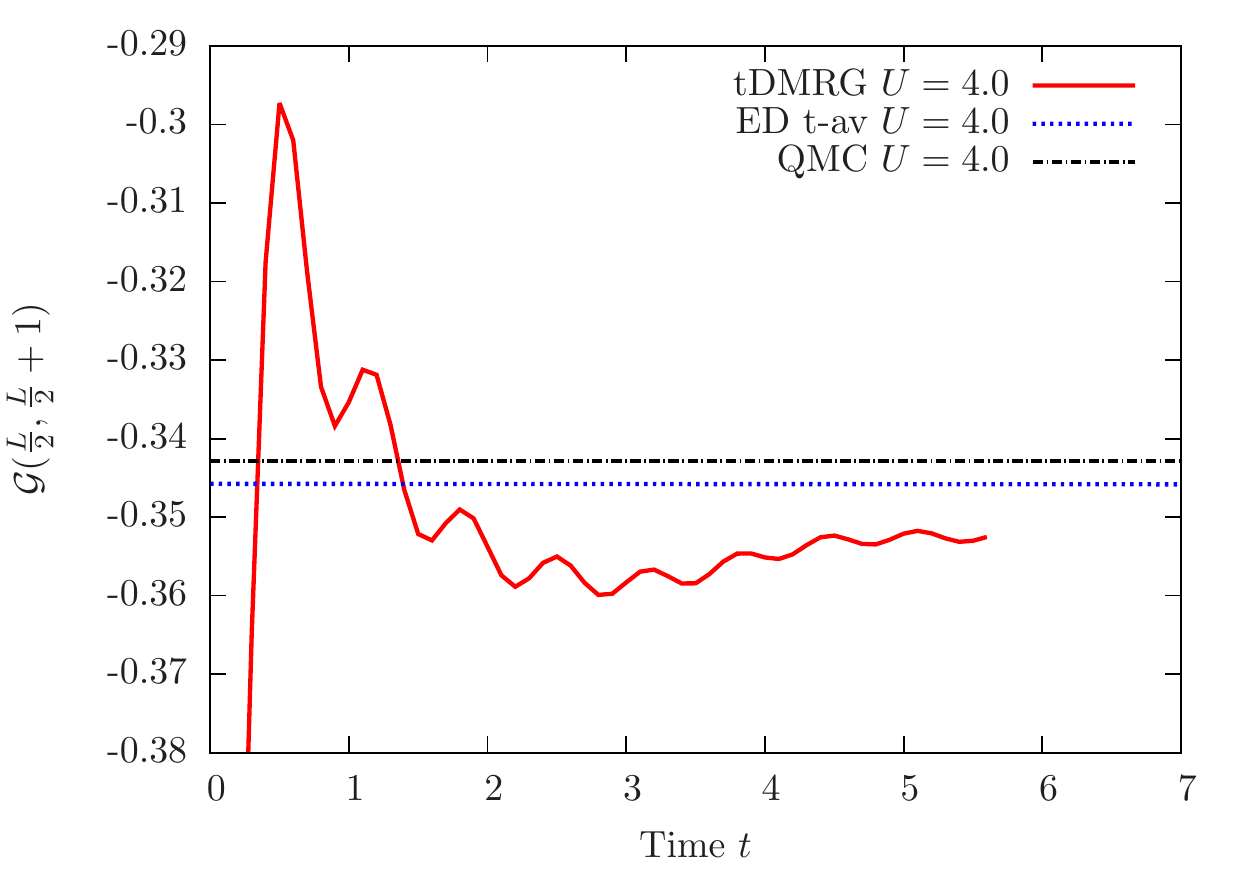}
\caption{(Color online) Comparison of the t-DMRG, time-averaged (t-av) ED and QMC results
for the nearest-neighbour Green's function at time $t$ after the quench
$\delta_i = 0.8\to\delta_f=0.4$ $U=0\to4$. t-DMRG and QMC simulations
are performed on the $L=100$ chain, whilst ED studies the $L=16$ chain.}
\label{fig:U4_thermcomp}
\end{figure}

\begin{figure}
\includegraphics[width=0.46\textwidth]{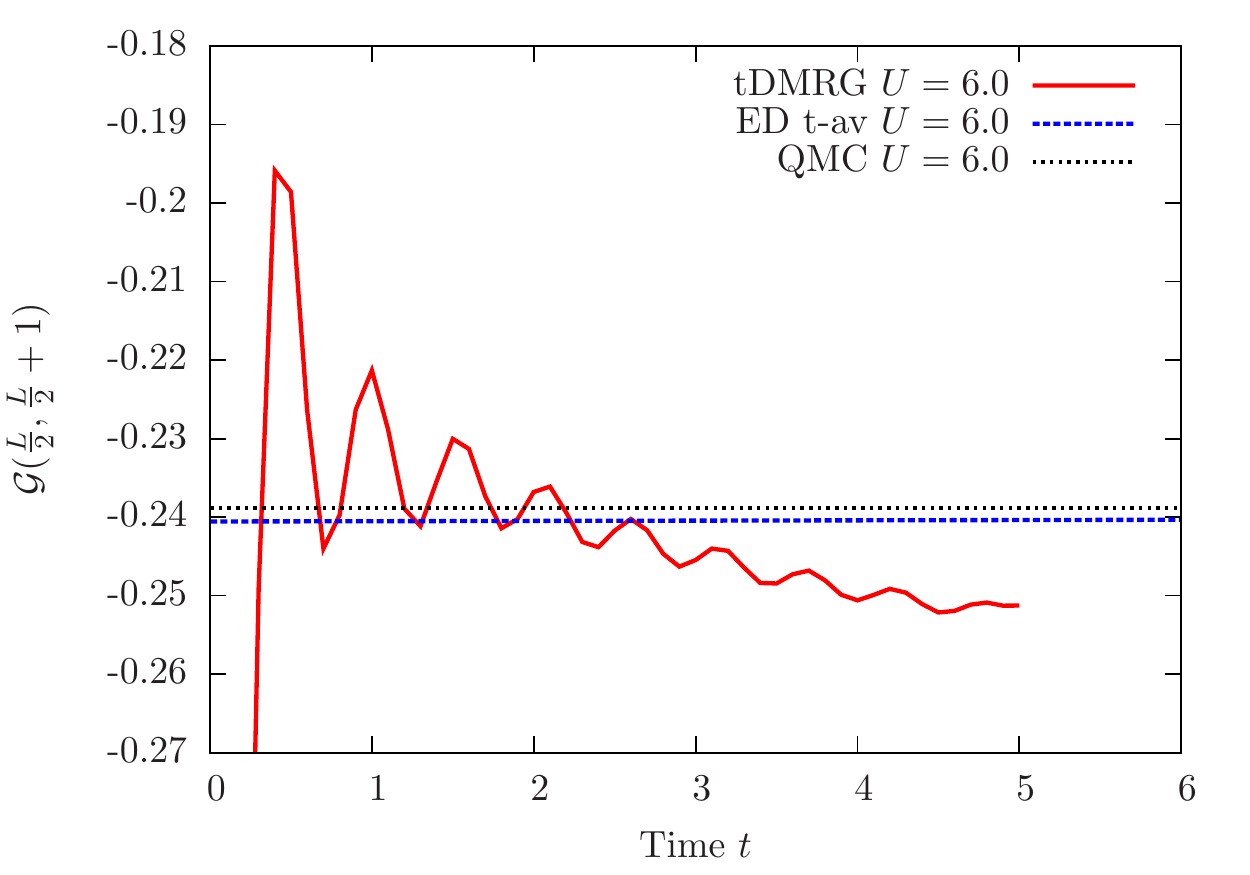}
\caption{(Color online) Comparison of the t-DMRG, time-averaged (t-av) ED and QMC results
for the nearest-neighbour Green's function at time $t$ after the quench
$\delta_i = 0.8\to\delta_f=0.4$ $U=0\to6$. t-DMRG and QMC simulations
are performed on the $L=100$ chain, whilst ED studies the $L=16$ chain.}
\label{fig:U6_thermcomp}
\end{figure}

One useful comparison is with the appropriate Gibbs ensemble
describing a putative thermal ensemble at late times. We have computed
these by quantum Monte Carlo (QMC) using the ALPS
collaboration~\cite{ALPS} directed loop stochastic series
expansion~\cite{SSE} code. Using the Jordan-Wigner transformation 
to map onto a spin model, the QMC calculations are performed in the
grand canonical ensemble; the chemical potential and the effective
temperature are fixed to ensure the correct energy and number
densities (within the QMC error): these are given in
Table~\ref{tab:Beta} (see also Figs.~\ref{fig:U04_thermcomp}--\ref{fig:U10_thermcomp}). 
\begingroup
\begin{table}[ht]
\begin{tabular}{|c|c|c|c|c|c|}
\hline
U	& $E/L$	&	$\beta$	& $\mu$ & ${\cal G}(\frac{L}{2},\frac{L}{2}+1)$	 & QMC  \\
& & & &  &  Error \\
\hline
0.4	& -0.664373 	&	3.0741 &	0.4 	& -0.46358 	&	$1.62\times10^{-3}$ \\
1 &	-0.589142 &	2.6494 &	1	& -0.46247 &	$2.98\times10^{-4} $\\
2 &	-0.463757 &	2.0437 &	2	& -0.44347 &	$6.94\times10^{-5} $\\
3 &	-0.338371	 & 1.5882 & 	3	& -0.40153 &	$6.49\times10^{-5} $ \\
4 &	-0.212986	 & 1.2175 &	4	& -0.34284 &	$3.06\times10^{-4}$  \\
6 &   0.037784 & 0.7250 & 	6	& -0.23885 & 	$1.34\times10^{-4}$ \\
8 & 	0.288550	& 0.4868 & 	8	& -0.17441 &	 $3.15\times10^{-4}$ \\
10& 0.539324	& 0.3591 &	10	& -0.13514 & 	$1.23\times10^{-4}$ \\
\hline
\end{tabular}
\caption{(Color online) Summary of the effective temperature $\beta$ and chemical
potential $\mu$ used in the QMC to calculate the Green's function
${\cal G}(\frac{L}{2},\frac{L}{2}+1)$ on the $L=100$ chain as
presented in
Figs.~\ref{fig:U04_thermcomp}--\ref{fig:U10_thermcomp}. The energy
density $E/L$ is found by taking the expectation value of the
interacting Hamiltonian $H(t>0)$ at $t = 0^+$.} 
\label{tab:Beta}
\end{table}
\endgroup
In the QMC simulations of the $L~=~100$ chain we perform $5\times10^7$
thermalization steps and perform measurements of the nearest-neighbour
Green's function after $1.5\times10^8$ sweeps. 

\subsubsection{Diagonal Ensemble}
A second useful benchmark is provided by the diagonal ensemble.
Given an initial state $|\Psi_0\rangle$ and a basis $\{|n\rangle\}$ of
energy eigenstates, the diagonal ensemble average of an observable
${\cal O}$ is defined as  
\be
\langle {\cal O}\rangle_{\rm DE}=\sum_n\langle n|{\cal
  O}|n\rangle\ |\langle n|\Psi_0\rangle|^2.
\ee
For finite systems this equals the long-time average (over many
recurrences). We compute the diagonal ensemble for a system of $L=16$
sites by exact diagonalization (ED). 

%%%%%%%%%%%%%%%%%%%%%%%%%%%%%%%%%%%%%%%%%%%%%%%%%%%%%%%%%%%%%%
\subsubsection{Difference between diagonal and Gibbs averages}
\begin{figure}
\includegraphics[width=0.48\textwidth]{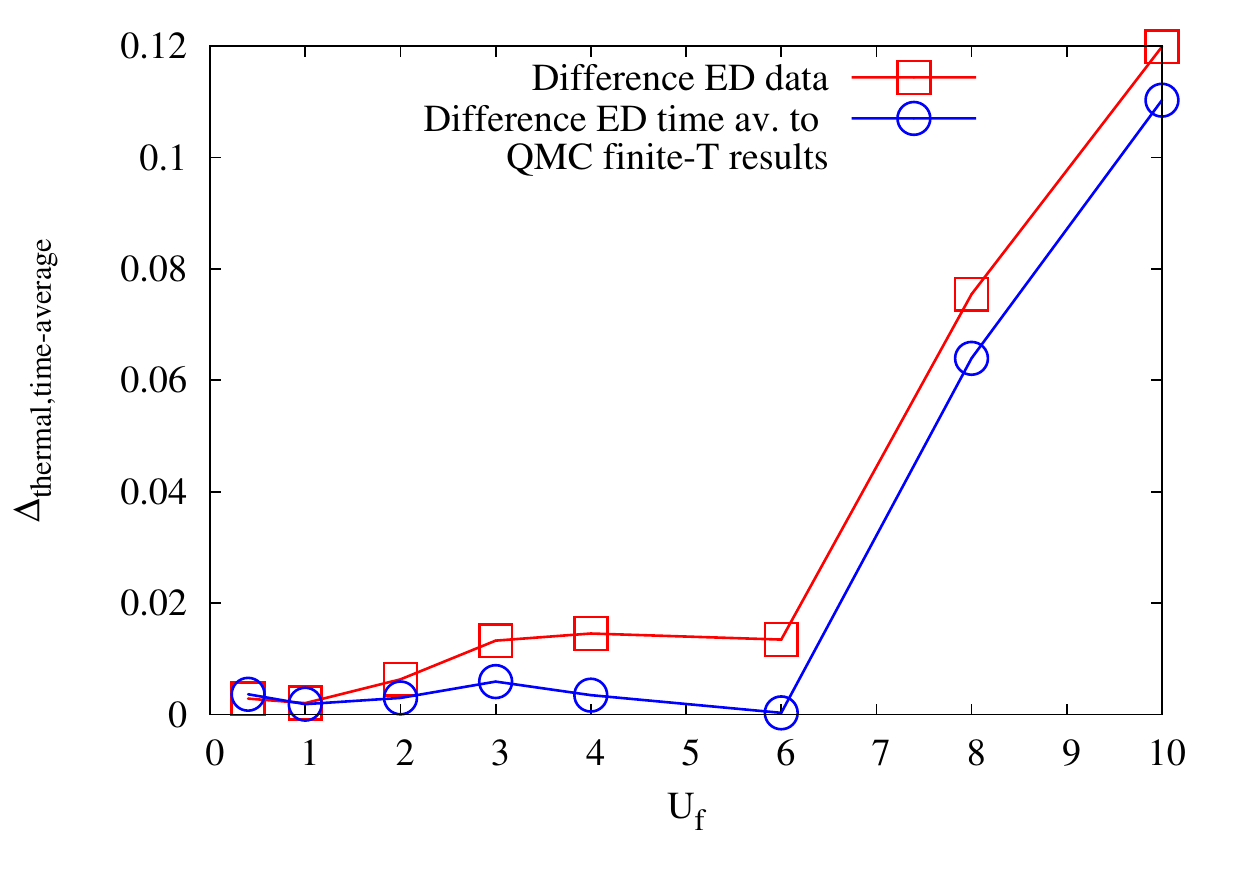}
\caption{(Color online) Difference in the value of $\mathcal G(L/2,L/2+1)$ between
finite temperature results obtained with QMC ($L=100$) or ED
($L=16$), respectively, to the time-average values obtained via ED
for $L=16$ for a quench with $\delta_i = 0.8 \to \delta_f = 0.4$ and
$U_i = 0 \to U_f$ as a function of $U_f$. Finite size effects are
less pronounced for small values of $U_f$, but prominent for $U_f >
1$. The intermediate region $1 \leq U_f < 8$ is the best candidate
to obtain thermalization on long time scales in this system.} 
\label{fig:differenceThermalTimeaverages}
\end{figure}

%%%%%%%%%%%%%%%%%%%%%%%%%%%%%%%%%%%%%%%%%%%%%%%%%%%%%%%%%%%%%%
In Fig.~\ref{fig:differenceThermalTimeaverages} we show the difference
between the expectations values of the nearest-neighbour Green's
function $\mathcal G(L/2,L/2+1)$ in the diagonal and Gibbs ensembles
respectively for different values of $U_f$. As the diagonal ensemble
is available only for system size $L=16$, we display the quantities
\bea
\langle c^\dagger_{L/2}c_{L/2+1}\rangle_{\rm DE,L=16}
-\langle c^\dagger_{L/2}c_{L/2+1}\rangle_{\rm Gibbs,L=16}\ ,\nn
\langle c^\dagger_{L/2}c_{L/2+1}\rangle_{\rm DE,L=16}
-\langle c^\dagger_{L/2}c_{L/2+1}\rangle_{\rm Gibbs,L=100}.
\eea
We see that for small values $U_f$ the two averages are close to one
another, but for large $U_f$ they become very different.

%%%%%%%%%%%%%%%%%%%%%%%%%%%%%%%
\subsubsection{Results}
\label{ssec:tdmrgresults}
\begin{figure}
\includegraphics[width=0.46\textwidth]{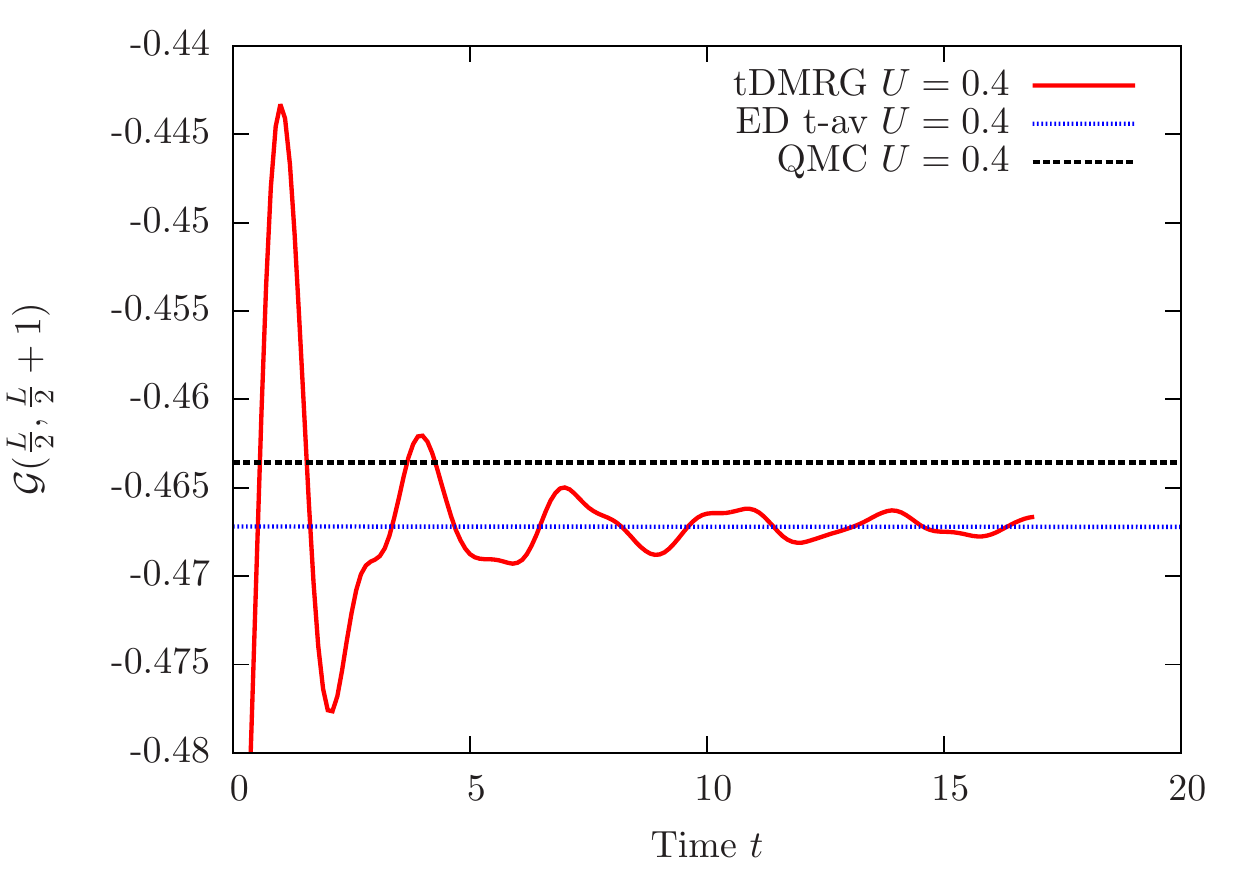}
\caption{(Color online) Comparison of the t-DMRG, time-averaged (t-av) ED and QMC results
for the nearest-neighbour Green's function at time $t$ after the quench
$\delta_i = 0.8\to\delta_f=0.4$ $U=0\to0.4$. t-DMRG and QMC simulations
are performed on the $L=100$ chain, whilst ED studies the $L=16$ chain.}
\label{fig:U04_thermcomp}
\end{figure}

\begin{figure}
\includegraphics[width=0.46\textwidth]{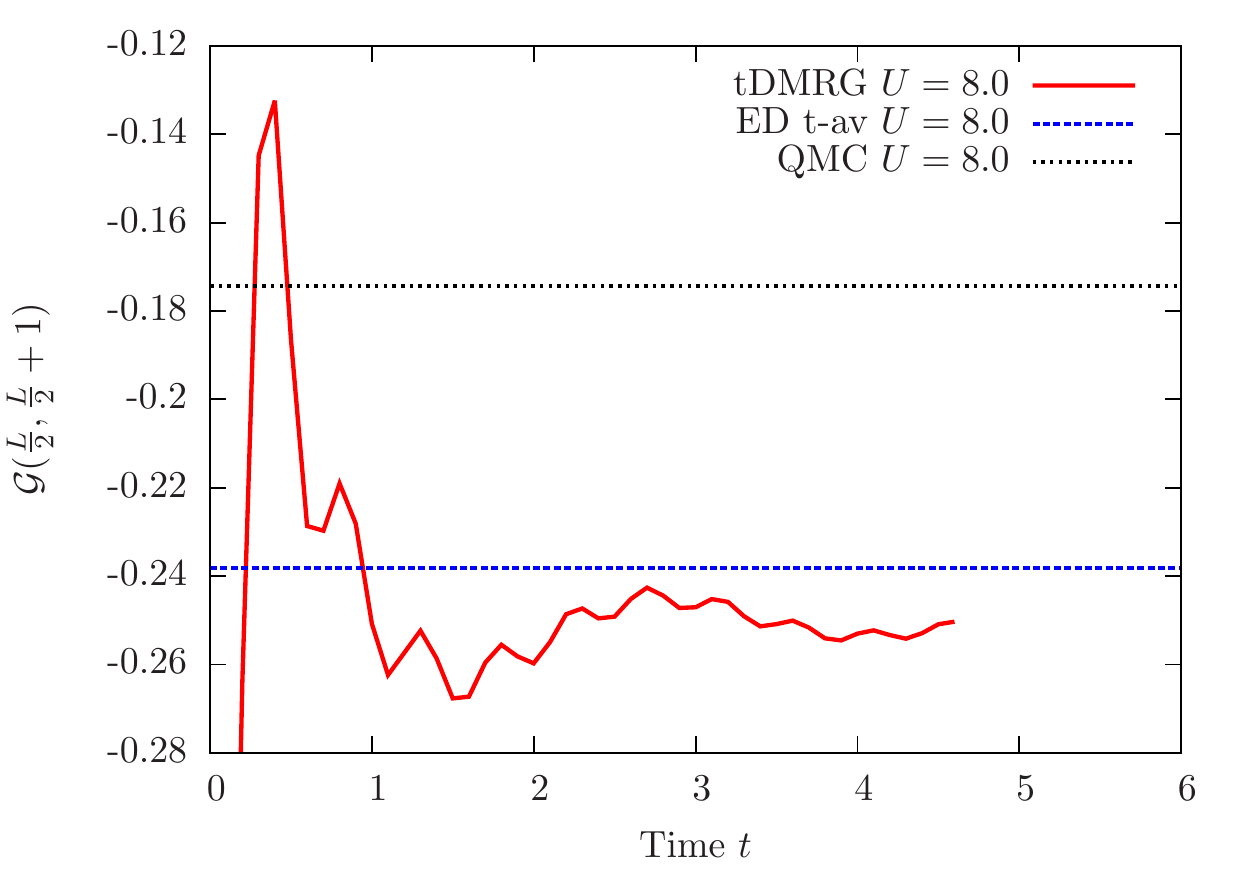}
\caption{(Color online) Comparison of the t-DMRG, time-averaged (t-av) ED and QMC results
for the nearest-neighbour Green's function at time $t$ after the quench
$\delta_i = 0.8\to\delta_f=0.4$ $U=0\to8$. t-DMRG and QMC simulations
are performed on the $L=100$ chain, whilst ED studies the $L=16$ chain.}
\label{fig:U8_thermcomp}
\end{figure}

\begin{figure}
\includegraphics[width=0.46\textwidth]{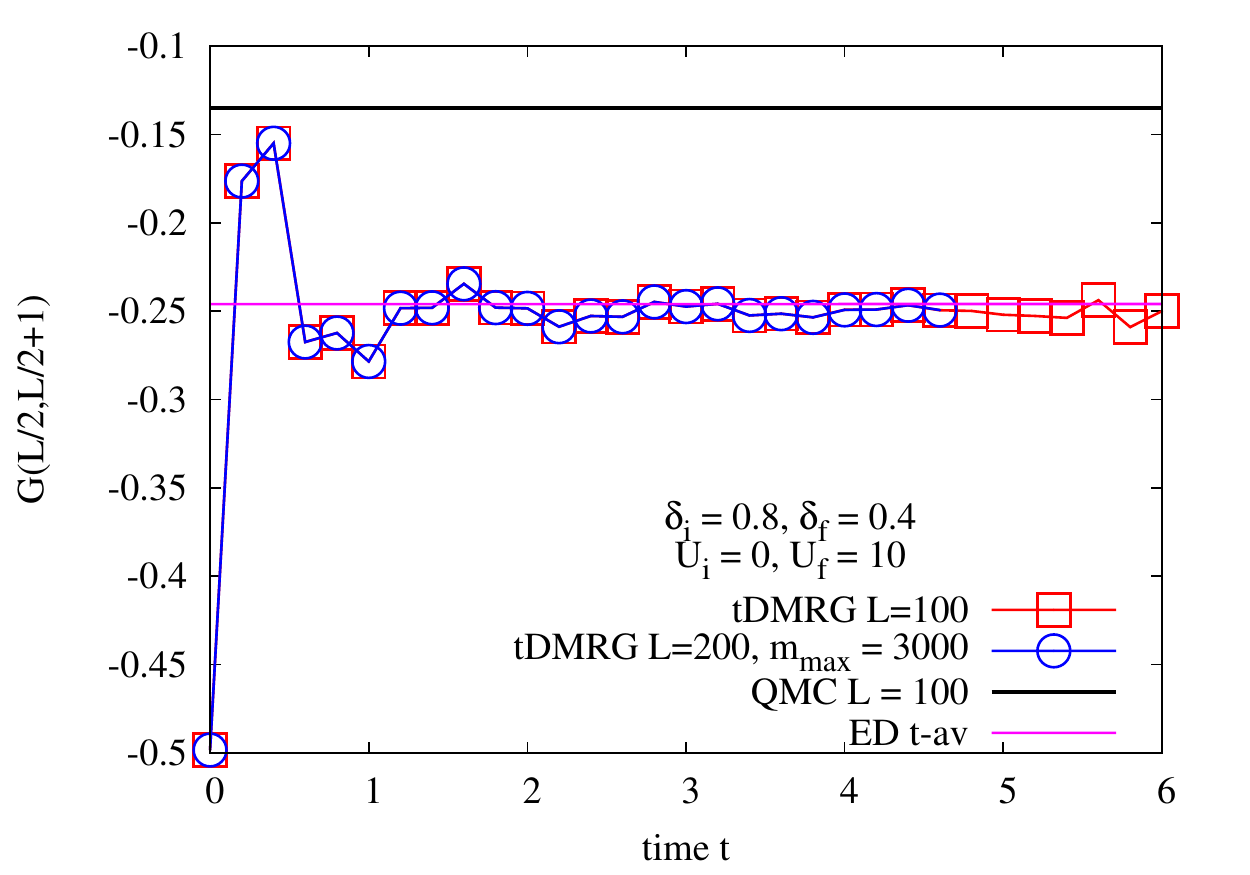}
\caption{(Color online) Comparison of the t-DMRG, time-averaged (t-av) ED and QMC results
for the nearest-neighbour Green's function at time $t$ after the quench
$\delta_i = 0.8\to\delta_f=0.4$ $U=0\to10$. t-DMRG and QMC simulations
are performed on the $L=100$ chain, whilst ED studies the $L=16$ chain.}
\label{fig:U10_thermcomp}
\end{figure}

%%%%%%%%%%%%%%%%%%%%%%%%%%%%%%%
As can be seen from Figs.~\ref{fig:U04_thermcomp},~\ref{fig:U8_thermcomp} 
and~\ref{fig:U10_thermcomp}, the nearest-neighbour Green's function 
approaches plateaus values at late times, which are compatible with the 
diagonal ensemble (given that the latter was calculated for L=16 we 
expect there to be finite-size effects), but not the Gibbs ensemble. 

On the other hand, the plateau for intermediate values $U\approx 2$
is compatible with a thermal ensemble on the accessible time
scales. We propose the following explanation for these observations:
\begin{enumerate}
\item{} The small-$U$ regime is described by a prethermalization
plateau as discussed in section \ref{Sec:Pretherm}. It can be
understood in terms of a deformation of the generalized Gibbs ensemble
characterizing the stationary state of the $U=0$ quench.
\item{} The large-$U$ regime is also described by a prethermalization
plateau, which now can be understood in terms of a deformation of the
generalized Gibbs ensemble characterizing the stationary state of the
$\delta_f=0$ quench. This corresponds to a quench to the Heisenberg
XXZ chain in the massive regime. Given that our initial state has a
short correlation length, GGE expectation values of local observables
could be calculated by the method of Ref.~\onlinecite{FE:13b}. In order to
test our interpretation, we have investigated the dependence of the
plateau value on $\delta_f$ ($\delta_f=0$ corresponding to an
integrable quench in the XXZ chain). 
\begin{figure}
\includegraphics[width=0.46\textwidth]{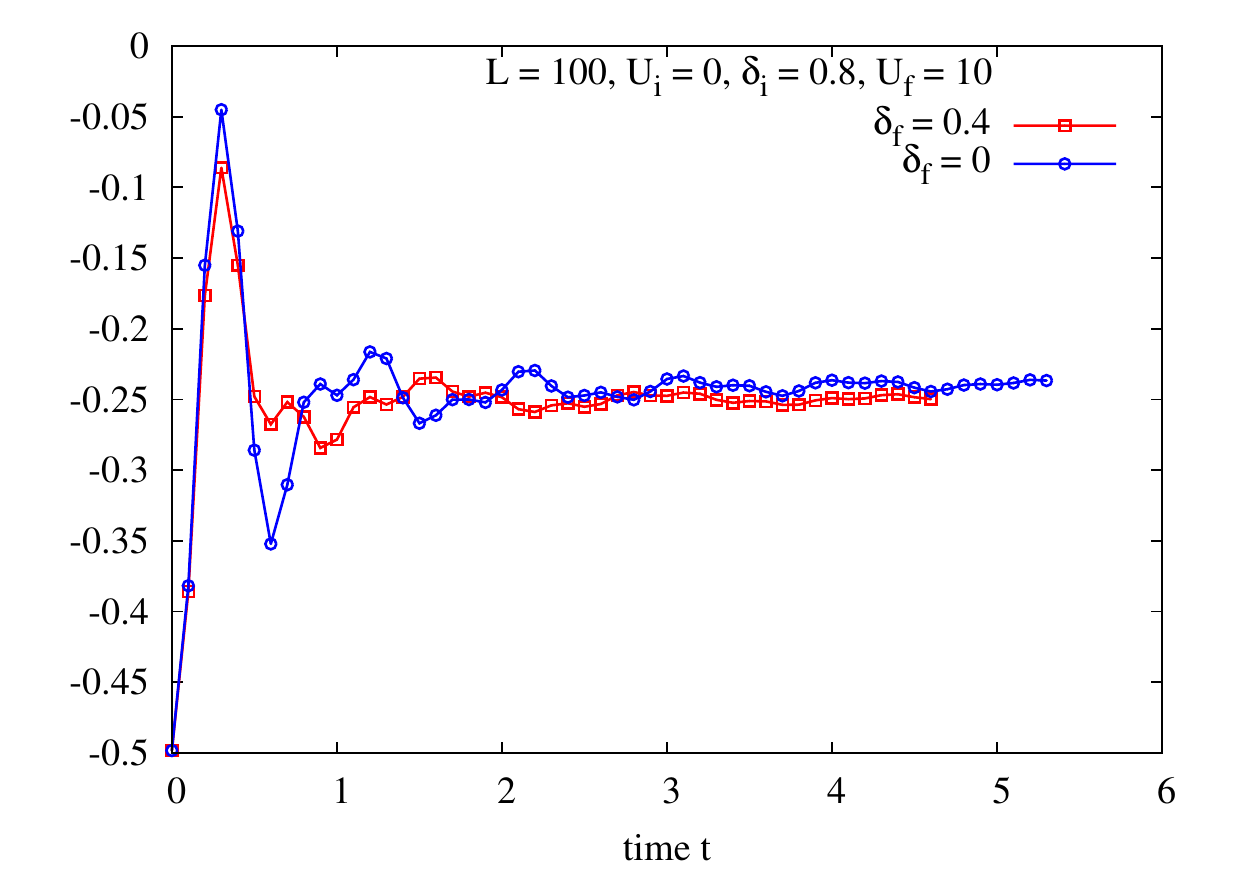}
\caption{(Color online) Comparison of t-DMRG results for the time evolution of $\mathcal G(L/2,L/2+1)$ for systems with $L=100$ sites for quenches with initial $U_i = 0, \, \delta_i = 0.8$ to values of $U_f = 10$ and $\delta_f = 0.4$ or $\delta_f = 0$, respectively. As can be seen, the expectation value for both cases is very similar.}
\label{fig:deltafzero}
\end{figure}
In Fig.~\ref{fig:deltafzero} we
show a comparison between quenches to $U_f \gg 1$ and $\delta_f = 0$ or
$\delta_f >0$, respectively. The correlator clearly approaches a
plateau, the value of which is only very weakly dependent on the
integrability-breaking parameter $\delta_f$, which supports our interpretation.
\item{} In the intermediate-$U$ regime there is no prethermalization
plateau, but the system relaxes slowly towards a Gibbs ensemble.
\end{enumerate}

%%%%%%%%%%%%%%%%%%%%%%%%%%%%%%%%%%%%%%%%%
\subsubsection{Initial-state dependence}
\begin{figure}
\includegraphics[width=0.48\textwidth]{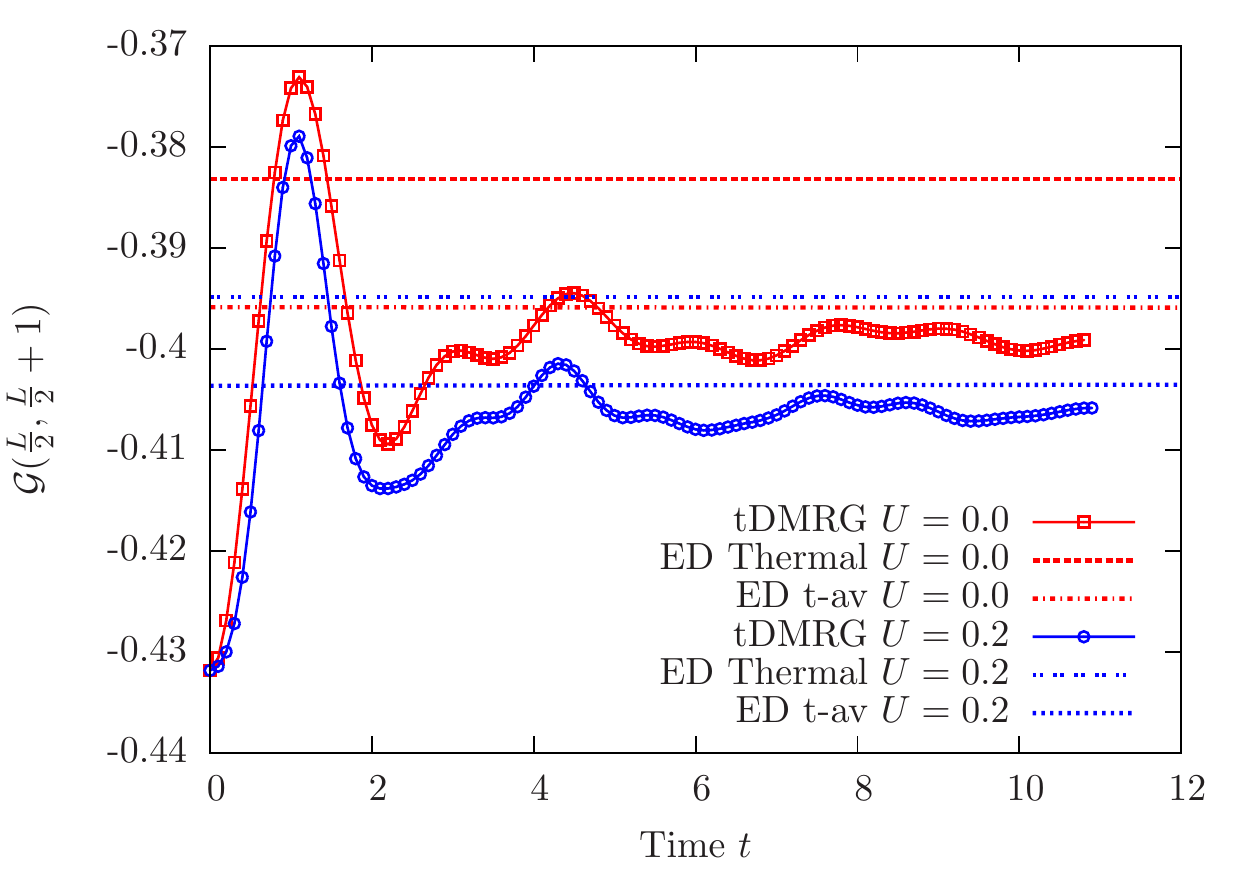}
\caption{(Color online)  Green's function results from t-DMRG and ED for the quench $\delta_i = 0.8 \to \delta_f = 0.4$ with $U_i = 5$ to
$U=0, 0.2$. As with Figs.~\ref{fig:U04_thermcomp}--\ref{fig:U10_thermcomp} we see that the time-averaged (t-av) ED
is compatible (up to finite size effects) with the t-DMRG plateau value, whilst the thermal expectation is not.}
\label{fig:initUstate1}
\end{figure}

\begin{figure}
\includegraphics[width=0.48\textwidth]{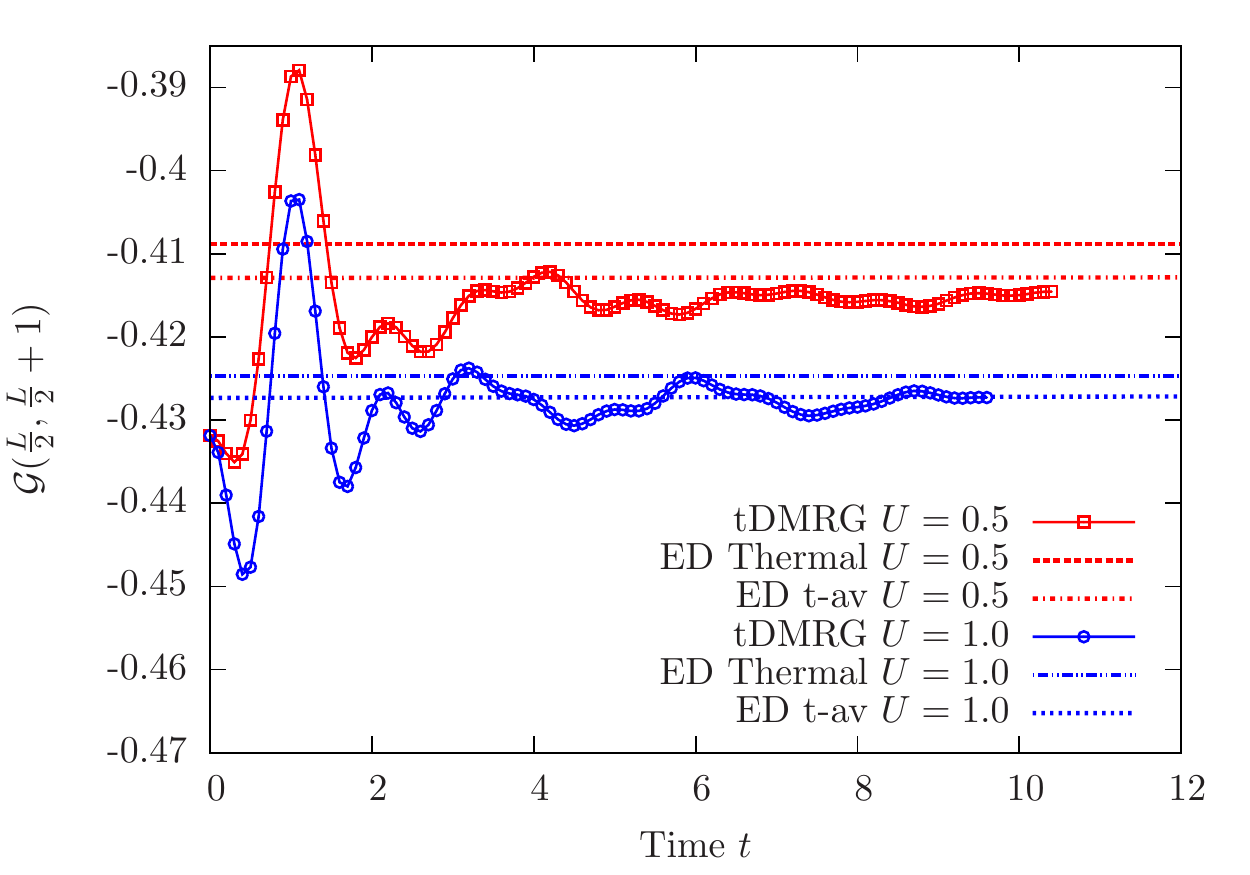}
\caption{(Color online) Green's function results from t-DMRG and ED for the quench $\delta_i = 0.8 \to \delta_f = 0.4$ with $U_i = 5$ to
 $U = 0.5, 1.0$. As with Figs.~\ref{fig:U04_thermcomp}--\ref{fig:U10_thermcomp} we see that the time-averaged (t-av) ED is compatible (up to finite size effects) with the t-DMRG plateau value.}
\label{fig:initUstate2}
\end{figure}

%%%%%%%%%%%%%%%%%%%%%%%%%%%%%%%%%%%%%%%%%
A final issue we would like to address is whether our findings
are sensitive to our particular choices of initial state. In order to
assess this question we have carried out t-DMRG computations for
quenches starting in the ground state of {\it strongly interacting} Peierls
insulators, i.e. Hamiltonians $H(\delta_i,U_i>0)$. Results for
quenches of the form
\be
(\delta_i=0.8,U_i=5)\longrightarrow(\delta_f=0.4,U_f)
\ee
with several values of $U_f$ are shown in
Figs.~\ref{fig:initUstate1}~\&~\ref{fig:initUstate2}. Here the
expectation values of both the diagonal and Gibbs ensembles have been computed
for $L=16$ site systems. Hence finite-size effects should be taken
into account when making comparisons to the t-DMRG data. 

The observed behavior is qualitatively very similar to that seen for
quenches starting in non-interacting ground states. Observables relax
to plateaus values that are incompatible with thermalization 
when $U_f$ is either small or large. 
%%%%%%%%%%%%%%%%%%%%%%%%%%%%
\section{Conclusions}
\label{Sec:Conc}
%%%%%%%%%%%%%%%%%%%%%%%%%%%%

Using a combination of anaytical calculations based on the
continuous unitary transform technique and time-dependent density
matrix renormalization group computations we have established the
existence of a robust prethermalization regime at intermediate times
after a quantum quench to the weakly nonintegrable interacting
Peierls insulator Hamiltonian~\fr{H0}. The combination of analytical
and numerical techniques we use to analyze this plateau enables us to
essentially eliminate finite-size effects. Our results thus represent
true ``bulk'' physics, and in particular are free from revival
effects. To the best of our knowledge, our work constitutes the first
one dimensional example of a robust prethermalization plateau in a
model with short-range interactions.

The CUT results allowed us to explicitly construct a ``deformed
generalized Gibbs ensemble'', which provides an approximate
statistical description of the prethermalization plateau. The deformed
GGE is constructed from charges ${\cal Q}_{\alpha}(k)$ cf
Eq.~\fr{newQ}, that form a mutually commuting set but {\it do not
  commute with the Hamiltonian}~\fr{Ham_Binf}. As such, the deformed
charges are not conserved at the operator level; only the expectation
values $\la{\cal Q}_{\alpha}(k)\ra$ with respect to the time-evolved
state $|\Psi(t)\ra$ are approximately conserved. Our construction is
therefore quite different from that of Ref.~\onlinecite{KollarPRB11}. 
We conjecture that the deformed GGE idea applies more widely
to quantum quenches in one-dimensional models with weak integrability
breaking. It would be interesting to test this conjecture for other
examples. 

We expect that at very late times the system will actually thermalize,
but we are not able to access sufficiently long times scales with
either the perturbative CUT approach or t-DMRG. A possible approach to
describe the dynamics at very late times might be through a quantum
Boltzmann equation (see, e.g., Refs.~\onlinecite{QuantumBoltz}). This
possibility is currently under investigation.

%%%%%%%%%%%%%%%%%%%%%%%%%%%%
\section*{Acknowledgements}
%%%%%%%%%%%%%%%%%%%%%%%%%%%%
We thank A. Chandran, M. Fagotti, M. Kolodrubetz, M. Rigol and S. Sondhi for
stimulating discussions. This work was supported by the EPSRC under
Grants No. EP/I032487/1 (F.H.L.E and N.J.R) and No. EP/J014885/1 (F.H.L.E). 
S. K. and S. R. M. acknowledge support through SFB 1073 of the Deutsche
Forschungsgemeinschaft (DFG). 

\appendix

%%%%%%%%%%%%%%%%%%%%%%%%%
\section{Local conservation laws for $H_0$}
\label{app:HCM}
%%%%%%%%%%%%%%%%%%%%%%%%%
To derive the local conservation laws for the non-interacting Hamiltonian $H(\delta,0)$
we follow Appendix C of Ref.~\onlinecite{FE_PRB13}. Below we give the local conservation laws 
and summarize the salient points of the derivation.

The Hamiltonian can be written in the form
\be
H_0 = \sum_{i,j=0}^{2L-1} a_i {\cal H}_{ij}a_j,\nonumber
\ee
where $a_i$ are Majorana fermions $\{ a_i, a_j \} = 2\delta_{i,j}$
defined by
\bea
a_{2n} &=& c\dg_n + c_n,\nn
a_{2n+1} &=& i(c\dg_n - c_n),\nonumber
\eea
and ${\cal H}$ is a skew-symmetric block-circulant matrix of the form
\bea
{\cal H} &=& \left[ \begin{array}{cccc} {\cal Y}_0 & {\cal Y}_1 & \ldots & {\cal Y}_{\tilde L-1} \\
{\cal Y}_{\tilde L-1} &{\cal Y}_0 & & \vdots \\
\vdots & & \ddots & \vdots \\
{\cal Y}_1 & \ldots & \ldots & {\cal Y}_0 \end{array} \right],\nonumber
\eea
where ${\cal Y}_n$ are $4\times4$ matrices with
${\cal Y}_n = - {\cal Y}_{\tilde L-n}^T$  and $\tilde L = L/2$. 
We define the Fourier transform of the block matrices as
\bea
\Big({\cal Y}_n\Big)_{jj'} &=& \frac{1}{\tilde L} \sum_{k=1}^{\tilde L} e^{\frac{2\pi i k}{\tilde L} n} \Big(Y_k\Big)_{jj'}
\nonumber
\eea
with $(Y_k)_{jn} = -(Y_{-k})_{nj}$. 

For free fermions a complete set of local conservation laws can be 
given by fermion bilinears
\be
I^{(r)} = \frac{1}{2}\sum_{l,n}a_l {\cal I}^{(r)}_{ln}a_n\ ,
\nonumber
\ee
where the matrices ${\cal I}^{(r)}$ must satisfy
\bea
\Big[ {\cal H},{\cal I}^{(r)}\Big] = 0
\quad{\rm and}\quad
\Big[ {\cal I}^{(r)}, {\cal I}^{(r')}\Big] = 0. \label{condition1}
\eea

The problem of deriving local conservation laws has now become the
problem of finding a set of mutually commuting matrices that also commutes
with the Hamiltonian matrix ${\cal H}$. At first sight the complexity of the problem
does not seem to have been reduced, but we can now utilize a useful property of
the Hamiltonian matrix ${\cal H}$: the projectors onto eigenvectors of block circulant 
matrices are themselves block circulant matrices. This means one can consider
${\cal I}^{(r)}$ that are block circulant:
\bea
{\cal I}^{(r)} = \left[ \begin{array}{cccc}
\bar{\cal Y}^{(r)}_0 & \bar{\cal Y}^{(r)}_1 & \ldots & \bar{\cal Y}^{(r)}_{\tilde L -1} \\
\bar{\cal Y}^{(r)}_{\tilde L -1} & \bar{\cal Y}^{(r)}_0 & & \vdots \\
\vdots & & \ddots &\vdots \\
\bar{\cal Y}^{(r)}_1 & \ldots & \ldots &  \bar{\cal Y}^{(r)}_0 \end{array}\right] \ .\nonumber
\eea
Imposing Eqs.~\fr{condition1}, we obtain the conditions (for all $k$)
\bea
\Big[Y_k , \bar Y_k^{(r)}\Big] = 0,\quad \Big[\bar Y_{k}^{(r)}, \bar Y_{k}^{(r')}\Big]=0,
\nonumber
\eea 
where $\bar Y_k^{(r)}$ is the Fourier transform of $\bar{\cal Y}^{(r)}$.

The construction of $\bar Y_k^{(r)}$ is straightforward as 
\be
Y_k = A_k \otimes \sigma^y, \nonumber
\ee
where
\bea
A_k &=& \Bigg[J(1+\delta)+J(1-\delta)\cos\bigg(\frac{2\pi k}{\tilde L}\bigg)\Bigg] \sigma^x\nn
&& -J(1-\delta)\sin\bigg(\frac{2\pi k}{\tilde L}\bigg)\sigma^y\ . \nonumber
\eea
So $\bar Y_k^{(r)}$ takes the form
\bea
\bar Y_k^{(r)} &=& \bar q_k^{(r)} A_k\otimes\sigma^y + q_k^{(r)} A_k \otimes \mathbb{1}_2\nn
&+& \bar\omega_k^{(r)} \mathbb{1}_2\otimes\sigma^y + \omega_k^{(r)}  \mathbb{1}_2\otimes\mathbb{1}_2\ ,\nonumber
\eea
where the functions $\omega_k^{(r)}$,  $\bar\omega_k^{(r)}$, $q_k^{(r)}$ and $\bar q_k^{(r)}$ are 
chosen such that the Fourier transform satisfies $(\bar Y_k)_{jn} = -(\bar Y_{-k})_{jn}$. 

The ambiguity in choice of functions leads to different representations of the conservation laws; 
following Ref.~\onlinecite{FE_PRB13} we make a particular choice that ensures there is a finite 
real-space range $r_0$ of the conservation laws: ${\cal I}^{(r)}_{ln} = 0$ for $|l-n|>r_0$.
We consider the conservation laws associated with each of the terms in $\bar Y_k^{(r)}$ 
separately and Fourier transforming back to real space we find
\begin{widetext}
\bea
I^{(r)}_1 &=& -\sum_{n=0}^{\tilde L-1}\frac{J}{2}(1+\delta)\bigg[ c\dg_{2n}c_{2n-2r+3} + c\dg_{2n}c_{2n+2r-1} + c\dg_{2n+1}c_{2n-2r+2} + c\dg_{2n+1}c_{2n+2r-2} + {\rm H.c} \bigg]\nn
&&- \sum_{n=0}^{\tilde L-1}\frac{J}{2}(1-\delta) \bigg[c\dg_{2n}c_{2n-2r+1} + c\dg_{2n}c_{2n+2r-3} + c\dg_{2n+1}c_{2n-2r+4} + c\dg_{2n+1}c_{2n+2r} + {\rm H.c.} \bigg],\nn
I^{(r)}_2 &=& -\sum_{n=0}^{\tilde L-1}\frac{J}{2}(1+\delta) \bigg[ i \Big(c\dg_{2n}c_{2n-2r+1}-c\dg_{2n}c_{2n+2r+1} + c\dg_{2n+1}c_{2n-2r}-c\dg_{2n+1}c_{2n+2r}\Big)+ {\rm H.c.} \bigg]\nn
&&- \sum_{n=0}^{\tilde L-1} \frac{J}{2}(1-\delta)\bigg[ i \Big( c\dg_{2n}c_{2n-2r-1}-c\dg_{2n}c_{2n+2r+1} + c\dg_{2n+1}c_{2n-2r+2} - c\dg_{2n+1}c_{2n+2r+2}\Big) +{\rm H.c.} \bigg], \nn
I_{3}^{(r)} &=& \sum_{n=0}^{\tilde L-1}\bigg[ i\Big( c\dg_{2n+2r+2}c_{2n} + c\dg_{2n+1}c_{2n+2r+3}\Big) + {\rm H.c.} \bigg],\nn
I^{(r)}_4 &=& \sum_{n=0}^{\tilde L-1}\bigg[ i\Big(c\dg_{2n+2r+2}c_{2n}-c\dg_{2n+1}c_{2n+2r+3} \Big)+{\rm H.c.}\bigg],\nonumber
\eea
\end{widetext}
where $r$ is a measure of the locality of the conservation laws
and takes values $1$ to $\tilde L$. 

The local conservation laws $I_{3}^{(r)}, I_{4}^{(r)}$ are independent of 
the microscopic parameters of the theory; they arise from the $\mathbb{1}_2\otimes\mathbb{1}_2$ 
and $\mathbb{1}_2\otimes\sigma^y$ terms in $\bar Y_k^{(r)}$. 
The remaining local conservation laws are dependent on the dimerization parameter $\delta$. Energy
conservation is also manifest in the set of local conservation laws with $I^{(1)}_1\propto H_0$. 

%%%%%%%%%%%%%%%%%%%%%%%
\section{Error estimate for the t-DMRG}
\label{appendix:tDMRGerrors}
\begin{figure*}
\begin{tabular}{lcl}
\includegraphics[width=0.45\textwidth]{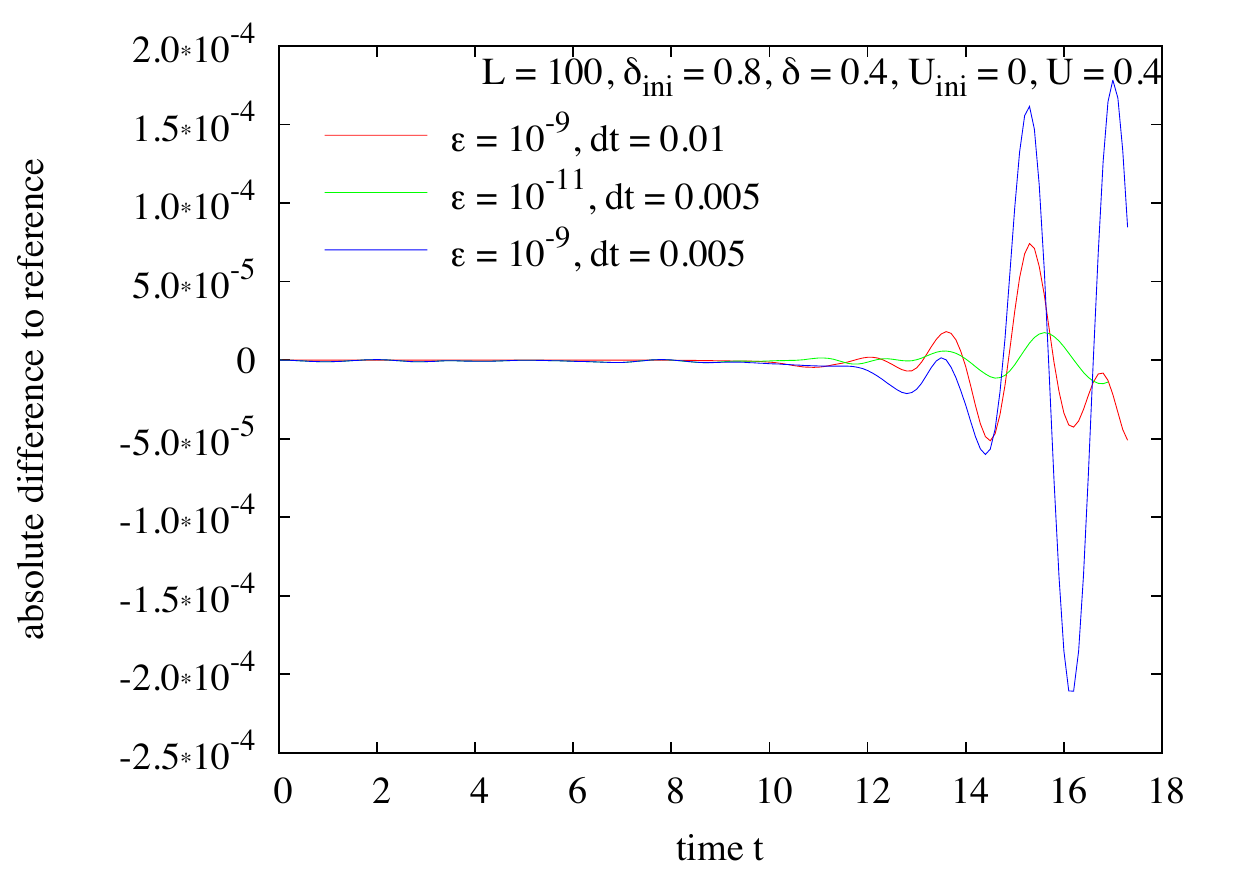} &
\includegraphics[width=0.45\textwidth]{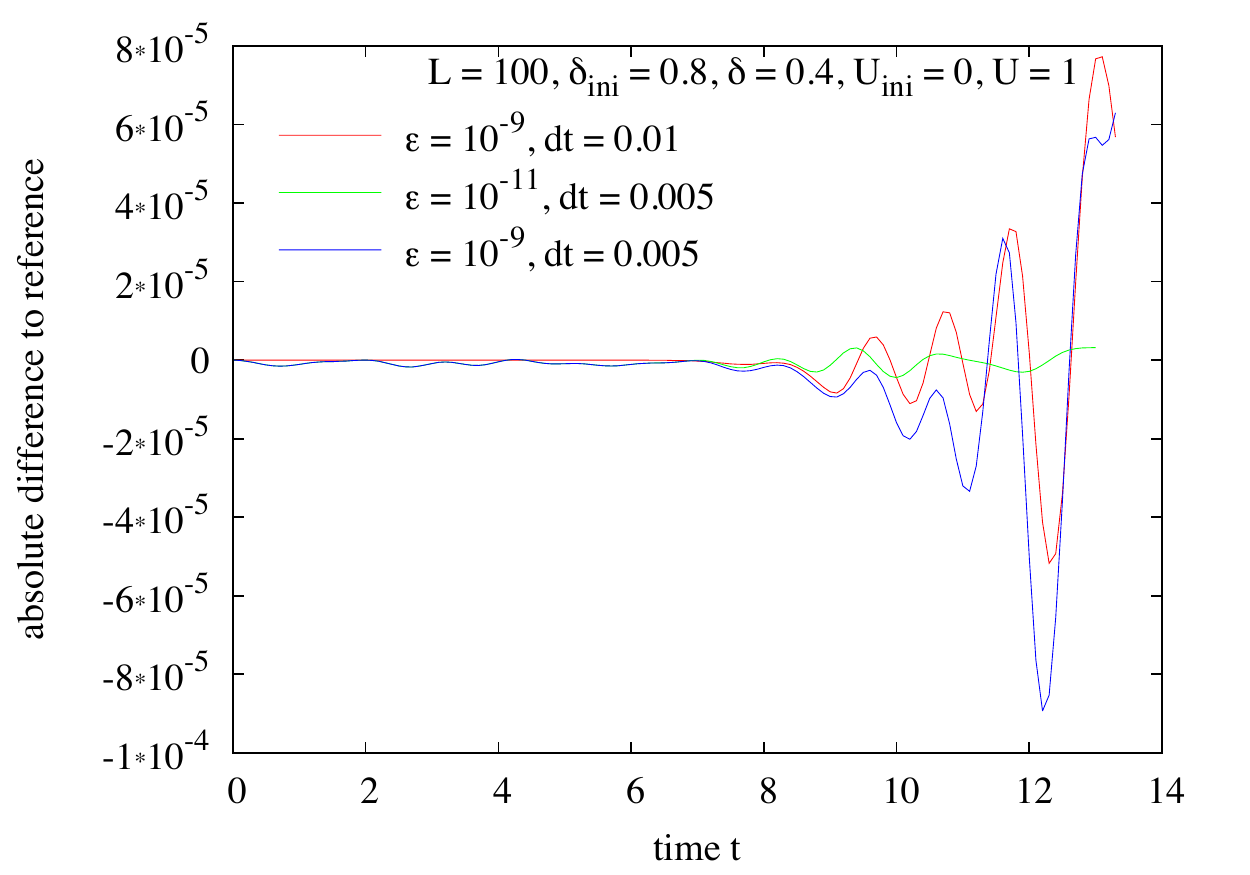} \\
\includegraphics[width=0.45\textwidth]{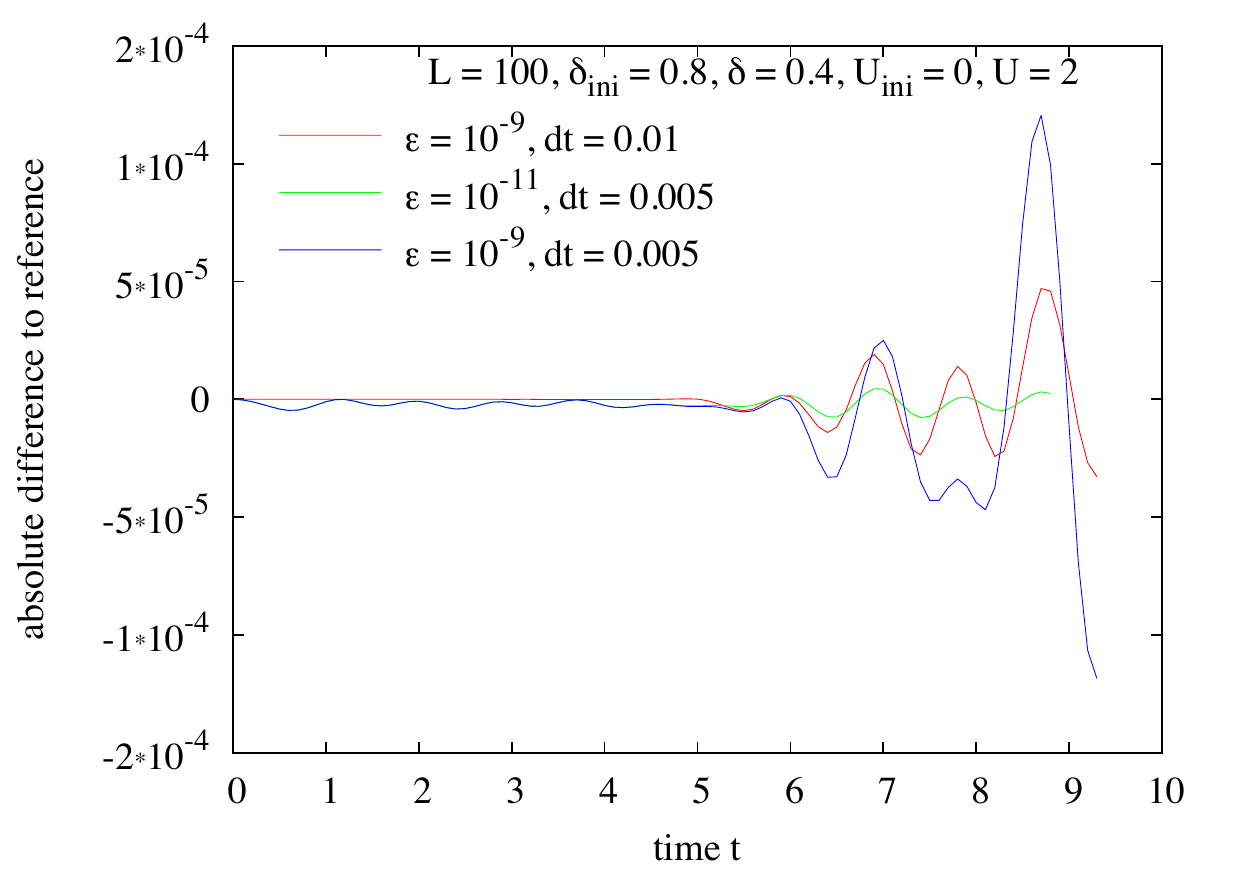} &
\includegraphics[width=0.45\textwidth]{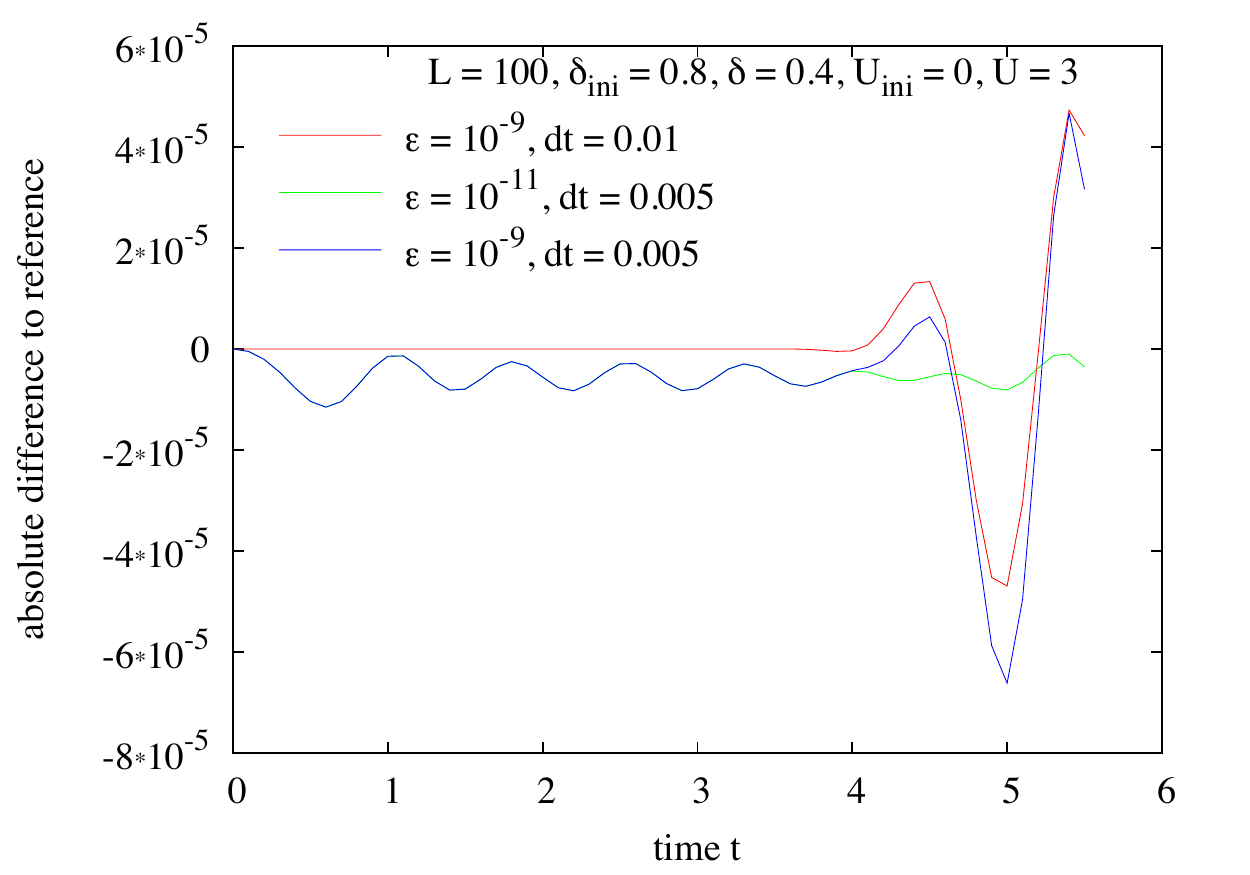} \\
\includegraphics[width=0.45\textwidth]{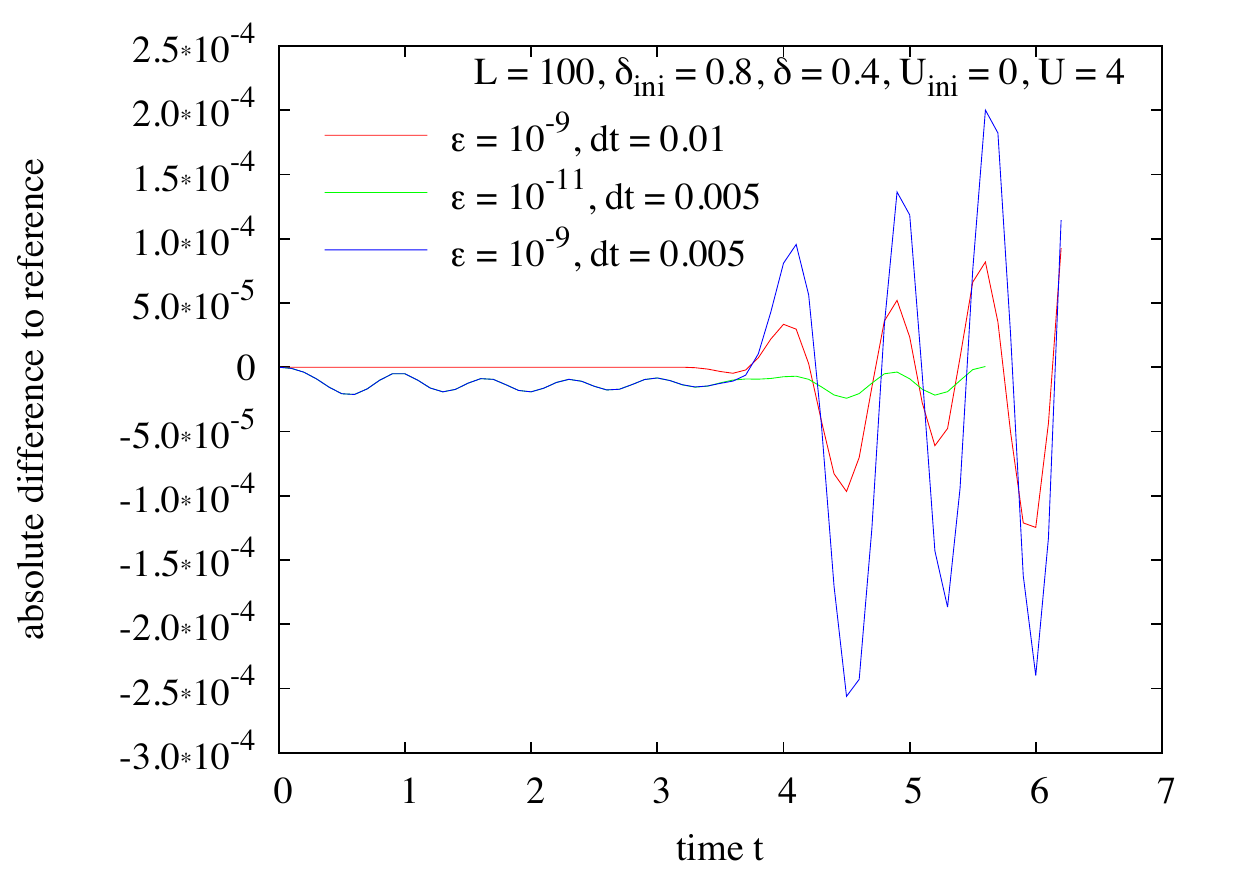} &
\end{tabular}
\caption{(Color online) Differences between runs with different parameters and different quenches ($L=100$ in all cases).}
\label{fig:tdmrgerrors}
\end{figure*}

\begin{figure}
\includegraphics[width=0.45\textwidth]{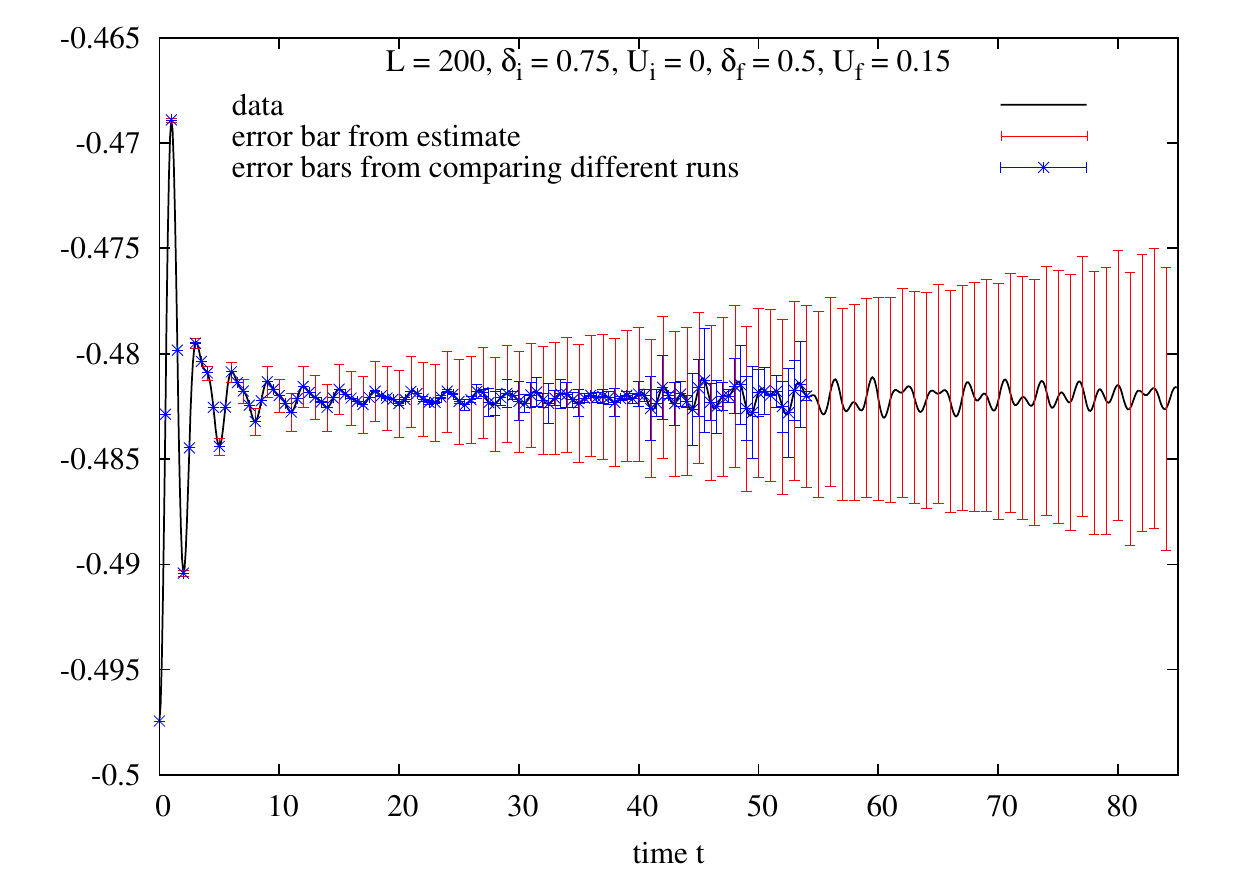}
\caption{(Color online) (Color online) Error estimates for t-DMRG results on the time evolution of $\mathcal G(L/2,L/2+1)$ for a system with $L=200$ sites and a quench $\delta_i = 0.75 \to \delta_f = 0.5$ and $U_i = 0 \to U_f = 0.5$. The data is obtained using a time step of $\delta t = 0.005$ and a target discarded weight of $\varepsilon = 10^{-9}$. The red error bars (lines) are obtained from the estimate discussed in the appendix, the blue ones (asterisks) are obtained by comparing to the results of a run with time step $\delta t = 0.01$. The error estimate appears to be larger, but of similar order of magnitude to the actual deviation between the results at times $\sim50$. From this estimate we obtain at the end of the time evolution a relative error of the order of 1.5\%.}
\label{fig:tdmrgerrors2}
\end{figure}

%%%%%%%%%%%%%%%%%%%%%%%
In this appendix, we estimate the error for the long time simulations. 
In principle, the error in a given observable can be estimated by the
discarded weight $\varepsilon$, and due to the variational nature of
the DMRG for {\it ground state} calculations, it is $\sim
\sqrt{\varepsilon}$ \cite{white2007}. At short times this provides a
reasonable estimate for time-evolved quantities as well. On longer time
scales a number of complications emerge.
1) Due to the entanglement growth, the discarded weight grows quickly
in time.\cite{gobert2005} This can be addressed by adjusting the number
of density matrix eigenstates, so that $\varepsilon$ is smaller than a
chosen threshold (in our case $10^{-9}$ or $10^{-11}$ for some
simulations, respectively).   
2) The error due to the Trotter decomposition becomes sizable.
3) Errors incurred in the sweeping procedure accumulate. In each DMRG
step, the change of basis needed during the sweeps introduces an error
$\sim \varepsilon$ as a result of the basis truncation. Hence, each
sweep introduces an error $\sim L \varepsilon$ for a system of size $L$.
This error is present at each time step. After a certain time $T$, a
simulation with a step size $dt$ leads to an error $\sim (T/dt) L\varepsilon$. 
This error is in addition to the error in the observable due to the density matrix truncation
discussed above.  At short times the error due to the basis truncation
$\sim \sqrt{\varepsilon}$ dominates, but at later times other error
sources can no longer be neglected. This can be seen by varying both the
target discarded weight and the time step. In
Fig.~\ref{fig:tdmrgerrors} we show the difference of runs with
different parameters to a reference run with  $\varepsilon = 10^{-11}$
and $dt = 0.01$. The error between the results with a target discarded
weight of $10^{-11}$ and $10^{-9}$ is seen to be roughly two orders of
magnitudes, as expected from the above estimate.  
The error bars shown in Figs.~\ref{fig:longtimes2}
and~\ref{fig:tdmrgerrors2} are estimated on the basis of the above
considerations. The error bars grow significantly towards the end of
the time evolution, but still permit us to make qualitative
statements. For the runs considered, this indicates that on the time
scales treated the quasistationary state does not change, i.e., the
prethermalization plateau is still present. Together with ED results
obtained for small systems for times up to $t=1000$, this indicates
that thermalization happens at much larger time scales ($\gg 100$), if
at all. 

%%%%%%%%%%%%%%%%%%%%%%%%%%%%%%%%%%%%%%%%%%%%%%%%%%%%%%%%%%%%%%%%%%%%%%%%%


\begin{thebibliography}{99}

%%%%%% Experiments %%%%%%%%%%
\bibitem{BlochNature02}
M. Greiner, O. Mandel, T. W. H\"ansch, and I. Bloch,
Nature {\bf 419}, 51 (2002).

\bibitem{WeissNature06}
T. Kinoshita, T. Wenger, D. S. Weiss, Nature {\bf 440}, 900 (2006); 
http://jila.colorado.edu/USJAPAN /pdf/Kinoshita.pdf.

\bibitem{SchmiedmayerNature07}
S. Hoerberth, I. Lesanovsky, B. Fischer, T. Schumm, and J. Schmiedmayer, 
Nature {\bf 449}, 324 (2007).

\bibitem{BlochNatPhys12}
 S. Trotzky Y.-A. Chen, A. Flesch, I. P. McCulloch, U. Sch\"ollwock, J. Eisert, and I. Bloch, 
Nature Physics {\bf 8}, 325 (2012).

\bibitem{KuhrNature12}
M. Cheneau, P. Barmettler, D. Poletti, M. Endres, P. Schauss, T. Fukuhara, C. Gross, I. Bloch, C. Kollath,
and S. Kuhr, 
Nature {\bf 481}, 484 (2012).

\bibitem{SchmiedmayerScience12} M. Gring, M. Kuhnert, T. Langen, T. Kitagawa, B. Rauer, M. Schreitl, I. Mazets, 
D. A. Smith, E. Demler, and J. Schmiedmayer, 
Science {\bf 337}, 1318 (2012). 

%%%%%%%%%%%%%%%%%%%%%%%%%%%%%%%%%%%%%%%%%%%%%%%%%%%

\bibitem{PSSV_RMP11}
A. Polkovnikov, K. Sengupta, A. Silva, and M. Vengalattore,
Rev. Mod. Phys. {\bf 83}, 863 (2011).


%%%%%%%%% GE in nonintegrable models %%%%%%%%%%%%%%%%%%%%%%%%%

\bibitem{nonint} J. M. Deutsch, Phys. Rev. A {\bf 43}, 2046 (1991);
  M. Srednicki, Phys. Rev. E {\bf 50}, 888 (1994); M. Rigol, V. Dunjko, and
M. Olshanii, Nature {\bf 452}, 854 (2008);
E. Canovi, D. Rossini, R. Fazio, G. Santoro, and A. Silva, 
New J. Phys. {\bf 14}, 095020 (2012).

%%%%%% GGE in integrable models %%%%%%%%%%%%%%%

\bibitem{BS_PRL08}
T. Barthel and U. Schollw\"ock, 
Phys. Rev. Lett. {\bf 100}, 100601 (2008).

\bibitem{CrEi}
M. Cramer, C.M. Dawson, J. Eisert, T.J. Osborne, Phys. Rev. Lett. {\bf
  100}, 030602 (2008);
M. Cramer, J. Eisert, New J. Phys. {\bf 12}, 055020 (2010).

\bibitem{CEF_PRL11} 
P. Calabrese, F.H.L. Essler, and M. Fagotti, Phys. Rev. Lett. {\bf
  106}, 227203 (2011); J. Stat. Mech. P07016 (2012);
J. Stat. Mech. P07022 (2012). 

\bibitem{FE_PRB13}
M.~Fagotti and F.H.L.~Essler, Phys. Rev. B {\bf 87}, 245107 (2013).

\bibitem{CC_JStatMech07} 
P. Calabrese and  J. Cardy, J. Stat. Mech. P06008 (2007).

\bibitem{IC_PRA09}
A. Iucci and M. A. Cazalilla, Phys. Rev. A {\bf 80}, 063619 (2009).

\bibitem{BKL_PRL10} 
G. Biroli, C. Kollath, and A.M. L\"auchli, Phys. Rev. Lett. {\bf 105},
250401 (2010). 

\bibitem{FM_NJPhys10}
D. Fioretto and G. Mussardo, New J. Phys. {\bf 12}, 055015 (2010). 

\bibitem{EEF:12}
F.H.L. Essler, S. Evangelisti and M. Fagotti,
Phys. Rev. Lett. {\bf 109}, 247206 (2012);

\bibitem{Pozsgay_JStatMech11} 
B.~Pozsgay, J. Stat. Mech. (2011) P01011.

\bibitem{CCR_PRL11}
A.~C. Cassidy, C.~W. Clark, and M.~Rigol.
Phys. Rev. Lett. {\bf 106}, 140405 (2011).

\bibitem{CIC_PRE11}
M. A. Cazalilla, A. Iucci, and M.-C. Chung, Phys. Rev. E {\bf 85}, 011133 (2012).

\bibitem{CK_PRL12}
J.-S. Caux and R.~M. Konik, Phys. Rev. Lett. {\bf 109}, 175301 (2012).

\bibitem{MC_NJPhys12}
J.~Mossel and J.-S. Caux, New J. Phys. {\bf 14}, 075006 (2012).

\bibitem{CE_PRL13} 
J.-S.~Caux and F.H.L.~Essler, Phys. Rev. Lett. {\bf 110}, 257203 (2013).

\bibitem{FE:13b}
M.~Fagotti and F.H.L. Essler, J. Stat. Mech. Theor. Exp. (2013) P07012.

\bibitem{Pozsgay:13a}
B. Pozsgay, J. Stat. Mech. Theor. Exp. P07003 (2013).

\bibitem{Mussardo_arXiv13} 
G.~Mussardo, arXiv:1304.7599 (2013).

\bibitem{CSC:13b}
M. Collura, S. Sotiriadis and P. Calabrese, 
Phys. Rev. Lett. {\bf 110}, 245301 (2013); J. Stat. Mech. (2013) P09025.

\bibitem{KCC:13a}
M. Kormos, M. Collura and P. Calabrese, Phys. Rev. A {\bf 89}, 013609 (2014).

\bibitem{KSCCI}
M. Kormos, A. Shashi, Y.-Z. Chou, J.-S. Caux and A. Imambekov,
Phys. Rev. B {\bf 88}, 205131 (2013).

\bibitem{GGE}
M. Rigol, V. Dunjko, V. Yurovsky,  and M. Olshanii,
Phys. Rev. Lett. {\bf 98}, 050405 (2007).

%%%%%%%%%%%%%%%%%%%%% integrability breaking %%%%%%%%%%%%%%%%%

\bibitem{Manmana0709} 
S. R. Manmana, S. Wessel, R. M. Noack, and A. Muramatsu, 
Phys. Rev. Lett. {\bf 98} 210405 (2007); Phys. Rev. B {\bf 79}, 155104 (2009);
C. Kollath, A. M. L\"auchli, and E. Altman; Phys. Rev. Lett. 98,
180601 (2007). 

\bibitem{Moeckel080910} 
M. Moeckel and S. Kehrein, Phys. Rev. Lett. {\bf 100}, 175702  (2008);
Ann. Phys. {\bf 324}, 2146 (2009); New J. Phys. {\bf 12}, 055016 (2010).

\bibitem{KollarPRB11}
M. Kollar, F. A. Wolf, and M. Eckstein,
Phys. Rev. B {\bf 84}, 054304 (2011).

\bibitem{thermalize}
M. Rigol, Phys. Rev. Lett. {\bf 103}, 100403 (2009); 
M. Rigol, Phys. Rev. A {\bf 80}, 053607 (2009);
L. F. Santos and M. Rigol, Phys. Rev. E {\bf 81}, 036206 (2010);
L. F. Santos and M. Rigol, Phys. Rev. E {\bf 82}, 031130 (2010).

\bibitem{MMGS_arXiv13}
M. Marcuzzi, J. Marino, A. Gambassi, and A. Silva,
Phys. Rev. Lett. {\bf 111}, 197203 (2013).

\bibitem{BCK_arXiv13}
G. Brandino, J.-S. Caux, and R. M. Konik,
arXiv:1301.0308 (2013). 

\bibitem{SchmiedmayerNJP11}
T. Kitagawa, A. Imambekov, J. Schmiedmayer and E. Demler,
New J. Phys. {\bf 13}, 073018 (2011).

\bibitem{SchmiedmayerEPJ13}
T. Langen, M. Gring, M. Kuhnert, B. Rauer, R. Geiger, D. A. Smith,
I. E. Mazets, J. Schmiedmayer 
Eur. Phys. J. Special Topics {\bf 217}, 43 (2013).

%%%%%%%%%%%%%%%%%%%%%%%%%%%%%%%%%%%%%%%%%%%%%%%%%%%%

\bibitem{SpinlessHubPeierls1}
V. Y. Krivnov and A. A. Ovchinnikov, Zh. Eksp. Teor. Fiz, {\bf 90},709 (1986); 
Sov. Phys. JETP {\bf 63} 414 (1986).

\bibitem{SpinlessHubPeierls2}
C. Schuster and U. Eckern, Eur. Phys. J. B {\bf 5}, 395 (1998).

\bibitem{SGM}
T. Nakano and H. Fukuyama, J. Phys. Soc. Jpn 50, 2489 (1981);
F.H.L. Essler and R.M. Konik, in ``From Fields to
  Strings: Circumnavigating Theoretical Physics'', ed. M. Shifman,
  A. Vainshtein, J. Wheater, World Scientific, Singapore (2005);
cond-mat/0412421.

\bibitem{normalization}
In practice we consider a very large system of size $L$ and take into
account $L$ conserved quantities. 

%%%%%%%%%%%%%%%%%%%%%%%%%%%%%%%%%%%%%%%%%%%%%%%
\bibitem{sirker13}
J. Sirker, N.P. Konstantinidis, F. Andraschko, N. Sedlmayr,
arXiv:1303.3064.

%%%%%%%%%%%%%%%%%%% CUTs %%%%%%%%%%%%%%%%%%%%%%

\bibitem{CUT} 
F. Wegner, 
Ann. Physik (Leipzig) {\bf 3}, 77  (1994); J. Phys. A {\bf 39} 8221 (2006);
S. D. Glazek and K. G. Wilson, 
Phys. Rev. D {\bf 48}, 5863 (1993); Phys. Rev. D {\bf 49}, 4214 (1994).

\bibitem{KehreinBook}
S. Kehrein, 
\textit{The Flow Equation Approach to Many-Particle Systems} (Springer, 2006).

\bibitem{CUT_NonEq}
S. Kehrein, 
Phys. Rev. Lett. {\bf 95}, 056602 (2005);
A. Hackl and S. Kehrein, 
Phys. Rev. B {\bf 78}, 092303 (2008).

%%%%%%%%%%%%%% ALPS / QMC %%%%%%%%%%%%%%%%%%%%%
\bibitem{ALPS}
B. Bauer {\textit et al}. (ALPS collaboration), J. Stat. Mech. P05001 (2011).

\bibitem{SSE}
A. W. Sandvik and J. Kurkij\"arvi, Phys. Rev. B {\bf 43}, 5950 (1991);
A. W. Sandvik, J. Phys. A {\bf 25}, 3667 (1992).

%%%%%%%%%%%%%%%%%%% Quantum Boltzmann %%%%%%%%%%%%%%%%%%%%%%

\bibitem{QuantumBoltz}
L. P. Kadanoff and G. Baym,
\textit{Quantum Statistical Mechanics}  (Benjamin, New York, 1962);
J. Rammer and H. Smith, Rev. Mod. Phys. {\bf 58}, 323 (1986);
M.L.R. F\"urst, C.B. Mendl and H. Spohn, Phys. Rev. E{\bf 86}, 031122
(2012);
M. Tavora and A. Mitra, Phys. Rev. B, 88, 115144 (2013).
%%%%%%%%%% Appendix Refs %%%%%%%%%%%%%%%

\bibitem{white2007}
S.R.~White and A.L.~Chernyshev,
Phys. Rev. Lett. {\bf 99}, 127004 (2007).

\bibitem{gobert2005}
D.~Gobert, C.~Kollath, U.~Schollw\"ock, and G.~Sch\"utz,
Phys. Rev. E {\bf 71}, 036102 (2005).

\end{thebibliography}
\end{document}